\documentclass[11pt,a4paper]{article}

\usepackage[affil-it]{authblk}
\usepackage[a4paper,textwidth=16.5cm,textheight=24cm,centering]{geometry}

\usepackage[]{amsmath}
	\numberwithin{equation}{section}
\usepackage{amsfonts}
\usepackage{amssymb,latexsym,cite}
\usepackage{amsfonts}
\usepackage{mathrsfs}
\usepackage{bbm}
\usepackage{bm}
\usepackage[T1]{fontenc}
\usepackage[matrix,arrow,color]{xy}

\def\ii{{\,{\rm i}\,}}

\newcommand{\ch}{{\rm ch}}

\def\e{{\,\rm e}\,}

\newcommand{\dd}{\mathrm{d}}

\newcommand{\IZ}{\mathbb{Z}}
\newcommand{\IC}{\mathbb{C}}
\newcommand{\IP}{\mathbb{P}}

\newcommand{\IR}{\mathbb{R}}
\newcommand{\IQ}{\mathbb{Q}}

\newcommand{\frg}{\mathfrak{g}}

\newcommand{\frt}{\mathfrak{t}}
\newcommand{\cI}{{\cal I}}
\newcommand{\cW}{{\cal W}}
\newcommand{\cN}{{\cal N}}

\newcommand{\cS}{{\cal S}}

\newcommand{\cO}{{\cal O}}

\newcommand{\cZ}{{\cal Z}}
\newcommand{\cL}{{\cal L}}

\newcommand{\cD}{{\cal D}}

\newcommand{\cG}{{\cal G}}
\newcommand{\PP}{\mathbb{P}}
\newcommand{\zed}{\mathbb{Z}}

\newcommand{\scrL}{\mathscr{L}}
\newcommand{\scrM}{\mathscr{M}}

\usepackage[utf8]{inputenc}
\usepackage[english]{babel}

\usepackage{hyperref}

\usepackage[pdftex]{graphicx}
\usepackage[usenames]{color}

\usepackage[toc,page]{appendix}
\usepackage[nottoc]{tocbibind}

\begin{document}


\thispagestyle{empty}
		\renewcommand*{\thefootnote}{\fnsymbol{footnote}}
	\title{Five-dimensional cohomological localization and \\ squashed 
         $q$-deformations of two-dimensional Yang-Mills theory}

	\author[a]{Leonardo Santilli\footnote{email: lsantilli@fc.ul.pt}}
	\author[b]{Richard J. Szabo\footnote{email: r.j.szabo@hw.ac.uk}}
	\author[c,a]{Miguel Tierz\footnote{email: mtpaz@iscte-iul.pt , tierz@fc.ul.pt}}
		
		\affil[a]{\small Grupo de F\'{\i}sica Matem\'{a}tica\\ Departamento de Matem\'{a}tica, Faculdade de Ci\^{e}ncias, Universidade~de~Lisboa, Campo Grande, Edif\'{\i}cio C6, 1749-016 Lisboa, Portugal.}
		\affil[b]{\small Department of Mathematics, Heriot-Watt University, Colin Maclaurin Building, Riccarton, Edinburgh EH14 4AS, United Kingdom; \textrm{and} Maxwell Institute for Mathematical Sciences, Bayes Centre, 47~Potterrow, Edinburgh EH8 9BT, United Kingdom; \textrm{and} Higgs Centre for Theoretical Physics, James~Clerk Maxwell Building, Kings Buildings, Edinburgh EH9 3JZ, United Kingdom.}
  		\affil[c]{\small Departamento de Matem\'{a}tica, ISCTE -- Instituto Universit\'{a}rio de Lisboa, Avenida das For\c{c}as Armadas, 1649-026 Lisboa, Portugal.}

{
	\maketitle
\thispagestyle{empty}

\begin{abstract}
\noindent
We revisit the duality between five-dimensional supersymmetric gauge theories and deformations of two-dimensional Yang-Mills theory from a new perspective. We give a unified treatment of supersymmetric gauge theories in three and five dimensions using cohomological localization techniques and the Atiyah-Singer index theorem. We survey various known results in a unified framework and provide simplified derivations of localization formulas, as well as various extensions including the case of irregular Seifert fibrations. We describe the reductions to four-dimensional gauge theories, and give an extensive description of the dual two-dimensional Yang-Mills theory when the three-dimensional part of the geometry is a squashed three-sphere, including its extension to non-zero area, and a detailed analysis of the resulting matrix model. The squashing parameter $b$ yields a further deformation of the usual $q$-deformation of two-dimensional Yang-Mills theory, which for rational values $b^2=p/s$ yields a new correspondence with Chern-Simons theory on lens spaces $L(p,s)$. 
\end{abstract}
}
	\clearpage
	{\thispagestyle{empty}
	\baselineskip=12pt
\tableofcontents
}
	\renewcommand*{\thefootnote}{\arabic{footnote}}
		\setcounter{footnote}{0}

	\clearpage
\setcounter{page}{1}

\section{Introduction}

Supersymmetric gauge theories in five and six dimensions have
undergone a surge of extensive and diverse investigations in recent
years. They have played critical roles in the understanding of the strong
coupling dynamics of quantum field theory, of M-theory where they can
be engineered, and of various problems in geometry. Their compactifications
generate many interesting theories in lower
dimensions, which can be used to elucidate novel features of
lower-dimensional quantum field theories from a geometric perspective.

Five-dimensional gauge theories with $\cN=1$ supersymmetry can be engineered from
compactifications of M-theory on Calabi-Yau
threefolds~\cite{Douglas:1996xp,Intriligator:1997pq,Bhardwaj:2019ngx}, which are related to the embedding
of topological string theory into M-theory~\cite{Gopakumar:1998jq}. In
six dimensions the worldvolume theory of coincident M5-branes is a
six-dimensional $\cN=(2,0)$ superconformal field
theory. Twisted compactification of this worldvolume theory on an
$n$-punctured Riemann surface $\Sigma_{h,n}$ of genus $h$ generically
engineers an $\cN=2$ superconformal field theory of class $\cS$ in four
dimensions~\cite{Gaiotto:2009we,Gaiotto2009}, which upon further twisted
compactification on a circle $S^1$ in the Schur
limit is conjecturally equivalent to two-dimensional $q$-deformed Yang-Mills theory
on $\Sigma_{h,n}$ in the zero area limit~\cite{Gadde:2011ik}. Because
of the lack of a Lagrangian description of this six-dimensional
theory, its twisted partition function is most easily computed by 
dimensionally reducing on a circle of radius $\beta$ and computing the
twisted partition function of five-dimensional $\cN=2$ supersymmetric
Yang-Mills theory with gauge coupling $g^2_{\textrm{\tiny YM}} =
2\pi\,\beta$~\cite{Douglas:2010iu,Lambert:2010iw}; the duality with two-dimensional
$q$-deformed Yang-Mills theory in this five-dimensional setting was checked by~\cite{Fukuda2012} using explicit supersymmetric localization techniques. 
From the four-dimensional perspective, the class $\cS$ theory is then equivalent to a three-dimensional theory on a certain squashed deformation $M_b$ of the three-dimensional part of
the compactification in six dimensions. In this paper we mostly focus on the round three-sphere $S^3$ and its squashed deformations~$S_b^3$. 

These dualities have by now been extensively discussed.
The purpose of the present paper is to survey and reinvestigate these
correspondences from a new and more detailed perspective; we give
extensive pointers to and comparisons with relevant previous works on the
subject as we go along. The main technique which we exploit in this
paper is cohomological localization, as pioneered by K\"all\'en for
three-dimensional theories in~\cite{Kallen2011}, and subsequently
extended to five dimensions by K\"all\'en, Qiu and Zabzine
in~\cite{Kallen2012,Kallen2012a}. This enables a unified treatment of
supersymmetric gauge theories in three and five dimensions using
solely topological techniques based on the Atiyah-Singer index
theorem, which extends and simplifies previous treatments based on
supersymmetric localization; this method is ultimately one of the main messages of the
present work. Compared to previous work on the subject,
we study full topologically twisted theories, instead of partially
twisted theories, with which we further extend general results in the
literature on cohomological localization. Our simplified treatment
also unavoidably comes with some limitations, and we shall extensively discuss
which backgrounds do not fit into our localization framework.

In three dimensions we provide different simplified derivations of
some known scattered results, treated in a unified framework. For example, the index theory calculations of localization on $S_b^3$ in
\cite{Drukker2012} uses a procedure which is different from that of
\cite{Kallen2011}, and their Lie derivative appearing in the square of the supercharge is not along
the Reeb vector field. We will explain this point thoroughly in this
paper.
When the supersymmetry transformations of \cite[Appendix
B]{Drukker2012} are put into the cohomological form of
\cite{Kallen2011}, index theorem calculations can be applied to obtain
the same results in a more economical way. Generally, our procedure of topological twisting rigorously justifies the localization calculations we employ at the field theory
level; in particular, it justifies the usage of the index formula and related techniques from~\cite{Kallen2012,Kallen2012a}.

In five dimensions the core of our work starts, where the reader
versed in localization techniques may begin. We adapt the formalism of~\cite{Kallen2012,Kallen2012a} to reobtain some known results in a different way
using a twisted gauge theory approach and also extend some general results (see
in particular our
localization formulas
\eqref{eq:ZvecgenM5reg} and \eqref{eq:ZvecM3xSigma}). 
In particular,
we rederive the results of \cite{Kawano2015,Willett2018}
through the Atiyah-Singer index theorem, first in the case of $S^3
\times \Sigma_{h,0}$ (which for
$h=0$ is a relatively straightforward adaptation
from the literature), and then
extending it to $S^3 _b \times \Sigma_{h,0}$. We also describe
the pushdown of these theories to the horizontal four-dimensional part
of the geometry, which was
justified in \cite{Willett2018} only for $h=0$, and is different from
\cite{Kallen2012a}. We study the resulting theory in detail and
describe its precise relation to $q$-deformed Yang-Mills theory on
$\Sigma_{h,0}$; we improve on various results in the literature (with
some overlap with \cite{Willett2018}), for example determining the
two-dimensional theory at non-zero area, and pay more attention to the
underlying matrix model, which has not been previously considered. For
non-trivial squashing parameters $b\neq1$, the resulting
two-dimensional gauge theory is new and we refer to it as a squashed
$q$-deformation of Yang-Mills theory on $\Sigma_{h,0}$. When
$b^2\in\IQ$, we show that
this theory is closely related to the $q$-deformations of Yang-Mills
theory considered in~\cite{Griguolo:2006kp} in a completely different setting, which in
turn is related to three-dimensional Chern-Simons theory on general lens spaces
$L(p,s)$. 

The outline of the remainder of this paper is as follows. We have
endeavoured throughout to give a relatively self-contained
presentation, while glossing over some well-known technical aspects for which we
refer to the pertinent literature; thus some of the earlier parts of this paper are somewhat expository in nature. We begin in Section~\ref{sec:prelim} by giving a more detailed introduction and background to the setting discussed briefly above, summarising the geometric settings and classifications, techniques and notation that will be used throughout this paper.

Section~\ref{sec:N=23d} is dedicated to the three-dimensional case, wherein we review the ideas behind K\"all\'en's cohomological localization technique, and explain how they are modified on squashed manifolds. We discuss the construction of the cohomological gauge theory, the computation of the one-loop determinants for the vector multiplet on the two possible classes of Seifert three-manifolds admitting $\cN=2$ supersymmetry, and the computation of the hypermultiplet one-loop determinants. We describe several explicit applications of our localization formulas.

Section~\ref{sec:N=15d} presents the five-dimensional analog of our considerations from Section~\ref{sec:N=23d}. We describe the cohomological field theory and derive the one-loop determinants of the vector multiplet in the two classes of Seifert five-manifolds admitting $\cN=1$ supersymmetry. We describe explicit applications of our localization formulas, and we briefly address the proof of Borel summability of the perturbative partition functions. We also explain in detail the relationship with a four-dimensional theory.

Section~\ref{sec:qYMfrom5d} explains the relation between our five-dimensional cohomological field theories and two-dimensional Yang-Mills theory. We derive the standard $q$-deformation of Yang-Mills theory on $\Sigma_{h,0}$ through a localization calculation on $S^3\times\Sigma_{h,0}$, and subsequently extend these considerations to the squashed deformations $S_b^3\times\Sigma_{h,0}$ where we obtain a new two-parameter deformation. We investigate the matrix model in detail and derive a new correspondence with Chern-Simons gauge theory on lens spaces $L(p,s)$. We conclude by briefly addressing how to obtain more general deformations of two-dimensional Yang-Mills theory through localization calculations in higher dimensions.

Three appendices at the end of the paper provide various technical details complementing some of the analysis in the main text. Appendix~\ref{app:notation} summarises our conventions and notation for spinors, which are adapted to treat the three-dimensional and five-dimensional cases, as well as the vector multiplets and matter hypermultiplets, in a unified way. Appendix~\ref{app:squashedspheres} provides mathematical details of the different types of squashings of spheres that preserve $\cN=2$ supersymmetry in three dimensions. 
Appendix~\ref{app:SE} provides some mathematical details on Sasaki-Einstein manifolds, which preserve $\cN=1$ supersymmetry in five dimensions, and we briefly review the formalism of \cite{Qiu2013}
to better explain why our results appear to be so different from those
of \cite{Qiu2013,Qiufact2013}.

\section{Preliminaries on superconformal field theories and localization}
\label{sec:prelim}

In this preliminary section we collect the relevant background
material that will be used throughout this paper. We begin with a
discussion of superconformal field theories in six dimensions, which
gives one of the primary motivations behind the present
investigations. We discuss how certain compactifications of these
theories on a Riemann surface suggest a duality between
four-dimensional superconformal theories and the standard
$q$-deformation of two-dimensional Yang-Mills theory. The purpose of
this paper is to investigate in detail how this duality is modified in
the case where the three-dimensional part of the compactification is a
squashed geometry, and we describe how cohomological localization
techniques for superconformal field theories on Seifert manifolds will
be applied to investigate the correspondence.

\subsection{Squashed geometry and six-dimensional superconformal field theories}
\label{sec:SCFT6d}

Consider the six-dimensional superconformal $(2,0)$ theory with gauge
group $G$ of ADE type on a twisted compactification of the form
\begin{equation}
M_6=S^1\times S^3\times \Sigma_h \ ,
\end{equation}
where $S^3$ is the standard round three-sphere and $\Sigma_h$ is a
compact oriented Riemann surface of genus $h$ without boundaries.
This setup dictates a remarkable duality: The correlators of a certain two-dimensional topological
quantum field theory on $\Sigma_h$ compute the partition function
of a four-dimensional $\cN=2$ field theory of class $\cS$ on $S^1\times S^3$, which is the
superconformal index~\cite{Gaiotto2012}
\begin{equation}
\cI(u,q,t)= \mathrm{Tr} \, (-1)^{\sf F}\, \Big(\frac{t}{u\,q}\Big)^{\sf r}\,
u^{{\sf J}_+}\, q^{{\sf J}_-}\, t^{\sf R}
 \ ,
\end{equation}
where ${\sf J}_\pm$ are the rotation generators in the two orthogonal planes
constructed from the Cartan generators of the Lorentz $SU(2)_{\rm L}\times
SU(2)_{\rm R}$ isometry group of $S^3$, the operator $\sf r$ is the $U(1)_r$
generator, and $\sf R$ is the $SU(2)_R$ generator of R-symmetries. The
superconformal index of four-dimensional $\cN=2$ superconformal field
theories was originally introduced in~\cite{Aharony:2003sx,Kinney:2005ej}, and it is a
highly non-trivial function of the three superconformal
fugacities $(u,q,t)$.

However, this duality is difficult to test because our current
understanding of the six-dimensional $(2,0)$ theory is rather
incomplete. Instead, we dimensionally reduce over $S^1$, which yields
five-dimensional supersymmetric Yang-Mills theory. The Yang-Mills
coupling in five dimensions has dimensions of length, $g^2 _{\textrm{\tiny YM}} =
2 \pi\, \beta$, where $\beta$ is the radius of $S^1$. 
The dimensional reduction over $S^1$ of the supersymmetric partition
function on $S^1\times S^3$ is achieved by assigning scaled chemical
potentials to the fugacities according to
\begin{equation}
q= \e^{2\pi\ii \beta \, \epsilon_1} \ , \qquad t= \e^{2\pi\ii\beta\,
  \epsilon_2} \qquad \mbox{and} \qquad u= \e^{2\pi\ii\beta} \ ,
\end{equation}
and taking the limit
$\beta\to 0$. Then the
four-dimensional index becomes a three-dimensional ellipsoidal
partition 
function~\cite{Dolan2011,Gadde2011red,Imamura2011}, i.e. the partition function on the squashed sphere $S_b^3$
with squashing parameter
\begin{equation}
b=\sqrt{\frac{\epsilon_1}{\epsilon_2}} \ .
\end{equation}
This deformation of the three-sphere of radius $r=\sqrt{\epsilon_1\, \epsilon_2}$ can be parametrized by the
ellipsoid in $\IC^2$ defined by
\begin{equation}
b^2\, |z_1|^2+ b^{-2}\, |z_2|^2 = r^2 \ ,
\end{equation}
which has isometry group $U(1)\times U(1)$.
In this setting the deformation parameters $(\epsilon_1,\epsilon_2)$
can also be interpreted as equivariant parameters
of the $\Omega$-background~\cite{Nekrasov2009}, which are the holonomies of the
twisted M-theory compactification defined by an $S^1$-bundle over $
M_6\times {\sf TN}$ realised as the quotient of
$\IR\times M_6\times {\sf TN}$ by the $\IZ$-action
\begin{equation}
n\cdot (\tau,x,z_1,z_2) = (\tau+2\pi\,\beta\,n,x,q^n\, z_1,t^{-n}\, z_2) \ ,
\end{equation}
where $n\in\IZ$, $\tau\in\IR$, $x\in M_6$ and $(z_1,z_2)\in\IC^2$ are
local coordinates on the Taub-NUT space ${\sf TN}$.

Hence we consider supersymmetric Yang-Mills theory on the five-manifold
\begin{equation}
M_5=S_b^3\times \Sigma_h \ .
\end{equation}
This theory sits in the infrared of a renormalization group flow
triggered by a relevant deformation of a five-dimensional
superconformal field theory in the ultraviolet~\cite{Seiberg1996}. For the ellipsoid
$S_b^3$ \cite{Hama2011} described above, the five-dimensional theory has four
supercharges and can alternatively be taken to directly descend from the
$\cN=(1,0)$ superconformal field theory on $S^1\times
S_b^3\times\Sigma_h$, with only one chiral Killing spinor in six
dimensions. On the other hand, from the present perspective it is more
natural to use the squashed sphere $S_b^3$ of~\cite{Imamura2011a} (see
Appendix \ref{app:squashedspheres} for details), for which the
five-dimensional theory descends from twisted compactification of the
$\cN=(2,0)$ superconformal field theory on $S^1\times
S^3\times\Sigma_h$ with round $S^3$. The geometric meaning of the squashing parameter $b$ is different in the two cases: $b>0$ for the ellipsoid, while $b \in \mathbb{C}$ with $\vert b \vert =1$ for the squashed sphere. From the point of view of the matrix ensemble we will find
in Section~\ref{app:mmqym}, the natural choice would be $b>0$. In
Section~\ref{sec:qYMfrom5d} we consider both cases.
We will find that, regardless of what choice we make for $S^3 _b$, the
partition function is the same and is a holomorphic function of $b$ in
the punctured complex plane $\mathbb{C} \setminus \left\{ 0 \right\}$, so we may start with either the squashed sphere or the ellipsoid and then analytically continue the result.

\subsection{Reductions to two and three dimensions}  \label{genasp}

By a localization calculation over the squashed sphere $S_b^3$, we
will identify the partition function of the two-dimensional gauge
theory dual on $\Sigma_h$ of the $\cN=2$ theory of class $\cS$. In
this paper we are exclusively interested in the slice of the
superconformal fugacity space defined by the Schur limit $u=0$,
$q=t$ of the superconformal index which is the Schur index
\begin{align}
\cI(q)= \mathrm{Tr} \, (-1)^{\sf F}\, q^{{\sf J}_-+{\sf R}}
 \ ,
\end{align}
where the trace is now restricted to states with $U(1)_r$ charge
${\sf r}={\sf J}_+$. In this case the index was computed in~\cite{Gadde:2011ik,Gadde2011} from a topological
quantum field theory on the Riemann surface $\Sigma_h$, which can be
identified with the
zero area limit ${\rm vol}(\Sigma_h)=0$ of the usual $q$-deformed two-dimensional Yang-Mills
theory; the duality with this two-dimensional gauge theory is
confirmed in~\cite{Fukuda2012} by an explicit localization computation on
$S^3\times\Sigma_h$ (i.e. for $b=1$). The $q$-deformed Yang-Mills theory is not topological when
${\rm vol}(\Sigma_h)\neq0$, but it still has a natural class $\cS$
theory interpretation as the supersymmetric partition function of the
$(2,0)$ theory on $S^1\times S^3\times\Sigma_h$ where the area of the
ultraviolet curve $\Sigma_h$ is kept finite~\cite{Tachikawa2012}. We shall
study this proposal explicitly via a localization calculation on the
five-manifold $S_b^3\times \Sigma_h$. The two-dimensional theory we
find is a further deformation by the squashing parameter $b$, that we
call `squashed' $q$-deformed Yang-Mills theory. For later use, let us
now briefly
review the standard $q$-deformation of two-dimensional Yang-Mills theory.

Let $\frg$ be the Lie algebra of a connected Lie group $G$. Let $\triangle$ be the root system of $\frg$ and $\triangle_+$ the system of
positive roots; similarly let $\Lambda\cong\IZ^{{\rm rank}(G)}$ be the weight lattice of
$\frg$ with dominant weights
$\Lambda_+$. We fix an invariant bilinear form $(-,-)$
on $\frg$, usually the Killing form. Let
\begin{align}
\delta=\frac12\, \sum_{\alpha\in \triangle_+}\, \alpha
\end{align}
be the Weyl vector of $\frg$. The vector $2\,\delta$ is always a
weight of $\frg$; if $G$ is semi-simple, then $\delta$ is also a
weight and we can identify it with an integer vector
$\delta\in\IZ^{{\rm rank}(G)}$.

The partition function for the $q$-deformation of 
Yang-Mills theory with gauge group $G$ on a closed oriented Riemann surface $\Sigma_h$ of
genus $h\geq0$ can be written as a
generalization of the Migdal heat kernel expansion given by~\cite{Aganagic:2004js}
\begin{align}
Z_{h,p}(q) = \sum_{\lambda\in\Lambda_+} \, \dim_{q}(R_\lambda)^{2-2h} \ q^{\frac p2\,
  (\lambda+2\,\delta,\lambda)} \ ,
\label{ZqtQpR}\end{align}
where $p\in\IZ$ is a discrete parameter and the sum runs over all
isomorphism classes of irreducible unitary representations
$R_\lambda$ of
$G$ which are parametrized by dominant weights $\lambda\in\Lambda_+$.
The deformation parameter
\begin{align}
q=\e^{-g_{\rm str}} 
\end{align}
is identified with the coupling constant $g_{\rm str}$ in topological string theory. 
The quantum dimension of the representation $R_\lambda$ labelled by $\lambda\in\Lambda_+$ is
\begin{align}
\dim_{q}(R_\lambda)= \prod_{\alpha\in \triangle_+}\,
\frac{\big[(\lambda+\delta,\alpha)\big]_q}{\big[(\delta,\alpha)
  \big]_q}
\ ,
\label{qtdim}\end{align}
where
\begin{align}
[x]_q=\frac{q^{x/2}-q^{-x/2}}{q-q^{-1}}
\label{qnumber}\end{align}
for $x\in\IR$ is a $q$-number. This theory is closely related to Chern-Simons theory on a principal $U(1)$-bundle of degree~$p$ over $\Sigma_h$~\cite{Blau2006}.

For many computations it is useful to have an explicit expression for the
partition function \eqref{ZqtQpR} in terms of highest weight variables. For this, we
define shifted weights $\vec k\in\IZ^{{\rm rank}(G)}$ by
\begin{align}
\vec k=\lambda+\delta \ ,
\label{nRbetarho}\end{align}
and use the Weyl reflection
symmetry of the summand of the
partition function \eqref{ZqtQpR} to remove the restriction to the fundamental chamber of the summation over
$\vec k$. Up
to overall normalization, the partition function \eqref{ZqtQpR} can thus be written
as
\begin{align}
Z_{h,p}(q)= \sum_{\vec k\in(\IZ^{{\rm rank}(G)})_{\rm reg}} \,
\Delta(g_{\rm str}\, \vec k\,)^{2-2h} \, \e^{-\frac{p\,g_{\rm str}}2\,
  (\vec k,\vec k\,)} \ ,
\label{ZqtQpn}\end{align}
where the Weyl determinant is given by
\begin{align}
\Delta(\vec x\,)= \prod_{\alpha\in \triangle_+}\, 2\sinh
\frac{(\alpha,\vec x\,)}2
\end{align}
for $\vec x=(x_1,\dots,x_{{\rm rank}(G)})\in\IC^{{\rm rank}(G)}$. The
sum in \eqref{ZqtQpn} is restricted to those shifted weights
where $\Delta(g_{\rm str}\,\vec k\,)$ is non-zero,
i.e. $(\alpha,\vec k\,) \neq 0$ for $\alpha\in \triangle_+$.

When the gauge group is the unitary group $G=U(N)$, this
two-dimensional gauge theory is conjecturally a non-perturbative
completion of topological string theory on the local Calabi-Yau
threefold which is the total space of the rank~$2$ holomorphic vector
bundle
\begin{align}
\cO_{\Sigma_h}(p+2h-2)\oplus\cO_{\Sigma_h}(-p)\longrightarrow\Sigma_h
  \ ,
\end{align}
with $N$ D4-branes wrapping the exceptional divisor
$\cO_{\Sigma_h}(-p)$ and D2-branes wrapping the base $\Sigma_h$~\cite{Aganagic:2004js}. In
turn, for $h=0$ the two-dimensional theory defines an analytical continuation of
Chern-Simons gauge theory on the lens space $L(p,1)$ to the case where
$q$ is not a root of unity~\cite{Caporaso2005}. In Section~\ref{sec:qYMfrom5d} we shall
find that five-dimensional cohomological localization over
$S_b^3\times\Sigma_h$ gives a squashed deformation of this theory at
$p=1$, which for genus $h=0$ and rational values $p/s$ of the squashing parameter
$b^2$ is an analytical continuation of Chern-Simons theory on the more
general lens spaces $L(p,s)\cong S_b^3$.

The correspondence with three-dimensional field theories has also been
studied from other perspectives. 
In \cite{Razamat2019}, the dimensional reduction of six-dimensional
theories on $S^1\times S^3 _b \times \Sigma_h$ is considered by
reducing to five dimensions as above, compactifying on $\Sigma_h$, and
then obtaining a three-dimensional theory on the squashed sphere; this
enables a comparison of the two possible compactification paths:
first along $S^1$ and then on $\Sigma_h$ to obtain a theory on $S^3
_b$, or first along $\Sigma_h$ to obtain a four-dimensional theory of
class $\mathcal{S}$ and then relating it by standard reasoning to the
three-dimensional partition function. In
\cite{Yagi2013,Lee2013,Gang2015} the five-dimensional theory on $M_3
\times S^2$ is obtained from dimensional reduction of the
six-dimensional superconformal field theory on $S^1 \times M_3 \times
S^2$ for more general three-manifolds $M_3$. The resulting theory on
$M_3$ is related to complex Chern-Simons theory. In these cases the
theory is partially twisted along $M_3$, and supersymmetric
localization on $S^2$ is used to reduce to a twisted three-dimensional
theory; this differs from the perspective of~\cite{Fukuda2012,Kawano2015}, where the partial twist is along
$\Sigma_h$ and localization over $S^3$ reduces the theory to two
dimensions. In contrast, in this paper we will consider the fully
twisted theory on $S_b^3\times\Sigma_h$. Finally, in~\cite{Cordova2013}, with squashed sphere
of~\cite{Imamura2011a}, the six-dimensional theory on $S^3 _b \times
M_3$ is reduced along the Hopf fibre of $S^3
_b$, then twisted along $M_3$ and localized to reduce
along $S^2$; the resulting three-dimensional theory is the same as in~\cite{Yagi2013,Lee2013}.

\subsection{Basics of cohomological localization}

In this paper we use techniques based on localization theorems in
equivariant cohomology, applied to supersymmetric quantum field theory, see e.g.~\cite{Szabo:2000fs,Pestun:2016jze} for introductions to the subject; we will now briefly sketch the main ideas that
will be used extensively in the remainder of this
paper. Supersymmetric localization is a technique which allows the
reduction of a supersymmetry preserving Euclidean path integral to an
integral over the smaller set of fixed points of a supercharge $\sf
Q$. To compute the partition function, one adds a $\sf Q$-exact term
${\sf Q}V$ to the action $S$ of the theory and computes the deformed
partition function $Z(t)$ defined by functional integration of the
Boltzmann weight $\e^{-S-t\,{\sf Q}V}$ for $t\in\IR$; then $Z(0)$ is
the original partition function we wish to compute. Supersymmetry of the path
integral then implies that $Z(t)$ is formally independent of the
parameter $t$, so that letting $t\to\infty$ and choosing the
localizing term $V$ to be positive semi-definite, the functional
integration reduces to a localization calculation around the fixed
points ${\sf Q}V=0$ of the supercharge $\sf Q$. In
general the set of fixed points is a superspace, with odd coordinates associated with
supersymmetric fermionic modes that have vanishing action in the
localizing term in the bosonic background.

In this paper we consider Euclidean manifolds preserving rigid
supersymmetry with four and eight supercharges; we restrict them to
non-trivial circle bundles $M_{2n+1}\to K_{2n}$, whose total spaces
have odd dimension $2n+1$, in order to avoid
dealing with fermionic fixed points. We shall derive the fixed point
loci based on cohomological forms of the (BRST) supersymmetry
transformations which are compatible with the $U(1)$-action on the
circle bundle~\cite{Kallen2011,Ohta2012}; this procedure is called
topological twisting and the resulting theory is called a
cohomological field theory (in the sense of equivariant cohomology). We shall also only consider 
localization on the Coulomb branch of the supersymmetric gauge theory,
where the path integral is reduced to a finite-dimensional integral
over a classical moduli space parameterized by scalars in
vector multiplets, and holonomies and fluxes of gauge fields around
non-contractible cycles in $M_{2n+1}$ (possibly together with other
continuous moduli). 

The localization calculation amounts to computing a ratio of one-loop
fluctuation determinants which is schematically given by~\cite{Pestun:2007rz}
\begin{align}
h(\phi) = \frac{\det\ii{\sf L}_\phi\vert_{{\rm coker}\,{\sf D}}}{\det\ii{\sf
  L}_\phi\vert_{\ker{\sf  D}}} \ ,
\end{align}
where $\sf D$ denotes differential operators entering the localizing terms
$V$, and ${\sf L}_\phi=-\ii{\sf Q}^2$ generates the geometric
$U(1)$-action and gauge transformations parametrized by $\phi$ on the fields
of the theory; the adjoint scalar $\phi$ is $\sf Q$-closed and does
not have a fermionic partner. Then the equivariant cohomology in the localization of the supersymmetric
gauge theory consists of gauge-invariant
states on the base space $K_{2n}$ of the circle bundle, together with
an infinite tower of 
Kaluza-Klein modes on the $S^1$ fibre. The effective contribution to this ratio
from the zero modes which remain after cancellation between fermionic and bosonic states is computed using the Atiyah-Singer index
theorem for transversally elliptic operators and the Atiyah-Bott
localization formula in equivariant cohomology for the $U(1)$-action
on $M_{2n+1}$. The schematic form of the localized partition function is then
given by
\begin{align}
Z(M_{2n+1}) = \int_\frg\, \dd\sigma \ \int_{\scrM_G ^{\textrm{\tiny BPS}}(M_{2n+1})}\,
  \dd\boldsymbol m \ \e^{-S_{\rm cl}(\boldsymbol m;\sigma)} \ Z_{\rm
  vec}(M_{2n+1}) \, Z_{\rm hyp}(M_{2n+1}) \ ,
\end{align}
where $\frg$ is the Lie algebra of the gauge
group $G$, and $\scrM_G  ^{\textrm{\tiny BPS}} (M_{2n+1})$ is the BPS locus inside a moduli space of $G$-connections on
$M_{2n+1}$ parametrized by moduli $\boldsymbol m$. The action $S_{\rm
  cl}$ is the classical bosonic action, while $Z_{\rm vec}$ and
$Z_{\rm hyp}$ are respectively the one-loop fluctuation determinants associated
with the vector multiplet and the matter hypermultiplets of the
supersymmetric gauge theory.

\subsection{Localization of $\cN=1$ gauge theories on Seifert manifolds}

In this paper we mainly focus on Seifert manifolds which admit a free
$U(1)$ action, so that they admit a $U(1)$ isometry. We shall comment
where appropriate on the extension to more general principal
$U(1)$-bundles over orbifolds, where the $U(1)$ action has fixed
points. 

\subsubsection*{Geometric setup}
\label{sec:setup}

Let $M_{2n+1}\xrightarrow{ \ \pi \ } K_{2n}$ be a circle bundle of
degree $p$ over a compact
K\"ahler manifold $(K_{2n},\omega)$ of real dimension $2n$ with
$[\omega]\in H^2(K_{2n},\IZ)$. The almost contact structure $\kappa\neq0$ on $M_{2n+1}$ can be chosen to be a connection one-form on this bundle which is locally written as
\begin{equation}\label{eq:kappa}
\kappa=\dd\theta+p\,\pi^*(a) \ ,
\end{equation}
where $\theta\in[0,2\pi\, r)$ is a local coordinate of the $S^1$ fibre and $a$ is a local symplectic potential for $\omega=\dd a$. Then
\begin{equation}
\label{eq:dkapparel}
\dd\kappa=p\,\pi^*(\omega) \ .
\end{equation}
In contrast to \cite{Kallen2012,Kallen2012a}, we will not assume $\kappa$ to be a K-contact structure on $M_{2n+1}$. Instead, our interest will mainly focus on product manifolds $M_{2n+1}=M_{2n-1} \times \Sigma_h$, where $M_{2n-1}$ is a compact contact manifold, but in general not K-contact, and $\Sigma_h$ is a compact Riemann surface of genus $h$. 
Then $\omega = \omega_{K_{2n-2}} + \omega_{\Sigma_h}$ is the
sum of the symplectic forms on the base $K_{2n-2}$ of the Seifert
fibration of $M_{2n-1}$ and $\Sigma_h$, and $\kappa\wedge(\dd\kappa)^{\wedge (n-1)}\neq0$ is proportional to the
volume form on $M_{2n-1}$ induced by a metric compatible with the contact structure. The canonical volume form on the total space $M_{2n+1}$ is 
\begin{equation}
\label{eq:volformcontact}
\dd\Omega_{M_{2n+1}} = \frac{(-1)^{n}}{2^{n-1}\,(n-1)!} \ \kappa\wedge(\dd\kappa)^{\wedge (n-1)} \wedge \omega_{\Sigma_h} \ .
\end{equation}

The Reeb vector
field $\xi$ is defined by the duality contraction
\begin{align}
\xi\,\llcorner\,
\kappa=1
\end{align}
and the invariance condition
\begin{align}
\cL_\xi\kappa=\xi\,\llcorner\,
\dd\kappa=0 \ , 
\end{align}
where $\mathcal{L}_\xi=\dd\,\xi\,\llcorner\,+\,\xi\,\llcorner\,\dd$ is
the Lie derivative along $\xi$. It is the generator of the
$U(1)$-action on $M_{2n+1}$, and in the coordinates \eqref{eq:kappa} it assumes the form
\begin{equation}
\xi=\frac{\partial \ }{\partial\theta} \ .
\end{equation}
A natural choice of $U(1)$-invariant metric on $M_{2n+1}$ is given by
\begin{align}
\dd s^2_{M_{2n+1}} = \pi^*\big(\dd s^2_{K_{2n}}\big) +
  \kappa\otimes\kappa \ ,
\end{align}
where $\dd s^2_{K_{2n}}$ is the K\"ahler metric on $K_{2n}$. Any
$k$-form $\alpha$ on $M_{2n+1}$ can be decomposed using the projector
$\kappa\wedge\xi\,\llcorner\,$ into horizontal and vertical components
as
\begin{align}
\alpha = \alpha_H+\alpha_V:=(1-\kappa\wedge\xi\,\llcorner\,)\alpha +
  \kappa\wedge\xi\,\llcorner\,\alpha \ ,
\end{align}
where $\xi\,\llcorner\,\alpha$ is the $k{-}1$-form component of
$\alpha$ along the fibre direction. 

The computation of the perturbative partition function of (twisted)
$\cN=1$ supersymmetric Yang-Mills theory on $M_{2n+1}$ is described
in~\cite{Kallen2012,Kallen2012a}, using equivariant localization
techniques with respect to the $U(1)$ action on $M_{2n+1}$ and the
maximal torus of the gauge group $G$. The relevant computations typically
involve the determinant of the square of the (twisted) supercharge
(equivariant differential), which is the kinetic operator
\begin{equation}
{\sf L}_\phi=\mathcal{L}_\xi+\cG_\phi
\end{equation}
acting on the tangent space to the space of fields, where $\cG_\phi$ denotes the action by an element $\phi$ valued in
the Cartan subalgebra of the Lie algebra $\frg$ of $G$; for fields in
the vector multiplet of the supersymmetric gauge theory,
$\cG_\phi={\rm ad}_\phi$ is the adjoint action. Here we assume
momentarily that the localization locus consists of constant field
configurations $\phi$; the case of non-constant $\phi$ is discussed
below. The operator ${\sf L}_\phi$ acts with the same eigenvalue on both
even and odd degrees in the spaces
$\Omega_H^{(0,\bullet)}(M_{2n+1},\frg)$ of horizontal anti-holomorphic
$\frg$-valued forms. The cancellation between
bosonic and fermionic fluctuation determinants in the localized path
integral is determined by the index of the Dolbeault complex of
$K_{2n}$ twisted by the line bundles $\scrL^{\otimes m}$ for $m\in\IZ$,
where $\scrL\to K_{2n}$ is the complex line bundle associated to the
circle bundle $M_{2n+1}\xrightarrow{ \ \pi \ } K_{2n}$ with first
Chern class
$c_1(\scrL)=p\,[\omega]$. Denoting the corresponding twisted Dolbeault
operators as $\bar\partial{}^{(m)}$, the Atiyah-Singer index theorem
gives the index as
\begin{equation}
{\rm index}\,\bar\partial{}^{(m)} = \int_{K_{2n}}\, \ch\big(\scrL^{\otimes m}\big)\wedge{\rm Td}\big(T^{1,0}K_{2n}\big) \ ,
\end{equation}
where $T^{1,0}K_{2n}$ is the holomorphic tangent bundle of $K_{2n}$,
while $\ch$ and ${\rm Td}$ respectively denote the Chern character and
the Todd
class.
	
\subsubsection*{Geometry from rigid supersymmetry}
	
We want to define a supersymmetric field theory on the backgrounds
$M_{2n+1}$ described above. In this paper we
focus on theories with four supercharges, which means $\mathcal{N}=2$ theories in three dimensions and
$\mathcal{N}=1$ theories in five dimensions; a thorough discussion on localization in five-dimensional superconformal
field theories can be found in \cite{Jafferis2012}. We follow the approach
of \cite{Festuccia2011}: we add a supergravity multiplet in flat spacetime, and then
take the rigid limit of the supergravity theory. This is done in two
steps: in the first step we set to zero all fermion fields, and in the
second step we set to zero the supersymmetry variations of the fermion
fields. The equation obtained by imposing the vanishing of the
gravitino variation is called the (generalized) Killing spinor
equation. The vector multiplet and hypermultiplets of the gauge theory
are then coupled to the background values of the supergravity
multiplet, and the theory is effectively put on a curved
spacetime. For the description of supersymmetric backgrounds in three
dimensions we mainly follow \cite{Closset2012,Alday2013}, and in five
dimensions we follow \cite{Pan2013,Alday2015}.

Let $\varepsilon$ be a solution to the Killing spinor equation for $M_{2n+1}$. We use it to define the vector field $v$ on $M_{2n+1}$ as
\begin{equation}
\label{eq:vpreferred}
	v^{\mu} = \varepsilon^{\dagger}\, \Gamma^{\mu} \varepsilon \ ,
\end{equation}
where $\Gamma^{\mu}$ are the gamma-matrices in either three or five
dimensions. The fact that $\varepsilon$ satisfies the Killing spinor
equation guarantees that $v$ is a nowhere vanishing Killing vector
field. In particular, its orbits foliate $M_{2n+1}$. 

As explained in
\cite[Section 5]{Closset2012} and in \cite[Section 4]{Alday2013}, in three dimensions there are two possibilities:
\begin{itemize}
	\item[(I)] The orbits of $v$ are closed. In this case $M_3$ is
          a Seifert manifold and $v$ coincides with the Reeb vector
          field $\xi$ of the $U(1)$ fibration of $M_3$. Particular examples
          belonging to this class are the round sphere $S^3$ and the lens spaces $L(p,s)$.
	\item[(II)] If the orbits of $v$ do not close, supersymmetry
          requires $M_3$ to have isometry group $U(1) \times U(1)$. In
          this case $M_3$ is a Seifert manifold but $v$ does not
          necessarily point along the $U(1)$-fibre. A particular
          example belonging to this class is the ellipsoid $S_b^3$ of \cite{Hama2011}.
\end{itemize}
We will henceforth refer to the manifolds belonging to the setting (I)
as \emph{regular}, and to those of setting (II) as
\emph{irregular}.\footnote{We will use the nomenclature ``regular
  geometry'' or ``regular fibration'', meaning that the integral
  curves of the Killing vector field have regular flow, and similarly
  for the ``irregular fibration''. In this paper we will not distinguish between
  irregular and quasi-regular cases.} Irregular geometries may or
may not admit a free $U(1)$ action of the Reeb vector field; see the
recent review \cite{Closset2019} for a thorough description of the
geometric approach to $\mathcal{N}=2$ supersymmetry on three-dimensional Seifert
manifolds.

In the five-dimensional case, we will focus on product
manifolds $M_3 \times \Sigma_h$, where $\Sigma_h$ is a closed 
Riemann surface of genus $h$. Killing spinor solutions in these
geometries are built from solutions on $M_3$, and thus an analogous
discussion applies; see the discussion at the end of \cite{Alday2015}
for further details about the difference in the approach we follow
here and that of \cite{Kallen2012a,Qiu2014}.

\subsubsection*{Examples: Regular vs irregular fibrations}
\label{sec:examplesregvsirr}

We write down some explicit examples of regular and irregular
five-dimensional manifolds. The round sphere $S^5$ and the product
$S^3 \times S^2$ are regular Sasaki-Einstein manifolds. The
Sasaki-Einstein manifolds $Y^{p,s}$ studied in
\cite{Qiu2013,Qiufact2013} are irregular (or quasi-regular). The
product manifolds $M_3 \times \Sigma_h$ where $M_3$ is either $S^3$,
$L(p,1)$ or the three-dimensional torus $T^3$ are regular, while if $M_3 = S^3 _b$ is the
ellipsoid of \cite{Hama2011} it is irregular. Among the irregular manifolds, $S^3 _b \times \Sigma_h$ (as well as replacing $S^3 _b$ with other squashed Seifert three-manifolds) admits a free $U(1)$ action, while $Y^{p,s}$ do not admit any free $U(1)$ action and are described as $U(1)$ fibrations over a warped product $S^2 \rtimes S^2$.
See Appendix \ref{app:squashedspheres} for a classification and
discussion of the different types of squashed three-spheres, and
Appendix~\ref{app:SE} for a discussion about cohomological localization on $Y^{p,s}$.

\subsubsection*{One-loop determinants}
	
For regular fibrations, it was shown in~\cite{Kallen2012,Kallen2012a} that the one-loop
contribution of the $\mathcal{N}=1$ vector multiplet to the
perturbative partition function on $M_{2n+1}$ is
\begin{equation}
\label{eq:zvecformula}
	Z_{\rm vec}(M_{2n+1})=\prod_{\alpha\in \triangle} \, \big(
        \,\mathrm{i}\, (\alpha,\phi) \big)^{d} \ 
        \prod_{m\neq 0} \, \Big( \frac{m}{r} + \mathrm{i}\, (\alpha,
          \phi) \Big)^{  \mathrm{index} \, \bar{\partial}^{(m)}} \ ,
\end{equation}
where as previously $\triangle$ is the root system of the Lie algebra $\frg$ and
$(\,\cdot\,,\,\cdot\,)$ is an invariant non-degenerate bilinear form 
on $\frg$; the power $d$ of the first
zero mode factor is given by
\begin{align}
d = {\rm index} \, \bar\partial - \dim H^0(M_{2n+1},\IR) \ ,
\end{align} 
the difference between the index of the ordinary (untwisted) Dolbeault
complex of $K_{2n}$ and the dimension of the space of harmonic functions on
$M_{2n+1}$. 
The one-loop contribution from an $\cN=1$ hypermultiplet in a
representation $R$ of the gauge group $G$ is given by
\begin{equation}
\label{eq:zhypformula}
	Z_{\rm hyp}(M_{2n+1}) = \prod_{\rho \in \Lambda_R} \ \prod_{m
          \in \mathbb{Z} } \, \Big( \frac{m}{r} + \mathrm{i} \,
        (\rho,\phi) + \frac{\Delta}{r} \Big) ^{- \mathrm{index} \,
          \bar{\partial}^{(m)}} \ ,
\end{equation}
where $\Lambda_R$ is the lattice of weights of $R$ and $\Delta$ is a constant determined by the conformal scalar field
coupling to the curvature in the gauge theory action.
In~\cite{Kallen2012a} these formulas are applied to $\cN=1$ gauge
theory on the five-sphere $M_5=S^5$, viewed as a circle bundle over
the projective plane $K_4=\mathbb{P}^2$, with $\Delta=\frac32$.

The extension of these formulas to the case of non-constant scalar fields
$\phi$ on $K_{2n}$ can be deduced from the prescription explained
in~\cite[Appendix~B]{Blau93}, at least in the case when the kinetic operator
${\sf L}_\phi$ is elliptic. In these instances one can apply the index
formula ``locally'' by moving the logarithms of the arguments of the
products into the integral and integrating against the index
density. For example, for the vector multiplet contribution this
prescription gives
\begin{align}
Z_{\rm vec}(M_{2n+1})= 
\exp\bigg(\, \int_{K_{2n}} \ & \sum_{m\in\IZ} \, \ch\big(\scrL^{\otimes m}\big)\wedge{\rm
  Td}\big(T^{1,0}K_{2n}\big) \ \sum_{\alpha\in \triangle} \, \log\Big(\frac mr
+\ii(\alpha,\phi) \Big) \nonumber \\ & - \dim H^0(M_{2n+1},\IR)\,
                                       \int_{K_{2n}}\,
                                       \frac{\omega^{\wedge n}}{n!} \ \sum_{\alpha\in \triangle} \, \log\big(\ii(\alpha,\phi) \big)
\, \bigg) \ .
\end{align}
For the cohomological localization we shall employ in this paper, the
further localization to constant $\phi$ in two dimensions will be
immediate (in contrast to the approach of~\cite{Fukuda2012}).

The expressions \eqref{eq:zvecformula} and \eqref{eq:zhypformula} are
proven in \cite{Kallen2012,Kallen2012a} in the case of
five-dimensional K-contact manifolds, with Killing vector field $v$
pointing along the Seifert fibre. In following sections we will review
the main steps in the proof, both in three and five dimensions, and
derive the corresponding expressions for the cases in which the
Killing vector field $v$ does not point in the direction of the $U(1)$
fibre. For this, we will have to introduce vector multiplets and
hypermultiplets, and then topologically twist the field content. For
the three-dimensional case we will follow the conventions of
\cite{Kallen2011}, while in five dimensions we follow
\cite{Kallen2012a,Qiu2016}.

\section{$\mathcal{N}=2$ cohomological gauge theories in three dimensions}
\label{sec:N=23d}

In this section we study $\mathcal{N}=2$ supersymmetric gauge theories
on three-dimensional manifolds. We will first present the theory and
its topological twist. Then we will reproduce the formula for the
one-loop determinants in the case of a Seifert fibration $M_3 \to K_2$
corresponding to closed orbits of the Killing vector field $v$. We shall
subsequently extend the formulas to the ellipsoid and ellipsoidal lens
spaces, corresponding to non-compact orbits of the Killing vector
field, by extending the application of the index theorem used in~\cite{Drukker2012}, for the ellipsoid, to any squashed Seifert manifold.
For the geometric setting we will follow \cite{Closset2012,Alday2013},
while for the topological twist and derivation of the one-loop
determinants, as well as for the normalization of the fields and supersymmetry variations, we will continue to follow \cite{Kallen2011}. 
Our conventions are summarized in Appendix \ref{app:notation}.

\subsection{Supersymmetric Yang-Mills theory and its cohomological formulation}
\label{sec:3dcohft}

As usual, we start by placing the gauge theory on flat Euclidean space $\IR^3$ and then
couple it to background supergravity, following
\cite{Festuccia2011,Closset2012}. Let $\varepsilon$ and $\widetilde{\varepsilon}$ be two Killing spinors with opposite R-charge, and define the Killing vector field
\begin{equation}
	v^{\mu} = \widetilde{\varepsilon}^{\,\,\dagger} \, \gamma^{\mu}
        \varepsilon \ .
\end{equation}

\subsubsection*{Vector multiplet}

The $\mathcal{N}=2$ vector multiplet in three dimensions consists of a
gauge connection $A$, a real scalar $\sigma$, a complex spinor
$\lambda$ (the gaugino), and an auxiliary real scalar $D$. The spinor
$\lambda = (\lambda^{I})$ and the real scalar $D = ( D^{I}{}_{J})$
carry $SU(2)_R$ indices.\footnote{By analogy with the five-dimensional setting, we work with $SU(2)_R$ R-symmetry. For a generic $\mathcal{N}=2$ theory with only $U(1)_R$ R-symmetry, simply neglect the indices $I,J$.} The supersymmetry transformations are
standard \cite{Kapustin2009,Kallen2011}. We denote by $\sf Q$ the
equivariant differential (supersymmetry generator) which is the sum of the two
independent supercharges $\frac12\,\big( \widetilde{\sf
  Q}_{\widetilde{\varepsilon}} + {\sf Q}_{\varepsilon} ^{\dagger} \big)$, and write
\begin{align}
	{\sf Q} A_{\mu} &= \tfrac{\mathrm{i}}{2}\, \big(
                         \widetilde{\varepsilon} ^{\,\,\dagger} _I\,
                         \gamma_{\mu} \lambda^{I} - \lambda^{\dagger}
                         _{I}\, \gamma_{\mu} \varepsilon^{I} \big) \ ,
                         \nonumber \\[4pt]
	{\sf Q} \sigma &= - \tfrac{1}{2}\, \big(
                        \widetilde{\varepsilon}^{\,\,\dagger} _I\,
                        \lambda^{I} + \lambda^{\dagger} _I\, \varepsilon
                        ^{I}  \big) \ , \nonumber \\[4pt]
	{\sf Q} \lambda^{I} &= - \tfrac{1}{2}\, \gamma^{\mu \nu}
                             \varepsilon^{I}\, F _{\mu \nu}  -
                             {D^{I}}_{J}\, \varepsilon ^{J} + \mathrm{i}\,
                             \gamma^{\mu} \varepsilon ^{I}\,({\sf D}_{\mu}
                             \sigma) \ , \nonumber \\[4pt]
	{\sf Q} \lambda^{\dagger} _I &= \tfrac{1}{2}\,
                                      \widetilde{\varepsilon}
                                      ^{\,\,\dagger} _{I} \gamma^{\mu \nu}\,
                                      F _{\mu \nu}  -
                                      \widetilde{\varepsilon}^{\,\,\dagger}
                                      _{J}\, {D^{J}}_{I}  - \mathrm{i}\,
                                      \gamma^{\mu}
                                      \widetilde{\varepsilon}^{\,\,\dagger}
                                      _{I}\, ({\sf D}_{\mu} \sigma) \ ,
                                      \nonumber \\[4pt]
	{\sf Q} {D_{I}} ^{J} &= \tfrac{\mathrm{i}}{2}\, \big(
                              \widetilde{\varepsilon}^{\,\,\dagger} _{I}\,
                              \gamma^{\mu} ({\sf D}_{\mu} \lambda^{J}) -
                              ({\sf D}_{\mu} \lambda^{\dagger} _{I} )\,
                              \gamma^{\mu} \varepsilon^{J} \big)  -
                              \tfrac{\mathrm{i}}{2}\, \big(  \widetilde{\varepsilon}^{\,\,\dagger} _{I}\, \big[
                              \sigma ,
                              \lambda^{J} \big] - \big[ \sigma
                              ,\lambda^{\dagger} _{I} 
                              \big]\varepsilon^{J} \big) + ( \ I \leftrightarrow
                              J \ ) \ ,
\end{align} 
where $F$ is the curvature of the gauge connection $A$, and ${\sf D}_{\mu}$ is the covariant derivative which involves the gauge
connection $A$ and also the spin connection when acting on the
dynamical spinor fields $\lambda$ and $\lambda^{\dagger}$. When the
theory is placed on a curved background $M_3$, one has to add curvature
terms proportional to $\frac1r$ to the supersymmetry variations
${\sf Q} \lambda$, ${\sf Q} \lambda^{\dagger}$ and ${\sf Q} D$. These
terms will also involve the spinor covariant derivative acting on the
Killing spinors from the supergravity background. This procedure is
standard and we do not review it here.

Following \cite[Appendix A]{Kallen2011}, we use the Killing vector field $v$
for the topological twist. 
We set $\widetilde{\varepsilon}= \varepsilon$, and rewrite the spinor
fields $\lambda$ and $\lambda^{\dagger}$ in the vector multiplet in
terms of an odd $\frg$-valued
one-form $\Psi$ and an odd $\frg$-valued zero-form $\chi$ defined as
\begin{equation}
	\Psi_{\mu} = \tfrac{1}{2}\, \big( \varepsilon^{\dagger}\,
        \gamma_{\mu} \lambda - \lambda^{\dagger}\, \gamma_{\mu}
        \varepsilon \big) \qquad \mbox{and} \qquad \chi =
        \varepsilon^{\dagger}\, \lambda - \lambda^{\dagger}\,
        \varepsilon \ ,
\end{equation}
which depend on the solution $\varepsilon$ of the Killing spinor
equation, and hence on the choice of contact structure, but not explicitly on the metric.
The field content of the vector multiplet is now written in a
cohomological form as
\begin{equation}
	A \in \Omega ^{1} (M_3, \mathfrak{g}) \ , \quad \sigma \in
        \Omega ^{0} (M_3, \mathfrak{g}) \ , \quad \Psi \in \Omega ^{1}
        (M_3, \mathfrak{g}) \qquad \mbox{and} \qquad \chi \in \Omega
        ^{0} (M_3, \mathfrak{g}) \ ,
\end{equation}
with $(A, \chi)$ treated as coordinates and $(\sigma, \Psi)$ as
conjugate momenta on field space. We do not include details about the
gauge fixing here, and again refer to \cite{Kallen2011} for the
technical details. It suffices to say that the bosonic ghost
coordinates are a pair of harmonic zero-forms and the fermionic ghost
coordinates are a pair of zero-forms. We use the localizing term
\begin{equation}
	{\sf Q} V \qquad \mbox{with} \quad V = \int_{M_3}\, \big(
        ({\sf Q} \lambda)^{\dagger}\, \lambda + \lambda^{\dagger}\,
        ({\sf Q} \lambda^{\dagger})^{\dagger} \big) \
        \dd\Omega_{M_3}
\end{equation}
in the path integral which brings the quantum field theory to the fixed point locus
\begin{equation}
 F=0 \qquad \mbox{and} \qquad \sigma = -D = \text{constant} \ .
\end{equation}

\subsubsection*{The kinetic operator}

Once the fields are in cohomological form, the supersymmetry transformation squares to
\begin{equation}
\label{eq:Lphi}
	{\sf Q}^2 = \mathrm{i}\, {\sf L}_{\phi} \qquad \mbox{with}
        \quad {\sf L}_{\phi} = \mathcal{L}_v  + \mathcal{G}_{ \phi} \ .
\end{equation}
Here ${\sf L}_{\phi}$ is the sum of a Lie derivative along $v$ and a gauge transformation $\mathcal{G}_{ \phi}$ with parameter 
\begin{equation}
	\phi = \mathrm{i}\, \sigma - v\,\llcorner\, A \ .
\end{equation}
At the end, we shall rotate $\sigma \mapsto \mathrm{i}\, \sigma_0$ and
integrate over real $\sigma_0 \in \mathfrak{g}$. The localization
locus consists of flat connections, and therefore
\begin{equation}
	\phi = - \big( \sigma_0 + v^{\mu}\, A_{\mu} ^{(0) } \big) \
        \in \ \mathfrak{g} \ ,
\end{equation}
where $ A^{(0) }$ is the point of the moduli space of flat
$G$-connections on $M_3$ around which we are expanding. If $M_3$ is
simply connected, the only point of the moduli space is the trivial
connection. Otherwise, expanding around $A^{(0)}=0$ gives the perturbative part of the partition function. In general, the full answer is given by integrating the partition function over the moduli space of flat $G$-connections $\scrM^0_G(M_3)$ supported on $M_3$, which is given by
\begin{align}\label{eq:scrMGM3}
\scrM^0_G(M_3) = {\rm Hom}\big(\pi_1(M_3),G\big)\big/G \ ,
\end{align}
where the quotient is taken by the conjugation action of $G$ on the
holonomy of a connection over representatives of elements in $\pi_1(M_3)$. When $M_3$
is a circle bundle of degree $p$ over a compact oriented Riemann
surface $C_g$ of genus $g$, there is an explicit presentation of
the fundamental group $\pi_1(M_3)$ with generators $a_i,b_i,\zeta$,
$i=1,\dots,g$ and the relation
\begin{align}
\prod_{i=1}^g\,[a_i,b_i]=\zeta^p \ ,
\end{align}
with all other pairwise combinations of generators commuting.
An explicit parametrization of the moduli space \eqref{eq:scrMGM3} in the case $G=U(N)$ can be found in~\cite[Section~6.2]{Caporaso:2006kk}.

\subsubsection*{Hypermultiplets}

The field content of an $\cN=2$ hypermultiplet in three dimensions consists of the complex scalars
$\mathtt{q} =(\mathtt{q} _I)$ with $SU(2)_R$ indices and a complex spinor $\psi$. These
fields are obtained by combining chiral and anti-chiral complex scalars and Weyl spinors. One also needs an auxiliary complex scalar. The supersymmetry transformations are
\begin{align}
	{\sf Q} \mathtt{q} _{I} & = - \mathrm{i}\, \widetilde{\varepsilon}_{I}
                       ^{\,\,\dagger}\, \psi \ , \nonumber \\[4pt]
	{\sf Q} \psi &=  \tfrac{1}{2}\, \gamma^{\mu} \varepsilon^{I}\,
                      ({\sf D}_{\mu} \mathtt{q} _{I}) + \tfrac{\mathrm{i}}{2}\,
                      \sigma\, \mathtt{q} _{I}\, \varepsilon^{I} \ , 
\end{align}
plus curvature corrections to be added when the theory is put on
$M_3$. The transformations of the conjugate fields $\mathtt{q}^{\dagger}, \psi^{\dagger}$ are the obvious
ones, with exchange $\varepsilon \leftrightarrow
\widetilde{\varepsilon}$. The topological twist of a hypermultiplet
was first performed in \cite[Appendix B]{Ohta2012}.

The Killing spinors are used to introduce a new spinor field
\begin{equation}
	\mathtt{q}^{\prime} = \mathtt{q}_{I}\, \varepsilon^{I} \ ,
\end{equation}
so that the physical fields are all reformulated in terms of
spinors. These fields are singlets under the action of $SU(2)_R$; this is instrumental to have the kinetic operator in the
desired form. One finds
\begin{equation}
	{\sf Q}^2 = \mathrm{i}\, {\sf L}_{\phi} \qquad \mbox{with}
        \quad {\sf L}_{\phi} = \mathcal{L}_v ^{\rm spin}  + \mathcal{G}_{ \phi} \ ,
\end{equation}
as in the vector multiplet. We used the notation $\mathcal{L}_v ^{\rm
  spin}$ to stress that the Lie derivative is twisted by the spin
covariant derivative when acting on spinors on curved manifolds.

To mimic the procedure of \cite{Kallen2011}, a further step is needed:
we rearrange the fields again in a cohomological form, combining
spinors into differential forms. For this, we need to define a
spin$^c$ structure on $M_3$, and use it to decompose the spinors $\psi_\pm$
and $\mathtt{q}^{\prime}$ into elements of
\begin{equation}
	\Omega^{0} _H (M_3,\frg) \oplus \Omega^{(0,1)} _H (M_3,\frg) \,
\end{equation}
where we decomposed $\psi$ according to the chirality operator as
\begin{equation}
	\psi= \psi_{+} + \psi_{-} \qquad \mbox{with} \quad \gamma_5
        \psi_{\pm} = \pm\, \psi_{\pm} \qquad \mbox{and} \qquad
        \gamma_5 = v^{\mu}\, \gamma_{ \mu} \ .
\end{equation}
The $SU(2)_R$ R-symmetry is not considered in \cite{Ohta2012}, and they
do not pass through the intermediate step of contracting the $SU(2)_R$
indices to form singlets; instead, they directly define anti-holomorphic forms.
In the present case, the cohomological field theory contains hypermultiplets only when the manifold $M_3$ admits a spin$^c$ structure.

\subsection{One-loop determinant of the vector multiplet in a regular background}
\label{sec:1loopvector3d}

Gaussian integration of the vector multiplet around the fixed point $\phi$ gives the ratio of fluctuation determinants
\begin{equation}
	h (\phi) = \sqrt{\frac{\det \mathrm{i}\, {\sf L}_{\phi}
            \vert_{\mathrm{f}} }{\det  \mathrm{i}\, {\sf L}_{\phi}
            \vert_{\mathrm{b}}} } = \sqrt{ \frac{ ( \det \mathrm{i}\,
            {\sf L}_{\phi} \vert_{\Omega^{0}(M_3,\frg)})^3 }{\det \mathrm{i}\,
            {\sf L}_{\phi} \vert_{\Omega^{1}(M_3,\frg)} \, ( \det \mathrm{i}\,
            {\sf L}_{\phi} \vert_{H^{0}(M_3,\frg)})^2  } } \ ,
\end{equation}
where the subscripts on the left-hand side refer to the operator
acting on fermionic or bosonic fields. Here $H^{0} (M_3,
\mathfrak{g})$ is the space of $\mathfrak{g}$-valued harmonic zero-forms,
and ${\sf L}_{\phi}$ is given in \eqref{eq:Lphi}. The numerator of
$h(\phi)$ includes the contributions from the fermionic field $\chi$ and the two
fermionic ghost fields, while the denominator includes the
contributions from the
bosonic field $A$ and the two bosonic ghost fields.

We use the Seifert structure of $M_3$ to decompose one-forms into
horizontal and vertical parts as
\begin{equation}
	\Omega^{1} (M_3, \mathfrak{g}) = \Omega^{1} _V (M_3, \mathfrak{g}) \oplus \Omega^{1} _H (M_3, \mathfrak{g}) \cong \Omega^{0} (M_3, \mathfrak{g}) \oplus \Omega^{1} _H (M_3, \mathfrak{g}) \ .
\end{equation}
We may also identify
\begin{equation}
	\Omega ^{2} _H (M_3,\frg) \cong \Omega^{0} (M_3,\frg)
\end{equation}
in three dimensions. The circle bundle structure
\begin{equation}
\xymatrix{
U(1) \ar[r] & M_3 \ar[d] \\ & K_2
}
\end{equation}
allows us to further decompose the spaces of zero-forms and horizontal one-forms as
\begin{align}
	\Omega^{0} (M_3, \mathfrak{g}) & = \Omega^{0}  (K_2 , \mathfrak{g}) \oplus \bigoplus_{m \ne 0} \,  \Omega^{0}  \big(K_2 , \mathscr{L}^{\otimes m } \otimes \mathfrak{g} \big) \ , \nonumber \\[4pt]
	\Omega^{1} _H (M_3, \mathfrak{g}) & = \Omega^{1}  (K_2 , \mathfrak{g}) \oplus \bigoplus_{m \ne 0} \,  \Omega^{1}  \big(K_2 , \mathscr{L}^{\otimes m } \otimes \mathfrak{g} \big) \ ,	
\label{eq:circledecomp}
\end{align}
where we recall that $\mathscr{L}$ is the line bundle associated to the $U(1)$ fibration of $M_3$. Using the short-hand notation
\begin{equation}
 \mathcal{D} ^{\bullet} _m  (\phi) = \det \mathrm{i}\, {\sf L}_{\phi} \vert_{\Omega^{\bullet} (K_2 , \mathscr{L}^{\otimes m } \otimes \mathfrak{g})} \ ,
\end{equation}
we obtain
\begin{equation}
	h(\phi) = \frac{1}{\big\lvert \det \mathrm{i}\, {\sf L} _{\phi} \vert_{H^{0} (M_3,\frg)} \big\rvert } \, \sqrt{ \frac{\mathcal{D}^0 _0  (\phi)\, \mathcal{D}^2 _0 (\phi) }{ \mathcal{D}^1 _0 (\phi) } } \ \prod_{m \ne 0 }\, \sqrt{ \frac{\mathcal{D}^0 _m  (\phi)\, \mathcal{D}^2 _m (\phi) }{ \mathcal{D}^1 _m  (\phi) } } \ .
\end{equation}

The crucial observation at this point is that when the Killing vector field $v$ points along the fibre direction, the decomposition \eqref{eq:circledecomp} corresponds to a decomposition in eigenmodes of the Lie derivative operator $\mathcal{L}_v$. 
The degeneracy of the action of the gauge transformations $\cG_\phi$ is resolved in the standard way \cite{Kapustin2009,Kallen2011}, by
decomposing the Lie algebra $\mathfrak{g} $ into its root system as
\begin{equation}
	\mathfrak{g} = \bigoplus_{\alpha \in \triangle}\,
        \mathfrak{g}_{\alpha} \ .
\end{equation}
We finally obtain
\begin{equation}
\label{eq:zvec3d}
\begin{tabular}{|c|}\hline\\
$\displaystyle{	Z_{\mathrm{vec}} (M_3) = \prod_{\alpha \in \triangle
  }\, \big( \mathrm{i}\, (\alpha, \phi) \big)^{\frac{1}{2}\,\chi (K_2)
  - \dim H^{0} (M_3,\IR)} \ \prod_{m \ne 0}\, \Big( \frac{m}{r} +
  \mathrm{i}\, (\alpha, \phi) \Big)^{\mathrm{index} \,
  \bar{\partial}^{(m)} } }
$\\\\\hline\end{tabular}
\end{equation}
where we used the fact that the number of remaining modes, after cancellation, is given by the index of the twisted Dolbeault complex
\begin{equation}
	\Omega^{0} (K_2, \mathscr{L}^{\otimes m} ) \xrightarrow{ \ \bar{\partial}^{(m)} \ } \Omega^{1} (K_2, \mathscr{L}^{\otimes m} ) \xrightarrow{ \ \bar{\partial}^{(m)} \ } \Omega^{2} (K_2, \mathscr{L}^{\otimes m} ) \ ,
\end{equation}
after identification of the complexified de~Rham differential with the
anti-holomorphic Dolbeault differential. This result agrees with
\cite{Closset2017}. The first multiplicative term
in \eqref{eq:zvec3d} is trivial for Seifert homology spheres.
The proof of this cohomological localization formula used the
properties that $M_3$ is a Seifert manifold and that $v$ is parallel
to the Reeb vector field $\xi$, and hence that $M_3$ is a K-contact manifold.

\subsection{One-loop determinant of the vector multiplet in an irregular background}

We now consider the case in which the orbits of the Killing vector field $v$ are not closed. The background geometry is required to have $U(1) \times U(1)$ isometry group in order to preserve supersymmetry \cite{Closset2012,Alday2013}. In this instance
$M_3$ is still a Seifert manifold but now $v$ does not point along the
fibre. In the spirit of Section~\ref{sec:1loopvector3d}, we calculate
the ratio of fluctuation determinants through the index theorem. A
similar calculation was performed in \cite{Drukker2012}, but there the supersymmetry
transformation squares to the sum of a Lie derivative along $v$, a
gauge transformation and a third transformation which is a sum of
R-symmetry and flavour symmetry transformations (determining a new
R-symmetry); this is not of the form described in
\cite{Kallen2011,Kallen2012}, and so our index theory calculations
cannot be applied directly in this framework.

We can express $v$ as a linear combination
\begin{equation}
	 v = a_1\, \xi + a_2\, \widetilde{\xi} 
\end{equation}
of the Reeb vector field $\xi$ and a vector field $\widetilde{\xi}$
which generates a residual $U(1)$ isometry. They are mutually
orthogonal and are linear combinations of the generators of the torus
isometry. Most of the calculation follows that of the round case from Section~\ref{sec:1loopvector3d}, particularly the decomposition \eqref{eq:circledecomp} of $\Omega^{\bullet} (M_3,\frg)$ according to the Seifert fibration. However, we now have to face the problem that the eigenmodes of $\mathcal{L}_{\xi}$ are no longer eigenmodes of $\mathcal{L}_v$.

In \cite{Drukker2012} it was shown how
one can exploit the fact that if the $U(1)$ action generated by the Reeb
vector field $\xi$ is free, then the problem can be reduced to the quotient
space $M_3 / U(1) \cong K_2$. Let us elaborate a bit more on this
point. The crucial observation is that the restriction of the
operator $\ii{\sf L}_\phi$ is no longer elliptic, but it is
transversally elliptic with respect to the isometry generated by
$\widetilde\xi\,$. Following \cite{Atiyah1974}, given a first order
transversally elliptic differential 
operator and a
subgroup which acts freely, the index can be computed on the quotient
space and the Atiyah-Bott localization formula localizes the
contributions to the fixed points of the action of
the subgroup generated by $\widetilde{\xi}\,$; this works even when $K_2$ possesses orbifold
points. We then decompose into eigenmodes corresponding to
$\widetilde{\xi}\,$, and the remaining modes after cancellation come
from the fixed points of the $U(1)$ action generated by
$\widetilde{\xi}\,$. Therefore for irregular Seifert manifolds we obtain
\begin{equation}
\label{eq:zvecsquash3d}
\begin{tabular}{|c|}\hline\\
$ \displaystyle{	Z_{ \mathrm{vec}} (M_3) = \prod_{\alpha \in \triangle }\,
        \big(\mathrm{i}\, (\alpha, \phi)\big) ^{ \frac{1}{2}\,\chi
          (K_2) - \dim H^{0}(M_3,\IR) }  \ \prod_f \ \prod_{m \ne 0}\,
  \Big( \frac{m}{\epsilon_f} + \mathrm{i}\, (\alpha, \phi )
  \Big)^{\frac{1}{2}\,  \mathrm{index} \, \bar{\partial} ^{(m)} }
} $\\\\\hline\end{tabular}
\end{equation}
where the second product runs over the fixed points of the $U(1)$
action generated by $\widetilde{\xi}$ on $K_2$, and $\epsilon_f$ is the radius of the circle fibre over the fixed point labelled by $f$.\footnote{The square roots of each fixed point contribution come from the square root of the original ratio of fluctuation determinants. In the regular case, the contributions are equal.} This formula gives the correct answer for ellipsoids~\cite{Hama2011,Alday2013}, and generalizes the result of \cite{Drukker2012} to any Seifert manifold which is not K-contact. Notice that an ellipsoid Seifert manifold is necessarily fibered topologically over a sphere $S^2$ with at most two punctures \cite[Section 3.5]{Closset2019}.

\subsection{One-loop determinant of a hypermultiplet}

For the Gaussian integration of a hypermultiplet around a fixed point $\phi$, we do not give all the details here since the computation is essentially the same as for the vector multiplet. One finds
\begin{equation}
\label{eq:oneloophyper3d}
	\sqrt{\frac{\det \mathrm{i}\, {\sf L}_{\phi} \vert_{\mathrm{f}} }{\det  \mathrm{i}\, {\sf L}_{\phi} \vert_{\mathrm{b}}} } = \sqrt{ \frac{ \det \mathrm{i}\, {\sf L}_{\phi} \vert_{\Omega^{(0,1)} _H(M_3,\frg)}  }{  ( \det \mathrm{i}\, {\sf L}_{\phi} \vert_{\Omega^{0} _H(M_3,\frg)} )^2  } } \ ,
\end{equation}
where we used the topological twist described in
Section~\ref{sec:3dcohft}. Recall that this formula only holds if we
are allowed to recombine the $SU(2)_R$ singlet spinors $\mathtt{q}^{\prime}$ into
anti-holomorphic differential forms. This fails for the ellipsoid $S^3
_b$ of \cite{Hama2011}, for instance, because a spin$^c$ structure is not guaranteed to exist when the dual one-form to the Killing vector $v$ is not a gauge connection for the $U(1)$ fibration.

For round Seifert manifolds, the eigenvalues of $\mathcal{L}_v ^{\rm spin}$ are
\begin{equation}
	- \frac{ \mathrm{i}\, m}{r} - \frac{ \mathrm{i}\, \Delta }{r} \ ,
\end{equation}
where $m\in\mathbb{Z}$ and $\Delta$ is the R-charge the hypermultiplet, which we assume equal to the conformal dimension, although more general assignments are possible in theories with only four supercharges. 
The number of remaining modes after cancellations in \eqref{eq:oneloophyper3d} is given by the index of the twisted Dolbeault differential, and we obtain
\begin{equation}
\begin{tabular}{|c|}\hline\\
$ \displaystyle{	Z_{\rm hyp} (M_3) = \prod_{\rho \in \Lambda_R}
  \ \prod_{m \in \mathbb{Z}} \, \Big( \frac{m + \Delta}{r} +
  \mathrm{i}\, (\rho, \phi) \Big)^{- \mathrm{index} \, \bar{\partial}
  ^{(m)}}
} $ 
\\\\\hline\end{tabular}
\end{equation}
where a shift $m \to m+ \Delta$ should be included when considering twisted boundary conditions along the fibre, but the contribution of this shift cancels in the computations.
This result agrees with \cite{Closset2017}.

\subsection{Applications of the cohomological localization formulas}
\label{sec:3dappl}

We shall now provide some simple examples illustrating how the
localization formulas obtained in this section work to give the correct known results in three dimensions.

\subsubsection*{Localization on $\boldsymbol{S^3}$}

As a first check, let us examine how to reproduce the three-dimensional
localization calculations of~\cite[Section~3]{Kapustin2009} in this
framework. We consider the three-sphere $M_3=S^3$ of radius $r$,
viewed as a circle bundle of degree one over the projective line $K_2= \IP^1$ via
the Hopf fibration, with Euler characteristic $\chi(\IP^1)=2$ and
$H^0(S^3,\IR)=\IR$. The total Chern class of the holomorphic tangent
bundle of $\IP^1$ is 
\begin{align}
c(T^{1,0}\IP^1)=(1+\omega)^{\wedge 2}=1+2\,\omega=1+c_1(T^{1,0}\IP^1) \ , 
\end{align}
and so the corresponding Todd class is given by
\begin{equation}
{\rm Td}(T^{1,0}\IP^1)= 1+\mbox{$\frac12$}\, c_1(T^{1,0}\IP^1)=1+\omega \ ,
\end{equation}
while the Chern characters of the line bundles $\scrL^{\otimes m}\to\IP^1$ are given by
\begin{equation}
\ch(\scrL^{\otimes m}) = 1+c_1(\scrL^{\otimes m})=1+m\, c_1(\scrL)=1+m\, \omega \ .
\end{equation}
The index of the corresponding twisted Dolbeault complex is thus given by
\begin{equation}
{\rm index}\, \bar\partial{}^{(m)}= \int_{\IP^1}\, (1+m\,\omega)\wedge(1+\omega)=\int_{\PP^1}\, (m+1)\, \omega= m+1 \ .
\end{equation}

The one-loop vector multiplet contribution is then computed to be
\begin{align}
Z_{\rm vec}(S^3) &= \prod_{\alpha\in \triangle} \ \prod_{m\neq 0}\,
\Big(\,\frac{m}{r} + \ii(\alpha,\phi)\, \Big)^{m+1} \nonumber \\[4pt]
&= \prod_{\alpha\in \triangle} \ \prod_{m=1}^\infty\,
\frac{\big(\frac mr-\ii(\alpha,\sigma_0 )\big)^{m+1}}{\big(-\frac
  mr-\ii(\alpha,\sigma_0)\big)^{m-1}} \nonumber \\[4pt]
  &= \prod_{\alpha\in \triangle_+} \ \prod_{m=1}^\infty\, \frac{\big(\frac
  {m^2}{r^2}+(\alpha,\sigma_0 )^2\big)^{m+1}}{\big(\frac{m^2}{r^2}
  +(\alpha,\sigma_0 )^2\big)^{m-1}} \nonumber \\[4pt]
  &=  \prod_{\alpha\in \triangle_+} \ \prod_{m=1}^\infty\, \frac{m^4}{r^4} \, \Big(1+\frac{r^2\,
  (\alpha,\sigma_0)^2}{m^2}\,\Big)^2 \nonumber \\[4pt]
&= \prod_{\alpha\in \triangle_+}\, \Big(\,
\frac{2\sinh\pi\, r\, (\alpha,\sigma_0)}{\pi\, r\, (\alpha,\sigma_0)} \, \Big)^2 \ ,
\end{align}
where as previously $\triangle_+$ is the system of
positive roots of the Lie algebra $\frg$, and we used the fact that
the roots come in positive-negative pairs. In the second line we used
the fact that the only flat connection on $S^3$ is trivial and
substituted $\phi=-\sigma_0$, and in the last line we evaluated the
infinite product using zeta-function regularization:
\begin{align}
\prod_{m=1}^\infty\, \Big(1+\frac{x^2}{m^2}\Big) =
  \frac{\sinh(\pi\,x)}{\pi\,x} \qquad \mbox{and} \qquad
  \prod_{m=1}^\infty\, \frac{m^2}{r^2} = 2\pi\, r \ .
\end{align}

The same calculation for a one-loop hypermultiplet determinant gives
\begin{align}
	Z_{\rm hyp}(S^3) &= \prod_{\rho \in \Lambda_R} \ \prod_{m \in \mathbb{Z}}\, \Big( \frac{ m + \Delta }{r} + \ii (\rho, \phi ) \Big)^{-1 -m} \nonumber \\[4pt]
	&= \prod_{\rho \in \Lambda_R} \ \frac{ \prod\limits_{m =0} ^{\infty}\, \big( \frac{ - m - 1 + \Delta }{r} - \ii (\rho, \sigma_0)  \big)^m }{ \prod\limits_{m =1} ^{\infty}\, \big( \frac{  m - 1 + \Delta }{r} + \ii (\rho, \sigma_0)  \big)^m } \nonumber \\[4pt]
	&= \prod_{\rho \in \Lambda_R} \ \prod_{m=1} ^{\infty}\, \bigg(  \frac{  \frac{  m + 1 - \Delta }{r} + \ii (\rho, \sigma_0)  }{  \frac{  m - 1 + \Delta }{r} - \ii (\rho, \sigma_0) } \bigg)^m \nonumber \\[4pt]
	&= \prod_{\rho \in \Lambda_R} \, {\tt s}_{1} \big( \ii (1 - \Delta ) - r\, (\rho, \sigma_0 ) \big) \ ,
\end{align}
where in the last line we inserted the definition of the double-sine
function which is the meromorphic function defined by the zeta-function regularized infinite
products~\cite{Quine1993}
\begin{align}\label{eq:doublesine}
{\tt s}_b(x) = \prod_{m,n=0}^\infty\, \frac{m\,b+n\,b^{-1}+\frac12\,
  (b+b^{-1}) - \ii x}{m\,b+n\,b^{-1}+\frac12\,(b+b^{-1})+\ii x}
\end{align}
evaluated at $b=1$.
These results all agree with the computations of~\cite[Section~3.2]{Kapustin2009} (see also~\cite[Section~3.2]{Drukker2012}).
	
\subsubsection*{Localization on ellipsoid $\boldsymbol{S^3 _b}$ and $\boldsymbol{L(p,1)_b}$}

We now consider the ellipsoid $S_b^3$ of \cite{Hama2011}, with squashing parameter $b>0$
and metric
\begin{equation}
	\dd s_{S_b^3}^2 = r^2\, \big( f ( \vartheta ) ^2\ \dd \vartheta\otimes\dd\vartheta + b^2\, \cos^2 \vartheta\ \dd \varphi_1\otimes\dd\varphi_1 + b^{-2}\, \sin^2 \vartheta\ \dd \varphi_2\otimes\dd\varphi_2 \big)
\end{equation}
induced from the standard metric on $\mathbb{C}^2$ restricted to the locus
\begin{equation}
	b^2\, \vert z_1 \vert^2 + b^{-2}\, \vert z_2 \vert^2 = r^2 \ ,
\end{equation}
where $f(\vartheta)=\sqrt{b^{2}\, \cos^2\vartheta+b^{-2}\,
  \sin^2\vartheta}$ and $(\vartheta,\varphi_1,\varphi_2)$ are Hopf
coordinates on the usual round sphere $S^3=S_{b=1}^3$ of radius
$r$. This defines an irregular fibration with isometry group $U(1)\times U(1)$: The Killing vector field $v$ takes the form
\begin{align}
v = \frac{b+b^{-1}}{2\,r}\, \xi + \frac{b-b^{-1}}{2\,r}\, \widetilde{\xi} \ ,
\end{align}
where $\xi=\frac{\partial \ }{\partial\theta}$ is the Reeb vector field of the Seifert fibration $S_b ^3 \to \PP^1$, with $\theta = \frac12\,(\varphi_1+\varphi_2)$, and $\widetilde{\xi}=\frac{\partial \ }{\partial\widetilde\theta}$ is the generator of the residual $U(1)$ isometry, with $\widetilde{\theta} = \frac12\,(\varphi_1-\varphi_2)$. The fixed points of $\widetilde{\xi}$ correspond to the north and south poles of the base $S^2\cong\PP^1$, with respective coordinates $\vartheta =0$, at which the fibre has radius $\epsilon_1 = r\,b$, and $\vartheta =\frac{\pi}{2}$, at which the fibre has radius $\epsilon_2 =r\,b^{-1}$. The index of the twisted Dolbeault complex is a topological invariant, and hence is the same as for round $S^3$. 

Altogether, the localization formula gives
\begin{align}\label{eq:3dvecellipsoid}
 Z_{\rm vec} (S_b ^3) & = \prod_{\alpha\in \triangle} \ \prod_{m\neq 0}\, \bigg( \Big( \frac{m\, b^{-1}}{r} - \ii (\alpha, \sigma_0) \Big)^{\frac{1}{2}}\, \Big( \frac{m\,b}{r} - \ii (\alpha, \sigma_0) \Big)^{\frac{1}{2}} \bigg)^{m+1} \notag \\[4pt]
 	&= \prod_{\alpha\in \triangle} \ \prod_{m =1} ^{\infty} \, \frac{ \big( \frac{m\,b^{-1} }{r}  - \ii (\alpha, \sigma_0) \big)^{ \frac{m+1}{2} }\, \big( \frac{m\,b }{r}  - \ii (\alpha, \sigma_0) \big)^{ \frac{m+1}{2} }  }{  \big( \frac{m\,b^{-1} }{r} + \ii (\alpha, \sigma_0) \big)^{ \frac{m-1}{2} }\, \big( \frac{m\,b }{r} + \ii (\alpha, \sigma_0) \big)^{ \frac{m-1}{2} } }  \notag \\[4pt]
 	&= \prod_{\alpha\in \triangle_+} \ \prod_{m =1} ^{\infty} \, \Big( \frac{m^2\, b^{-2} }{r^2} + (\alpha, \sigma_0)^2 \Big)\, \Big( \frac{m^2\, b^{2} }{r^2} + (\alpha, \sigma_0)^2 \Big) \notag \\[4pt]
 	&= \prod_{\alpha\in \triangle_+}\, \frac{ \sinh \big( \pi\, b\, r\, (\alpha , \sigma_0 )\big)\, \sinh \big( \pi\, b^{-1}\, r\, (\alpha , \sigma_0 )\big) }{ \pi^2\, r^2\, (\alpha, \sigma_0 )^2 } \ ,
\end{align}
which agrees with \cite[eq.~(5.33)]{Hama2011} and
\cite[eq.~(4.24)]{Drukker2012}. This result is independent of the
particular form of the smooth squashing function $f(\vartheta)$ and depends
only on its values at the fixed points
$\vartheta=0,\frac\pi2$; it can therefore be
extended to a larger class of backgrounds with the same topology~\cite{Closset2013}.

This result straightforwardly extends to the ellipsoid lens spaces
$L(p,1)_b$, with the induced metric on the quotient $S^3 _b /
\mathbb{Z}_p$ and associated line bundle $\scrL\to\PP^1$ of degree $p$, so
that now $c_1(\scrL)=p\,\omega$; see the review \cite{Willett2016} for a
description of localization on $L(p,1)_b$. The modifications are the
same as for the round case, and amount to a shift $\sigma_0 \mapsto
\sigma_0 + v^{\mu}\, A_{\mu} ^{(0)}$, where $A ^{(0)}$ is an isolated
point of the moduli space of flat $G$-connections on $L(p,1)$. Such
flat connections are classified by conjugacy classes of embeddings of the fundamental group
$\pi_1 (L(p,1)_b) = \mathbb{Z}_p$ in the gauge group $G$, and are in
one-to-one correspondence with arrays $\vec{m} \in (\mathbb{Z}_p)^{
  \mathrm{rank}(G)}$ modulo Weyl symmetry. For the vector multiplet,
the localization formula then gives
\begin{align}
 Z_{\rm vec} \big(L(p,1)_b\big) = \prod_{\alpha\in \triangle_+}\,
  \frac{ \sinh \big( \frac{\pi\, b\, r}p\, (\alpha ,
  \sigma_0+\ii\vec m)\big)\, \sinh \big( \frac{\pi\, b^{-1}\,
  r}p\, (\alpha , \sigma_0+\ii\vec m)\big) }{ \frac{\pi^2\,
  r^2}{p^2}\, (\alpha, \sigma_0+\ii\vec m)^2 }  \ .
\end{align}
With similar modifications, one can extend these calculations to any ellipsoid Seifert manifold.

There is no spin$^c$ structure on $S^3 _b$ that can be used to apply our index theory formalism to the hypermultiplet contributions. Nevertheless, the one-loop fluctuation determinant can still be calculated in this case. For example, one could split the twisted Lie derivative as
\begin{equation}
	\mathcal{L}_v ^{\rm spin} = \frac{2\,b}r\, \mathcal{L}_{\frac{\partial \ }{\partial\varphi_1}} ^{\rm spin} + \frac{2\,b^{-1}}r\, \mathcal{L}_{\frac{\partial \ }{\partial\varphi_2}} ^{\rm spin}
\end{equation}
and decompose the fields into eigenmodes of the Lie derivatives in the
two orthogonal toroidal directions $\frac{\partial \ }{\partial\varphi_1}$
and $\frac{\partial \ }{\partial\varphi_2}$, whose corresponding
eigenvalues are then of the form $\frac{m + \Delta}{\epsilon_1} +
\frac{n + \Delta}{\epsilon_2} $ where $m,n\in\mathbb{Z}$. Then the
one-loop contribution of a hypermultiplet in a representation $R$ of the gauge group $G$ is given by~\cite[Section~4]{Drukker2012}
\begin{align}
Z_{\rm hyp}(S_b^3) &= \prod_{\rho
                                               \in \Lambda_R} \ \prod_{m,n=0}^\infty\, \frac{m\,b+n\,b^{-1}+\big(b+b^{-1}\big)\,\big(1-\frac\Delta2\big)+\ii r\,(\rho,\sigma_0)}{m\,b+n\,b^{-1}+\big(b+b^{-1}\big)\,\frac\Delta2-\ii r\,(\rho,\sigma_0)} \nonumber \\[4pt] &= \prod_{\rho
                                               \in \Lambda_R} \, {\tt s}_{b}
                                               \big(
                                               \tfrac{\mathrm{i}}{2} \,
                                               (b + b^{-1})
                                               \, (1 - \Delta) -
                                               r\, ( \rho, \sigma_0 )
                                               \big)
 \ .
\label{eq:3dhypellipsoid}
\end{align}
Note that here the squashing parameter $b$ serves as a 
zeta-function regulator in the infinite product formula for the double-sine function \eqref{eq:doublesine}
in the hypermultiplet contribution. For the $\cN=1$ adjoint
hypermultiplet with $\Delta=1$, the product over $\Lambda_R=\triangle$ can be
split into contributions from positive-negative pairs of roots and the
one-loop 
contribution is trivial: $Z_{\rm hyp}(S_b^3)=1$.
 
\subsection{Extension to $S^1\times C_g$}
\label{sec:ext3d}

One may easily adapt the previous prescriptions to other
three-dimensional geometries, which are related to but so far not included in our discussion.
As a very basic example, we can consider the three-manifold $M_3=S^1
\times C_g$, where the circle $S^1$ has radius $\epsilon$ and $C_g$ is a closed 
Riemann surface of genus $g\geq0$ of area ${\rm
  vol}(C_g)= 4\pi\,r^2$. These manifolds are of course examples of regular fibrations, 
but admit a different localization locus than those considered thus far.

The fixed point locus in this case consists of $D=0$ and covariantly constant $\sigma$, while the curvature of the gauge connection vanishes along $S^1$ and is further constrained by
\begin{equation}
	v \,\llcorner\, \ast F = - 2\, \sigma \ ,
\end{equation}
where $\ast$ is the Hodge duality operator constructed from the metric
of $M_3$.
Then after gauge fixing, the set of localization equations is solved by connections $A$ whose 
component $A_{\theta}$ along the fibre is a non-vanishing constant $a_0$, which we take to lie in
a chosen Cartan subalgebra of $\frg$. 
Expressing the restriction of the curvature to $C_g$ in terms of the
symplectic form of $C_g$ as $F_{C_g}= f_H\, \omega_{C_g}$, we must impose $[a_0, f_H]=0$ to satisfy the condition that $F$ has vanishing components along $S^1$. This implies $\sigma=- \frac{1}{2}\, f_H$ is constant, 
and can be conjugated into the same Cartan subalgebra of $\mathfrak{g}$ as $a_0$.
This gives a flux $\vec{m}\in \Lambda $ for the restriction $F_{C_g}$ to the base
$C_g$, where $\Lambda$ is a co-root lattice of $\mathfrak{g}$ in the same Cartan subalgebra to which $a_0$ belongs. 
Finally the localization locus is
parameterized by $\vec{m}$ and $a_0$ as
\begin{align}
A=a_0\ \dd\theta + \vec m\,A_{C_g} \qquad \mbox{and} \qquad \sigma=-\frac{\vec
  m}{r} \ ,
\end{align}
where $A_{C_g}$ is a monopole gauge field on $C_g$ of unit first
Chern class, $\dd A_{C_g}=2\pi\,\omega_{C_g}$. 
On the other hand, the circle
bundle over $K_2=C_g$ is trivial, hence the associated line bundle $\scrL$ has a
trivial first Chern class contribution to the index and the
topological dependence of the
localization formula is solely through the Euler
characteristic $\chi(C_g)=2-2g$. 

Including such
modifications, the cohomological localization formula yields the
one-loop contribution from the vector multiplet as\footnote{We keep the $n=0$
  contribution, and use the overall harmonic dimension factor coming from the ghosts to
  cancel the Jacobian in the measure for integration over $a_0$ in the
  localized path integral after
  gauge rotation into a Cartan subalgebra of $\frg$.}
\begin{align}
	Z_{\rm vec} ( S^1 \times C_g ) & = \prod_{\alpha \in
                                            \triangle} \ \prod_{n \in
                                            \mathbb{Z} }\,
                                            \bigg(\frac{n}{\epsilon} -
                                            \frac{ \ii (\alpha, \vec{m}) }{r}  - (\alpha, a_0) \bigg)^{1-g}  \notag \\[4pt]
		& =  \prod_{\alpha \in \triangle}\, \bigg( 2 \sinh \Big(  \frac{  \epsilon\, (\alpha, a_0) }{2} +  \frac{ \mathrm{i}\, \epsilon\, (\alpha, \vec{m}) }{2\, r} \Big) \bigg)^{1-g} \notag \\[4pt]
		& = \prod_{\alpha \in \triangle_+ }\, \e^{ (1-g)\,
                  \frac{\epsilon}{r}\, ( \alpha , \vec{m} ) } \, \Big(
                  1- \e^{-\mathrm{i}\, \epsilon\, (\alpha, a_0) }
                  \e^{- \frac{\epsilon}{r}\, ( \alpha , \vec{m} ) }
                  \Big)^{1-g}\, \Big( 1- \e^{\mathrm{i}\, \epsilon\,
                  (\alpha, a_0) } \e^{-\frac{\epsilon}{r}\, ( \alpha , \vec{m} ) } \Big)^{1-g} \notag \\[4pt]
		& = \prod_{\alpha \in \triangle}\,  q^{ -\frac12\, (1-g)\,
                  \lvert  ( \alpha , \vec{m} ) \rvert }\,  \Big(
                  1- \e^{- \mathrm{i}\, \epsilon\, (\alpha, a_0) }\,
                  q^{ \lvert  ( \alpha , \vec{m} ) \rvert }
                  \Big)^{1-g}
\end{align}
where in the last line we defined $q= \e^{- \frac{\epsilon }{r}}$. This reproduces the result of \cite[Section~2.2]{Benini2015}, see also \cite[Section~5.2]{Willett2016}.

\section{$\mathcal{N}=1$ cohomological gauge theories in five dimensions}
\label{sec:N=15d}
	
In this section we derive one-loop fluctuation determinants in various five-dimensional geometries using the Atiyah-Singer index theorem. We first present the supersymmetric gauge theory and its topologically twisted version, and then derive expressions for the one-loop determinants in the regular and irregular cases separately. We subsequently apply the general formalism to some explicit examples, mainly focusing on five-manifolds of the form $M_5=M_3 \times \Sigma_h$, with $M_3$ one of the three-dimensional geometries studied in Section~\ref{sec:N=23d} and $\Sigma_h$ a closed Riemann surface of genus $h$.
In this section we adopt the conventions and normalization of \cite{Kallen2012a}. This differs from the rest of the literature on the topic, and in particular the expressions here will only involve anti-holomorphic Dolbeault differentials.\footnote{The normalization in \cite{Kallen2012a} uses the opposite sign for the contact structure, compared to other literature. After the topological twist, some fields will come with additional minus signs, and in particular a two-form in five dimensions, which is usually taken to be self-dual, becomes anti-self-dual here; with our convention, self-dual 2-forms descend to anti-instantons in four dimensions, and \emph{vice versa}. In practice, the cohomological complex in five dimensions that we will work with only involves the anti-holomorphic Dolbeault differential, while in previous works (see \cite{Qiu2016} for a review) the contributions from both holomorphic and anti-holomorphic forms are included. The final results will of course be the same in either convention, but the intermediate steps will slightly differ.
	The only motivation for our choice is to achieve a unified treatment in three and five dimensions. Furthermore, the topological twist of the hypermultiplets involves only anti-holomorphic forms, so this choice also puts the vector multiplet and the hypermultiplets on the same footing.\label{foot:selfduality}}

Topologically twisted gauge theories on the five-sphere were studied in \cite{Kallen2012,Kallen2012a}, which ignited the stream of activity in this area. The next examples considered were the Sasaki-Einstein manifolds $Y^{p,s}$ in \cite{Qiu2013,Qiufact2013,Schmude2014}. Sasaki-Einstein manifolds are backgrounds which admit $\mathcal{N}=2$ supersymmetry; see Appendix \ref{app:SE} for a brief review.
In \cite{Pan2013}, further five-dimensional geometries preserving $\mathcal{N}=1$ and $\mathcal{N}=2$ supersymmetry were obtained, following the idea of \cite{Festuccia2011,Closset2012} and adapting it to five dimensions. In particular, one can put an $\mathcal{N}=2$ supersymmetric gauge theory on $S^3 \times \Sigma_h$ and $T^3 \times \Sigma_h$, where $T^3 $ is a three-dimensional torus. On the other hand, the manifolds $L(p,1) \times \Sigma_h$ only admit $\mathcal{N}=1$ supersymmetry.
Further geometries admitting Killing spinor solutions, and hence admitting supersymmetric field theories, were obtained in \cite{Alday2015} starting from a holographic setting and taking the rigid limit of supergravity. Sasaki-Einstein manifolds, products $M_3 \times \Sigma_h$ with $M_3$ a Seifert three-manifold, and more general $U(1)$ fibrations over products $C_g \times \Sigma_h$ with $C_g$ and $\Sigma_h$ Riemann surfaces, possibly with orbifold points, are all examples of manifolds studied in \cite{Alday2015}.

\subsection{Supersymmetric Yang-Mills theory and its cohomological formulation}
	
Consider five-dimensional $\mathcal{N}=1$ supersymmetric Yang-Mills
theory. We define the theory on flat Euclidean spacetime $\IR^5$ and
then, by coupling it to background supergravity fields, the theory is
put on curved manifolds $M_5$. For manifolds admitting two Killing
spinors, the $\mathcal{N}=2$ vector multiplet is described in the
$\mathcal{N}=1$ superspace formalism by an $\mathcal{N}=1$ vector multiplet and an $\mathcal{N}=1$ adjoint hypermultiplet. The required modifications to the supersymmetry transformations are described in detail in \cite{Qiu2014}.
Let $\varepsilon$ be the five-dimensional Killing spinor on $M_5$ (see Appendix \ref{app:notation} for our notation), and define the vector field $v$ through\footnote{This differs by a sign from other definitions in the literature, see Footnote \ref{foot:selfduality}.}
\begin{equation}
	v^{\mu} = \varepsilon^{\dagger}\, \Gamma^{\mu} \varepsilon \ .
\end{equation}
It is a nowhere vanishing Killing vector on $M_5$.

\subsubsection*{Vector multiplet}
\label{sec:5dvmtwist}

The five-dimensional $\mathcal{N}=1$ vector multiplet consists of a
gauge connection $A$, a scalar $\sigma$, a symplectic Majorana spinor
$\lambda$ and an auxiliary real scalar $D$, where $\lambda = (\lambda^{I})$ is a $SU(2)_R$ doublet and $D = ({D^{I}}_{J})$ is a $SU(2)_R$ triplet. The supersymmetry transformations in flat space are
\begin{align}
	{\sf Q}_{\varepsilon} A_{\mu} &= \mathrm{i}\, \varepsilon^{\dagger} _{I}\, \Gamma_{\mu} \lambda^{I} \ , \nonumber \\[4pt]
	{\sf Q}_{\varepsilon} \sigma &= \varepsilon^{\dagger} _{I}\,  \lambda^{I} \ , \nonumber \\[4pt]
	{\sf Q}_{\varepsilon} \lambda ^{I} &= - \tfrac{1}{2}\, \Gamma^{\mu \nu} \varepsilon^{I}\, F_{\mu \nu} - {D^{I}}_{J}\, \varepsilon^{J} + \mathrm{i}\, \Gamma^{\mu} \varepsilon^{I}\, ({\sf D}_{\mu} \sigma) \ , \nonumber \\[4pt]
	{\sf Q}_{\varepsilon} {D_{I}}^{J} &= \mathrm{i}\, \varepsilon^{\dagger} _{I}\, \Gamma^{\mu} ({\sf D}_{\mu} \lambda^{J}) - \mathrm{i}\, \big[ \sigma , \varepsilon^{\dagger} _{I}\, \lambda^{J} \big] + ( \ I \leftrightarrow J \ ) \ ,
\end{align}
where $F$ is the curvature of the gauge connection $A$ and ${\sf D}_{\mu}$ is the covariant derivative, which includes the gauge connection $A$ and also the spin connection when acting on dynamical spinors $\lambda$. Curvature corrections proportional to $\frac1r$ must be added to these flat space transformations when the field theory is put on $M_5$.

At this point we perform the topological twist. We introduce the one-form $\Psi$ and the horizontal anti-self-dual two-form $\chi$ according to\footnote{We are using the same Greek letter $\chi$ for a two-form here and for a zero-form in Section~\ref{sec:N=23d}. There should not be any confusion.}
\begin{equation}
	\Psi_{\mu} = \varepsilon^{\dagger} _{I}\, \Gamma_{\mu} \lambda^{I} \qquad \mbox{and} \qquad \chi_{\mu \nu} =  \varepsilon^{\dagger} _{I}\, \Gamma_{\mu \nu} \lambda^{I} - \eta_{\mu}\,  \varepsilon^{\dagger} _{I}\, \Gamma_{\nu} \lambda^{I} + \eta_{\nu}\,  \varepsilon^{\dagger} _{I}\, \Gamma_{\mu} \lambda^{I} \ ,
\end{equation}
where $\eta$ is the one-form dual to the Killing vector $v$. We regard $M_5$ as a $U(1)$ fibration over a compact K\"{a}hler manifold $K_4$, and when $v$ coincides with the Reeb vector field $\xi$ of the Seifert fibration, then $\eta$ coincides with the K-contact structure $\kappa$ of $M_5$. For squashed geometries, however, $\eta \ne \kappa$.

\subsubsection*{Contact structure and localization locus}

The localizing term we add to the action is the standard one:
	\begin{equation}
	{\sf Q}_\varepsilon V \qquad \mbox{with} \quad V = \int_{M_5}\,
        ({\sf Q}_\varepsilon \lambda)^{\dagger}\, \lambda \ \dd\Omega_{M_5} \ ,
\end{equation}
which in the path integral brings the quantum field theory to the fixed point locus
\begin{equation}
\label{eq:locuseqs}
 	v \,\llcorner\, \ast F= F \ , \quad {\sf D} \sigma = 0 \qquad
        \mbox{and} \qquad D = - \sigma\, \otimes\, 
        \left({}^{1}_{0} \ {}^{ \ 0}_{-1}\right) \ ,
\end{equation}
where $\ast$ is the Hodge duality operator constructed from the metric of $M_5$.

It is important at this point to stress a major distinction in our setting from that of \cite{Kallen2012} and subsequent work. When we work with a product of a three-dimensional contact manifold and a Riemann surface, $M_5=M_3 \times \Sigma_h$, there is a crucial difference: the contact structure $\kappa$ lives on $M_3$, and $\kappa \wedge \mathrm{d} \kappa$ is a volume form on $M_3$, as is clear from \eqref{eq:kappa}, but it need not be a contact structure on $M_3 \times \Sigma_h$.
This is important for a choice of compatible metric. For the supersymmetry transformations to be those of a cohomological field theory, one requires the Lie derivative $\mathcal{L}_v$ to commute with the Hodge duality operator. Equivalently, we need $v$ to generate an isometry. The Hodge duality operator on $\Omega^\bullet(M_3\times\Sigma_h)$ takes the form $\ast_{M_3 \times \Sigma_h} = (-1)^{\bullet}\, \ast_{M_3}  \wedge\, \ast_{\Sigma_h}  $. This is an important simplification in studying the localization locus on product manifolds.

\subsubsection*{The kinetic operator}

The supersymmetry transformation squares to ${\sf Q}_\varepsilon^2 = \mathrm{i}\, {\sf L}_{\phi}$, with
\begin{equation}
\label{eq:defLphi}
	{\sf L}_{\phi} = \mathcal{L}_v + \mathcal{G}_{\phi}
\end{equation}
the sum of the Lie derivative along $v$ and a gauge transformation with parameter $\phi = \mathrm{i}\, \sigma - v \,\llcorner\, A$. Since at the end the integration contour for $\sigma$ must be rotated to the imaginary axis, $\sigma \mapsto \mathrm{i}\, \sigma_0$, we are eventually led to
\begin{equation}
	\phi = - \big(  \sigma_0 + v^{\mu}\, A^{(0)} _{\mu} \big) \ \in \ \mathfrak{g} \ ,
\end{equation}
with $A^{(0)}$ a connection whose curvature is a solution to the first
fixed point equation in \eqref{eq:locuseqs}. Setting $A^{(0)}=0$
retains the perturbative partition function, while contributions from
non-trivial solutions are related to instantons on the horizontal
submanifold.

\subsubsection*{Hypermultiplets}

The field content of a five-dimensional $\cN=1$ hypermultiplet consists of a complex scalar $\mathtt{q}= (\mathtt{q}_{I})$, which forms an $SU(2)_R$ doublet, and a complex spinor $\psi$. These fields are obtained by combining chiral and anti-chiral complex scalars and Dirac spinors. The supersymmetry transformations are
\begin{align}
	{\sf Q}_{\varepsilon} \mathtt{q}_{I} &= - 2\, \mathrm{i}\, \varepsilon^{\dagger} _I\, \psi \ , \nonumber \\[4pt]
	{\sf Q}_{\varepsilon} \psi &= \Gamma^{\mu} \varepsilon^{I}\, ({\sf D}_{\mu} \mathtt{q}_{I}) - \sigma\, \mathtt{q}_{I}\, \varepsilon^{I} \ .
\end{align}
When coupled to background supergravity fields, additional terms proportional to $\frac{1}{r}$ are to be included. 

The topological twist in \cite{Kallen2012a} is then achieved in two steps. 
First, contract all the $SU(2)_R$ indices, and therefore define the $SU(2)_R$ singlet spinor $\mathtt{q}^{\prime}$ from the scalar $\mathtt{q}_{I}$ as
\begin{equation}
	\mathtt{q}^{\prime} = \mathtt{q}_{I}\, \varepsilon^{I} \ .
\end{equation}
The square of the supersymmetry transformation, which equals the kinetic operator in the action, is
\begin{equation}
	{\sf Q}_{\varepsilon} ^2 = \mathrm{i}\, {\sf L}_{\phi} \qquad \mbox{with} \quad {\sf L}_{\phi} = \mathcal{L}_v ^{\rm spin} + \mathcal{G}_{\phi} \ ,
\end{equation}
where we indicated explicitly that the Lie derivative is twisted by the spin connection on $M_5$.

The second step consists in defining a spin$^c$ structure on $M_5$. For this, in~\cite{Kallen2012a} (see also~\cite[Section 3]{Qiu2016}) the following assumption is made. Let $\eta$ be the dual one-form to the Killing vector field $v$. Then, according to \cite{Alday2015}, the most general metric on the Seifert fibration $M_5\to K_4$ admitting supersymmetry is of the form
\begin{equation}
	\dd s^2_{M_5} = \eta\otimes\eta + \dd s^2_{K_4} \ ,
\end{equation}
with transverse Hermitian metric on the K\"ahler surface $K_4$. If $\eta$ is proportional to the contact structure defined by the Seifert fibration, then one can define a canonical spin$^c$ structure on $M_5$. This condition is equivalent to requiring the orbits of $v$ to be all closed. Manifolds supporting $\mathcal{N}=2$ supersymmetry belong to this class \cite{Pan2013}, and the index theorem can be applied in that case.

The spin$^c$ structure identifies, through the action of a representation of the Clifford algebra, spinors with elements of
\begin{equation}
	\Omega^{(0, \bullet)} _H (M_5,\frg) \ ,
\end{equation}
so the hypermultiplet is put in cohomological form. See \cite{Kallen2012a,Qiu2016} for further details.

After the standard localizing term is added to the action, one has to
compute the localization locus. If one considers the trivial solution
$A^{(0)}=0$ in the vector multiplet, then the localization locus
consists in setting all hypermultiplet fields to zero. It was proven
in \cite{Qiu2013} that this holds for any solution
$A^{(0)}$ in the localization locus of the vector multiplet,
as long as $\dd s^2_{M_5}$ is a Sasaki-Einstein
metric.

\subsubsection*{Supersymmetric Yang-Mills action at the localization locus}

Evaluating the full gauge theory action at the fixed point locus on $M_5=M_3\times\Sigma_h$ gives
\begin{equation}\label{eq:5daction}
	S_{\rm cl} (F, \sigma_0) = \frac{1}{2\,g_{\textrm{\tiny YM}}^2}
        \, \int_{M_5}\, \Big( \big(F \stackrel{\wedge}{,} \ast F\big) + \frac{1}{ r}\,
        (\sigma_0,F) \wedge  \kappa \wedge \mathrm{d} \kappa 
        + \frac{1}{r^2}\, (\sigma_0,\sigma_0) \,  \kappa\wedge\dd\kappa \wedge \omega_{\Sigma_h}  \Big) \ ,
\end{equation}
where $g_{\textrm{\tiny YM}}$ is the Yang-Mills coupling constant. Here $\kappa$ is the contact structure on the Seifert three-manifold $M_3$, $\omega_{\Sigma_h}$ is the symplectic structure on the Riemann surface $\Sigma_h$ and $\frac{1}{2}\, \kappa \wedge \kappa \wedge \omega_{\Sigma_h}$ is the volume form \eqref{eq:volformcontact} on $M_3 \times \Sigma_h$.

\subsection{One-loop determinant of the vector multiplet in a regular background}
\label{sec:1loopvecreg5d}

After the topological twist performed in Section \ref{sec:5dvmtwist}, all fields of the vector multiplet are in a cohomological form
\begin{equation}
	A \in \Omega ^{1} (M_5, \mathfrak{g}) \ , \quad \sigma \in \Omega ^{0} (M_5, \mathfrak{g}) \ , \quad \Psi \in \Omega ^{1} (M_5, \mathfrak{g}) \qquad \mbox{and} \qquad \chi \in \Omega ^{2} _{H,-} (M_5, \mathfrak{g}) \ ,
\end{equation}
where by $\Omega^{2} _{H, \pm} (M_5,\frg)$ we denote the spaces of
self-dual and anti-self-dual horizontal two-forms with values in the
Lie algebra $\frg$. Here we assume that $M_5$ is a regular background,
so that contraction by $v$ separates the horizontal and vertical parts
of forms. The gauge connection $A$ is our even coordinate and $\chi$ is the odd coordinate on the space of fields. We also have to introduce ghosts, and we refer to \cite{Kallen2012,Kallen2012a} for the procedure. For our purposes, it suffices to say that these give two even harmonic scalars and two odd scalars.

Gaussian integration of the vector multiplet around the fixed point gives the ratio of fluctuation determinants
\begin{equation}
	h (\phi) = \sqrt{\frac{\det \mathrm{i}\, {\sf L}_{\phi} \vert_{\mathrm{f}} }{\det  \mathrm{i}\, {\sf L}_{\phi} \vert_{\mathrm{b}}} } = \sqrt{ \frac{  \det \mathrm{i}\, {\sf L}_{\phi} \vert_{\Omega^{2} _{H,-}(M_5,\frg)}\,  ( \det \mathrm{i}\, {\sf L}_{\phi} \vert_{\Omega^{0}(M_5,\frg)} )^2 }{\det \mathrm{i}\, {\sf L}_{\phi} \vert_{\Omega^{1}(M_5,\frg)}\, ( \det \mathrm{i}\, {\sf L}_{\phi} \vert_{H^{0}(M_5,\frg)} )^2  } } \ ,
\end{equation}
where $\vert_{\mathrm{f}}$ (respectively $\vert_{\mathrm{b}}$) refers to the operator acting on fermionic (respectively bosonic) fields. Here $H^{0} (M_5, \mathfrak{g})$ is the space of $\mathfrak{g}$-valued harmonic zero-forms on $M_5$, and the differential operator ${\sf L}_{\phi}$ is given in \eqref{eq:defLphi}. The numerator of $h(\phi)$ involves the contributions from the fermionic coordinate $\chi$ and the two fermionic ghost coordinates, while the denominator involves the contributions from the bosonic coordinate $A$ and the two bosonic ghost coordinates.

We split
\begin{align}
	\Omega^{2} _H (M_5,\frg) &= \Omega^{2} _{H,+} (M_5,\frg)  \oplus \Omega^{2} _{H,-} (M_5,\frg) \ , \nonumber \\[4pt]
 \Omega^{2} _{H,+} (M_5,\frg) &=  \Omega^{(2,0)} _H (M_5,\frg) \oplus  \Omega^{(0,2)} _H (M_5,\frg) \oplus \Omega^{(1,1)} _{\mathrm{sympl}} (M_5,\frg) \ ,
\end{align}
where $\Omega^{(1,1)} _{\mathrm{sympl}} (M_5,\frg) $ are the $\frg$-valued horizontal two-forms proportional to the symplectic structure on the base K\"ahler manifold $K_4$. Then 
\begin{equation}
	\Omega^{(1,1)} _H (M_5,\frg) = \Omega^{(1,1)} _{\mathrm{sympl}} (M_5,\frg) \oplus \Omega^{2} _{H,-} (M_5,\frg) \cong \Omega^{0} (M_5,\frg) \oplus \Omega^{2} _{H,-} (M_5,\frg) \ .
\end{equation}
Here we consider the regular case, in which $v$ is parallel to the Reeb vector field $\xi$. With our choices, $v=-\xi = - \frac\partial{\partial\theta}$, where $\theta \in [0, 2 \pi\, r)$ is the coordinate along the circle fibre. We can decompose horizontal forms according to the fibration structure of $M_5\to K_4$ as
\begin{equation}
	\Omega^{(\bullet,\bullet)} _H (M_5, \mathfrak{g}) = \Omega^{(\bullet,\bullet)}  (K_4 , \mathfrak{g})\oplus \bigoplus_{m \ne 0}\,   \Omega^{(\bullet,\bullet)}  (K_4 , \mathscr{L}^{\otimes m } \otimes \mathfrak{g} ) \ .
\end{equation}

The crucial step now is to recognise that, according to this
splitting, the Lie derivative along the Killing vector field $v$ acts
on a form  $\alpha_m \in \Omega^{(\bullet,\bullet)}  (K_4 , \mathscr{L}^{\otimes m } \otimes \mathfrak{g} )  $ as
\begin{equation}
	\mathcal{L}_v \alpha_m = - \mathcal{L}_{\xi} \alpha_m = - \frac{ \mathrm{i}\,{m}}{r}\, \alpha_m \ .
\end{equation}
The fact that the orbits of $v$ coincide with the orbits of the Reeb vector field $\xi$ is essential here. 
The action of $\mathrm{i}\, {\sf L}_\phi$ on each Kaluza-Klein mode
labelled by $m \in \mathbb{Z}$ also includes the gauge transformation
$\cG _{\phi}$, whose eigenmodes are found decomposing the Lie algebra
$\mathfrak{g}$ into its root system
\begin{align}
\mathfrak{g} = \bigoplus_{\alpha \in \triangle}\,
  \mathfrak{g}_{\alpha} \ .
\end{align} 
We are therefore ready to evaluate
\begin{align}
	h(\phi) & = \frac{1}{\big\lvert \det \mathrm{i}\,{\sf L}_{\phi} \vert_{H^0(M_5, \mathfrak{g})} \big\rvert } \ \sqrt{ \frac{ \det \mathrm{i}\,{\sf L}_{\phi} \vert_{\Omega^2 _{H,-} (M_5, \mathfrak{g})}\, \det \mathrm{i}\,{\sf L}_{\phi} \vert_{\Omega^0(M_5, \mathfrak{g})} }{ \det \mathrm{i}\,{\sf L}_{\phi} \vert_{\Omega^{1} _{H} (M_5, \mathfrak{g})} } } \nonumber \\[4pt]
	&= \frac{1}{\big\lvert \det \mathrm{i}\,{\sf L}_{\phi} \vert_{H^0(M_5, \mathfrak{g})} \big\rvert } \ \sqrt{ \frac{ \mathcal{D}_0 ^{(1,1)} (\phi)}{\cD_0^{(1,0)}(\phi) \, \mathcal{D}_0 ^{(0,1)} (\phi) }} \ \prod_{m \ne 0} \,  \sqrt{ \frac{ \mathcal{D}_m ^{(1,1)} (\phi)}{  \mathcal{D}_m ^{(1,0)} (\phi) \,  \mathcal{D}_m ^{(0,1)}  (\phi) }     }  \ ,
\end{align}
where in the second line we denoted
\begin{equation}
	\mathcal{D} ^{(\bullet,\bullet)} _m (\phi) = \det \mathrm{i}\,{\sf L}_{\phi} \vert_{\Omega^{(\bullet,\bullet)}  (K_4 , \mathscr{L}^{\otimes m } \otimes \mathfrak{g} )  }  \ .
\end{equation}
Standard manipulations at this point~\cite{Kallen2012a} (see also~\cite[Appendix C]{Kallen2012}) finally lead to the cohomological localization formula
\begin{equation}
\label{eq:ZvecgenM5reg}
\begin{tabular}{|c|}\hline\\
$ \displaystyle{	Z_{\rm vec} (M_5) = \prod_{\alpha \in
  \triangle }\, \big( \mathrm{i}\, (\alpha, \phi)
  \big)^{\frac{1}{12}\,( c_2 (K_4) + c_1 (K_4)^2) - \dim H^0(M_5,\IR)}
  \ \prod_{m \ne 0}\,  \Big( \frac{m}{r} + \mathrm{i}\, (\alpha, \phi)
  \Big)^{\mathrm{index}\, \bar{\partial}^{(m)}}
}
$\\\\\hline\end{tabular}
\end{equation}
For a $U(1)$ bundle $M_5 \to C_g \times \Sigma_h$ over the product of two Riemann surfaces of genera $g$ and $h$, the power of the first multiplicative factor is $(1-g)\,(1-h) - 1$.

\subsection{One-loop determinant of the vector multiplet in an irregular background}
\label{sec:irreg5d}

We now consider the alternative case of an irregular fibration, whereby $v$
does not point along the $U(1)$ fibre of $M_5$. Let
$\eta$ be the dual one-form to the Killing vector field $v$. It is
an almost contact structure on $M_5$; if it is a contact structure,
then we are in the situation of Section~\ref{sec:1loopvecreg5d}
above. For the present discussion, we assume that $M^{g,h}_5 \to C_g \times
\Sigma_h$ is a $U(1)$ fibration over a direct product of two Riemann
surfaces, both compact and closed. Rotations along the
circle fibre are assumed to act freely\footnote{The Sasaki-Einstein
  manifolds $Y^{p,s}$ do not belong to this class, see Appendix
  \ref{app:SE} and Section \ref{sec:examplesregvsirr}.} on $M^{g,h}_5$. This
means that, although the gauge theory could be put on
$M_5^{g,h}$ preserving $\mathcal{N}=1$ supersymmetry when $M^{g,h}_5 /U(1)$ admits orbifold points \cite{Alday2015}, the
procedure we describe below does not apply to that case. Since we are
in the irregular setting, we need an additional $U(1)$ isometry on
$C_g \times \Sigma_h$. In practice, this restricts our considerations to $C_0= S^2$ or $C_1=T^2$.

Most of the procedure is exactly the same as in
Section~\ref{sec:1loopvecreg5d}, particularly the decomposition of
differential forms in terms of the Reeb vector field
$\xi$. Nonetheless, we have to face two problems. First, as for the
irregular three-dimensional case, we have to bear in mind that forms
$\alpha_m \in \Omega^{(\bullet,\bullet)} (C_g\times\Sigma_h, \mathscr{L}^{\otimes m}\otimes\frg )$ are
no longer eigenmodes of $\mathcal{L}_v$. The other important issue is that
now the conditions following from the definition of the two-form $\chi$,
\begin{equation}
	v\,\llcorner\,\chi = 0 \qquad \mbox{and} \qquad v
        \,\llcorner\,\ast \chi = - \chi \ ,
\end{equation}
cannot be interpreted as saying that $\chi$ is a horizontal
anti-self-dual two-form. If we express $v$ as a linear combination 
\begin{align}
v=
a_1\, \xi + a_2\, \widetilde{\xi} \ , 
\end{align}
where $\xi$ is the Reeb vector
field and $\widetilde{\xi}$ is a vector field orthogonal to $\xi$ generating a $U(1)$ action, we find that the vertical part of $\chi$ may not
vanish, but it lies in the subspace orthogonal to
$\widetilde{\xi}\,$. However, more is true: we can decompose $\chi$ in terms of a two-form $\chi_T$ and a one-form $\chi_P$. Explicitly
\begin{equation}
	\chi = \Big( \kappa - \frac{a_1}{a_2}\, \widetilde{\kappa}
        \Big) \wedge \chi_P + \chi_T  \ ,
\end{equation}
where $\kappa$ is the contact structure and $\widetilde{\kappa}$ is
the one-form dual to $\widetilde{\xi}\,$, and
with further anti-self-duality relations imposed on $\chi_P$ and
$\chi_T$. In the end, we are left with the same number of degrees of
freedom as for a horizontal anti-self-dual two-form.
From the more geometric perspective of transverse holomorphic
foliations, the most natural point of view is to consider now the
index of a \emph{new} Dolbeault-like operator
$\widetilde{\partial}\,^{(m)}$, whose cohomological complex is a deformation
of the cohomological complex of the regular case according to the
deformation of the transverse holomorphic foliation of $M^{g,h}_5$ as
described in \cite[Section 5]{Closset2013} (see also \cite[Section
7]{Closset2013} and \cite[Section 5]{Closset2019} for a discussion
about the particular case of ellipsoids).

At this point, we can again follow the approach of
\cite{Drukker2012} and extend it to five dimensions. The action of the
Reeb vector field is free, and we can reduce the computations to the
quotient space $ C_g \times \Sigma_h$. In this way we arrive at the
cohomological localization formula
\begin{equation}
\begin{tabular}{|c|}\hline\\
$ \displaystyle{	Z_{\rm vec} \big(M^{g,h}_5\big) = \prod_{\alpha \in \triangle }\, \big(
        \mathrm{i}\, (\alpha, \phi) \big)^{g\,h-g-h} \ \prod_{f} \
        \prod_{m \ne 0}\, \Big( \frac{m}{\epsilon_f} + \mathrm{i}\,
        (\alpha, \phi ) \Big)^{\frac{1}{2}\, \mathrm{index} \,
        \bar{\partial} ^{(m)} }
}
$ \\\\\hline\end{tabular}
\end{equation} 
with an extra product over the fixed points of the additional
$U(1)$ action on $C_g \times \Sigma_h$. The length parameter
$\epsilon_f$ is the radius of the circle fibre over the fixed
point labelled by $f$. 

\subsubsection*{One-loop determinant of the vector multiplet on
  $\boldsymbol{M_3 \times \Sigma_h}$}

We will now specialise the present discussion to product manifolds $M_5^{g,h}=M_3 \times
\Sigma_h$, where the classification reduces to the discussion of
\cite{Closset2012} about
the orbits of the three-dimensional Killing vector field on $M_3$. We shall
explicitly compute the Atiyah-Singer index in this case. The K\"ahler
surface $K_4=C_g\times\Sigma_h$ is endowed with the product K\"ahler
structure $\omega=\omega_{C_g}+\omega_{\Sigma_h}$ and the
$U(1)$-bundle projection $\pi$ is the product of the Seifert fibration
$M_3\to C_g$ and the identity map on $\Sigma_h$. The integer
cohomology of $K_4$ has generators $[\omega_{C_g}]\in H^2(C_g,\IZ)$
and $[\omega_{\Sigma_h}]\in H^2(\Sigma_h,\IZ)$ in this case, and the first Chern class of the line bundle $\scrL\to K_4$ associated to the circle fibration is given by $c_1(\scrL) = {\rm deg}(\scrL)\, \omega_{C_g}$. One has
$c(T^{1,0}\Sigma_h)=1+\chi(\Sigma_h)\,\omega_{\Sigma_h}=1+c_1(T^{1,0}\Sigma_h)$
with $\chi(\Sigma_h)= 2-2h$ the Euler characteristic of the Riemann
surface $\Sigma_h$, and similarly for $C_g$. The total Chern class is thus
\begin{align}
c (T^{1,0} K_4) & = c (T^{1,0} C_g ) \wedge c (T^{1,0}  \Sigma_h ) \notag \\[4pt]
	&= \big( 1 + \chi (C_g)\, \omega_{C_g} \big) \wedge \big( 1 + \chi (\Sigma_h)\, \omega_{\Sigma_h} \big) \notag \\[4pt]
	&= 1 + 2\,\big( (1-g)\, \omega_{C_g} + (1-h)\, \omega_{\Sigma_h} \big)  + 4\,(1-g)\,(1-h)\, \omega_{C_g} \wedge  \omega_{\Sigma_h} \notag \\[4pt]
	&= 1+c_1(T^{1,0}K_4)+c_2(T^{1,0}K_4) \ ,
\end{align}
and the corresponding Todd class is
\begin{align}
	\mathrm{Td} (T^{1,0}K_4 ) &= 1+\mbox{$\frac12$}\, c_1(T^{1,0}K_4)+ \mbox{$\frac1{12}$}\, \big(c_1(T^{1,0}K_4)\wedge c_1(T^{1,0}K_4)+c_2(T^{1,0}K_4)\big) \nonumber \\[4pt]
	&= 1 + (1-g) \, \omega_{C_g} + (1-h)\, \omega_{\Sigma_h}  + (1-g)\,(1-h)\, \omega_{C_g} \wedge \omega_{\Sigma_h} \ .
\end{align}
Using $c_1(\scrL^{\otimes m})=m\,{\rm deg}(\scrL)\, \omega_{C_g}$, the corresponding Chern character is found to be
\begin{equation}
	\mathrm{ch} ( \mathscr{L}^{\otimes m } ) = 1 + c_1  ( \mathscr{L}^{\otimes m } ) + \tfrac12\,c_1  ( \mathscr{L}^{\otimes m } )\wedge c_1  ( \mathscr{L}^{\otimes m } ) = 1 + m\, {\rm deg}( \mathscr{L} )\, \omega_{C_g} \ .
\end{equation}

The index of the Dolbeault complex in this case is thus given by
\begin{align}
	\mathrm{index}\, \bar{\partial } ^{(m)} &= \int_{K_4}\, \big( 1 + m\,{\rm deg}( \mathscr{L} )\,\omega_{C_g}\big) \wedge \big( 1 + (1-g)\,  \omega_{C_g} +(1-h)\, \omega_{\Sigma_h} \notag \\ & \hspace{6cm} +  (1-g)\,(1-h)\, \omega_{C_g} \wedge \omega_{\Sigma_h}  \big) \notag \\[4pt]
		&= \int_{K_4}\, \big( (1-g)\,(1-h) + m\, (1-h)\, {\rm deg}( \mathscr{L} )\big)\, \omega_{C_g} \wedge \omega_{\Sigma_h} \notag \\[4pt]
		&= (1-h)\, \int_{C_g }\, \big( m\,  {\rm deg}( \mathscr{L} ) + 1-g\big)\, \omega_{C_g} \notag \\[4pt]
		&= (1-h) \, \big[ m\, \mathrm{deg} (\mathscr{L}) + 1-g \big] \ .
\label{eq:indexcalc}
\end{align}
The term in square brackets is the index of the twisted Dolbeault complex associated to the circle bundle $M_3 \to C_g$, and we finally get the localization formula
\begin{equation}
\label{eq:ZvecM3xSigma}
\begin{tabular}{|c|}\hline\\
$ \displaystyle{	Z_{\rm vec} (M_3 \times \Sigma_h) =  Z_{\rm vec} (M_3)^{1-h}  \ \prod_{\alpha \in \triangle_{+} }\, ( \alpha, \mathrm{i}\, \sigma )^{-2h } }
$ \\\\\hline\end{tabular}
\end{equation}
Again, the multiplicative factor in the general localization formula, which includes also contributions from the ghosts, is essential for cancelling the denominator.
This formula can be extended to the case in which $C_g$ has orbifold points, as reviewed in \cite[Section 3.5]{Closset2019}.

Notice that the localization formula for the one-loop determinants on $M_3 \times \Sigma_h$ lifts the \emph{perturbative} three-dimensional partition function to the \emph{perturbative} five-dimensional partition function.
However, while the full (non-perturbative) partition function on $M_3$ receives contributions from flat connections on $M_3$, the full partition function on $M_3 \times \Sigma_h$ includes connections $A^{(0)}$ that descend to instantons on $C_g \times \Sigma_h$. The moduli spaces over which we integrate are different. In fact, the pullback to $M_5$ of flat connections on $M_3$ are not generally solutions to the fixed point equation~\eqref{eq:locuseqs}. This is a major difference from the partially twisted theory, in which the BPS configurations decompose into flat connections on $M_3$ and unconstrained connections on $\Sigma_h$.

\subsection{One-loop determinant of a hypermultiplet}

For the contribution of a hypermultiplet, we will only consider a
regular background here, due to the issues encountered with the vector multiplet discussed in Section~\ref{sec:irreg5d}. Furthermore, we consider only the one-loop determinant in the perturbative partition function, that is we set $A^{(0)}=0$, hence $\phi = - \sigma_0$. We want to calculate the ratio of fluctuation determinants
\begin{equation}
	 \sqrt{\frac{\det \mathrm{i}\, {\sf L}_{- \sigma_0} \vert_{\mathrm{f}} }{\det  \mathrm{i}\, {\sf L}_{- \sigma_0} \vert_{\mathrm{b}}} } = \sqrt{ \frac{\det \mathrm{i}\, {\sf L}_{- \sigma_0} \vert_{\Omega^{(0,1)} _H(M_5,\frg)}   }{   \det \mathrm{i}\, {\sf L}_{- \sigma_0 } \vert_{\Omega^{0} _H(M_5,\frg) } \, \det \mathrm{i}\, {\sf L}_{- \sigma_0 } \vert_{\Omega^{(0,2)} _H(M_5,\frg) } } } \ .
\end{equation}
Applying the same strategy as with the vector multiplet, that is, decomposing the horizontal forms according to the tensor powers of the line bundle $\mathscr{L}$ associated to the $U(1)$ fibration, one arrives at the cohomological localization formula
\begin{equation}\label{eq:5dhyperpert}
\begin{tabular}{|c|}\hline\\
$ \displaystyle{	Z_{\rm hyp} ^{\rm pert}(M_5) = \prod_{\rho \in \Lambda_R} \ \prod_{m \in \mathbb{Z}}\, \Big( \frac{m + \Delta}{r} - \mathrm{i}\, (\rho, \sigma_0) \Big)^{- \mathrm{index} \, \bar{\partial} ^{(m)} } }
$ \\\\\hline\end{tabular}
\end{equation}

We briefly comment on the cohomological formulation of the hypermultiplet in a squashed or ellipsoid background. The topological twist depends, in general, on the geometric data. 
However, as discussed in \cite{Qiu2014} and also in \cite[Section 3.3]{Qiu2016}, we can turn on the squashing and continuously deform 
the contact structure, so that the Reeb vector field associated to the new contact structure stays parallel to the Killing vector field $v$. Then, the cohomological localization applies to the vector multiplet, 
although using a notion of `horizontal' which differs from that on the round manifold we started with. 
For the topological twist of the hypermultiplet, however, a choice of spin$^{c}$ structure is needed, and hence additional assumptions on the geometry of the base of the $U(1)$ fibration are required, usually that it is K\"{a}hler-Einstein. 
The pragmatic solution of \cite{Qiu2014} was to take the cohomological form of the hypermultiplet as a \emph{definition} in a squashed Sasaki-Einstein geometry. This is not, however, a continuous deformation of the hypermultiplet in the round geometry, and we do not follow this strategy here.

\subsection{Perturbative partition functions}
\label{sec:pertN=2}

We now come to the first applications of our cohomological localization formulas
in five dimensions. Pan proved in \cite{Pan2013} that the product manifolds $M_3 \times
\Sigma_h$ admit $\mathcal{N}=2$ supersymmetry if $M_3 = S^3$ or $M_3=T^3$,
the sphere or the torus. The second choice is not included in the discussion so far, and we will consider it later in Section \ref{sec:5dDirectProd}. 
We will work out the full
perturbative $\mathcal{N}=2$ partition functions on $S^3 \times
\Sigma_h$. We shall then write down the
perturbative $\cN=1$ partition functions on more general product
five-manifolds $M_3\times\Sigma_h$.

\subsubsection*{$\boldsymbol{\cN=2}$ perturbative partition
  functions on $\boldsymbol{S^3\times\Sigma_h}$}

Regard the three-sphere $S^3$ as the Hopf fibration of degree one over
$C_0=S^2$. 
We use the localization formula \eqref{eq:ZvecM3xSigma} with
$A^{(0)}=0$ together with
\eqref{eq:5dhyperpert} which gives
\begin{align}
	Z_{\rm vec}^{\rm pert} (S^3 \times \Sigma_h ) =  Z_{\rm vec} (S^3) ^{1-h}
  \qquad \mbox{and} \qquad
	Z_{\rm hyp}^{\rm pert} (S^3 \times \Sigma_h ) =  Z_{\rm hyp} (S^3) ^{1-h}
  \ .
\end{align}
The full perturbative partition function is given by taking
the product of the one-loop vector multiplet determinant with products
of the one-loop hypermultiplet determinants over all $\cN=2$
hypermultiplets $a$ of conformal dimensions $\Delta_a$ in
representations $R_a$ of the gauge group $G$. We then multiply by the Boltzmann weight
of the classical action \eqref{eq:5daction} evaluated at the trivial
solution $A^{(0)}=0$, and integrate over the remaining scalar moduli
$\sigma_0\in\frg$ using the localization formulas of
Section~\ref{sec:3dappl}. We can conjugate $\sigma_0$ into a Cartan
subalgebra $\frt\subset\frg$ and use the Weyl integral formula to perform the resulting integral with the measure
\begin{align}\label{eq:Weylformula}
\dd\mu(\sigma_0) = \dd\sigma_0 \ \prod_{\alpha\in\triangle_+}\,
  (\alpha,\sigma_0)^2 \ ,
\end{align}
where $\dd\sigma_0$ is the Lebesgue measure on $\frt=\IR^{{\rm
    rank}(G)}$. The applicability of the Weyl integral formula is
restricted to elements $\sigma_0\in\frt$ for which the determinant in
\eqref{eq:Weylformula} is non-vanishing; these are called regular
elements, and they form an open dense subset $\frt_{\rm reg}\subset\frt$. 
After cancelling the Jacobian
in the integration measure with the denominator of the vector multiplet one-loop determinant, we obtain the perturbative partition function in the background of \cite{Pan2013}:
\begin{align}
Z_{\mathcal{N}=2} ^{\rm pert} (S^3 \times \Sigma_h) =
                                                      \int_{\mathfrak{t}_{\rm reg}}\,
                                                      \dd
                                                      \tilde{\sigma}\,
                                                      \e^{-\frac{
                                                      4\pi^2\,{\rm vol}(\Sigma_h)} {r\,g^2
                                                      _{\textrm{\tiny
                                                      YM}}}\, (\tilde\sigma,\tilde\sigma)} \ &
                                                      \prod_{\alpha
                                                      \in
                                                      \triangle_+}\,
                                                      \sinh \big(
                                                      \pi\, (\alpha,
                                                      \tilde{\sigma})
                                                      \big)^{2-2h}
                                                      \nonumber \\ &
                                                                     \times
                                                                     \ 
                                                                     \prod_{a}
                                                                     \
                                                                     \prod_{\rho_a
                                                                     \in
                                                                     \Lambda_{R_a}
                                                                     }\,
                                                                     {\tt
                                                                     s}_{1}
                                                                     \big(
                                                                     \mathrm{i}\,
                                                                     (1-
                                                                     \Delta_a)
                                                                     -
                                                                     (\rho_a,
                                                                     \tilde{\sigma}
                                                                     )
                                                                     \big)^{1-h}
                                                                     \
                                                                     , 	
\label{eq:N=2pertS3}\end{align}
where we defined the variable $\tilde{\sigma}= r\,
\sigma_0$ in the Cartan subalgebra $\mathfrak{t}_{\rm reg} \subset
\mathfrak{g}$. 
The $\mathcal{N}=1$ partition function on $S^3 \times \Sigma_h$
without matter and including instanton contributions will be analysed
in Section~\ref{sec:qYMS3xSigma}.

\subsubsection*{$\boldsymbol{\cN=1}$ perturbative partition functions
  on $\boldsymbol{M_3\times\Sigma_h}$}

We shall now consider the $\mathcal{N}=1$ perturbative partition
functions on more general $U(1)$ fibrations over $C_g \times
\Sigma_h$, where $C_g$ is a Riemann surface of genus $g$, focusing on
the case $M_5=M_3\times\Sigma_h$ where only $M_3\to C_g$ is fibered
over $C_g$. Proceeding as above using the index formula
\eqref{eq:indexcalc}, we arrive at
\begin{align}
Z_{\mathcal{N}=1} ^{\rm pert} (M_3 \times \Sigma_h)  =
  \int_{\mathfrak{t}_{\rm reg}}\, \mathrm{d} \tilde{\sigma}\, \e^{- \frac{\pi\,
  \mathrm{vol} (C_g)\, \mathrm{vol} (\Sigma_h) }{r^3\,g^2
  _{\textrm{\tiny YM}}}\, (\tilde{\sigma},\tilde\sigma) } & \ \prod_{\alpha \in \Delta_{+}}\, \sinh \big( \pi\, (\alpha, \tilde{\sigma} ) \big)^{2\,(1-h)\,(1-g) }  \\
	 & \times \, \prod_{a} \ \prod_{\rho_a \in \Lambda_{R_a} }\,
           {\tt s}_{1} \big( \mathrm{i}\, (1- \Delta_a) - (\rho_a ,
           \tilde{\sigma} ) \big)^{(1-h) \deg (\mathscr{L}) } \ . \notag
\end{align}
The products $L(p,1) \times \Sigma_h$ are particular
examples~\cite{Pan2013} with $C_0=S^2$, for which $\deg(\scrL)=p$. The
formalism should also apply when $M_3$ is a more general Seifert
homology sphere which admits a contact structure, and in particular
for the lens spaces $M_3=L(p,s)$. It would also be interesting to extend the formalism to the
case in which $\Sigma_h$ has punctures.

\subsection{Contact instantons and their pushdown to four dimensions}
\label{sec:continst-pushdown}

We shall now work out solutions to the fixed point equation \eqref{eq:locuseqs} on $M_5=M_3 \times \Sigma_h$, and then study their pushdown to four dimensions.

\subsubsection*{Regular fibrations}
\label{sec:solFreg}

We first focus on regular Seifert manifolds. We want to solve the equation
	\begin{equation}
	\label{eq:contactinstreg}
		v \,\llcorner\, \ast F = F
	\end{equation}
on $M_3 \times \Sigma_h$, where $v=-\xi$ is the Killing vector field with $\xi$ the Reeb vector field of the Seifert fibration $M_3\to C_g$. 
On K-contact five-manifolds, the solutions to this equation are
refered to as contact instantons~\cite{Kallen2012}, and their moduli
spaces are studied in~\cite{Baraglia:2014gma}. We can rewrite \eqref{eq:contactinstreg} as
\begin{equation}
\ast F = \kappa \wedge F 
\end{equation}
or equivalently
\begin{equation}\label{eq:loceqequiv}
v \,\llcorner\, F =0 \qquad \text{and} \qquad F_{H,-} =0 \ ,
\end{equation}
where $F_{H,\pm}$ denote the self-dual and anti-self-dual horizontal parts of the curvature two-form $F$.
Let us decompose the gauge connection as
\begin{equation}
	A = A_{\theta}\, \kappa + A_{H} \ ,
\end{equation}
where $\kappa$ is (minus) the contact structure on $M_3$ dual to $v$, so that
\begin{equation}
F =  \big( A_{\theta}\, \mathrm{d} \kappa +  \mathrm{d} A_{\theta} \wedge \kappa + \mathrm{d} A_H  \big) -\mathrm{i}\, \big( A_H \wedge A_H + [A_{\theta}, A_H ] \wedge \kappa  \big) \ .
\end{equation}

We partly follow the treatment of \cite[Section 3.2]{Kallen2012a}. The first equation $v \,\llcorner\, F =0$ reads
\begin{equation}\label{eq:Athetacovconst}
{\sf D}_H A_\theta = \mathrm{d}A_\theta + \ii [A_{H}, A_{\theta}] =0 \ ,
\end{equation}
so $A_{\theta}$ is covariantly constant along $C_g \times \Sigma_h$. At this point, it is useful to prove that both the $\frg$-valued function $A_{\theta}$ and one-form $A_H$ are invariant under translations along the fibre, generated by $v$. For this, we choose the gauge\footnote{We avoid formal considerations involved in the gauge fixing procedure. The details are exactly as in \cite{Kallen2012,Kallen2012a}.}
\begin{equation}
\label{eq:choicegauge}
	\mathcal{L}_v A = 0 \ .
\end{equation}
Then
\begin{align}
\mathcal{L}_v A_{\theta} = v \,\llcorner\, \mathrm{d} A_{\theta} = -\ii v \,\llcorner\,  [A_H, A_{\theta}] = 0 \qquad \mbox{and} \qquad
	\mathcal{L}_v A_{H} = v \,\llcorner\, \mathrm{d} A_{H} +  \mathrm{d} (v  \,\llcorner\, A_{H}) = 0 \ ,
\end{align}
where $v \,\llcorner\, \mathrm{d} A_{H}=0$ follows from the gauge fixing condition \eqref{eq:choicegauge}.

From \eqref{eq:Athetacovconst} it follows that the curvature $F$ has only a horizontal part given by
\begin{align}
F=F_H = A_{\theta}\, \mathrm{d} \kappa + \mathrm{d} A_{H}  - \mathrm{i}\, A_{H} \wedge A_H \ .
\end{align}
At this point we use the second equation $F_{H,-} =0$.
The surviving self-dual part belongs to the vector space
\begin{equation}
	F_{H,+} \ \in \ \Omega^{(1,1)}_{\rm sympl} (C_g \times \Sigma_h,\frg) \oplus \Omega^{(2,0)} (C_g \times \Sigma_h,\frg) \oplus \Omega^{(0,2)} (C_g \times \Sigma_h,\frg) \ .
\end{equation}
This implies that $\dd A_H$ is proportional to the K\"ahler two-form $\omega_{C_g}+\omega_{\Sigma_h}$ on the base, and recalling the relation \eqref{eq:dkapparel} between $\mathrm{d} \kappa $ and the K\"{a}hler form on $C_g \times \Sigma_h$, 
altogether we arrive at a curvature which is of the form
\begin{equation}
	F = f_H \, \big( \omega_{C_g} + \omega_{\Sigma_h}\big) + F^{(2,0)} + F^{(0,2)} \ ,
\end{equation}
where
\begin{equation}
	F^{(2,0)}  = - \mathrm{i}\, \big[ (A_{H})_y , (A_{H})_z
        \big]\, \mathrm{d} y \wedge \mathrm{d} z \qquad \mbox{and}
        \qquad F^{(0,2)} = - \mathrm{i}\, \big[ (A_{H})_{\bar{y}} ,
        (A_{H})_{\bar{z}} \big]\, \mathrm{d} \bar{y} \wedge \mathrm{d}
        \bar{z} \ ,
\end{equation}
and we have chosen local complex coordinates $(y,\bar y) \in C_g$ and
$(z,\bar z) \in \Sigma_h$. The function $f_H\in\Omega^0(C_g\times\Sigma_h,\frg) $ is a purely four-dimensional quantity. 

To summarise, we arrive at a solution $A = A_{\theta}\, \kappa + A_H$
on $M_3 \times \Sigma_h$, where $A_H$ is a connection on $C_g \times
\Sigma_h$, and $A_{\theta}$ is a $\frg$-valued function which is constant along the fibre and covariantly constant on $C_g \times \Sigma_h$ with respect to $A_H$. The curvature $F$ of $A$ lives on $C_g \times \Sigma_h$ and is self-dual (from the five-dimensional point of view).
The Yang-Mills action evaluated at these connections gives
\begin{align}
S_{\rm YM} (F)&=\frac1{2\,g_{\textrm{\tiny YM}}^2}\, \int_{M_5}\, \big(F \stackrel{\wedge}{,} \ast F\big)
\notag \\[4pt] &= \frac1{2\,g_{\textrm{\tiny YM}}^2}\, \int_{M_5}\, \kappa \wedge \big(F_{H,+}
\stackrel{\wedge}{,} F_{H,+}\big) \notag \\[4pt]  &= - \frac{\pi\,
                                                    r}{g_{\textrm{\tiny
                                                    YM}}^2}\, \int_{C_g
  \times \Sigma_h}\, \big(F_{H,+} \stackrel{\wedge}{,} F_{H,+}\big)
\notag \\[4pt] &= \frac{8 \pi^3\, r\, (\vec{m} ,
                 \vec{n})}{g_{\textrm{\tiny YM}}^2} \ ,
\end{align}
where we integrated over the circular fibre, of
radius $r$, and used the fact that $F_{H,+}$ is independent of the
fibre direction. The integer vectors $\vec m,\vec n\in\IZ^{{\rm
    rank}(G)}$ are the gauge fluxes through $C_g$ and $\Sigma_h$,
respectively, which can be identified with weights of the Lie algebra
$\frg$. Then $(\vec{m} , \vec{n} ) \in \mathbb{Z}$ is proportional to
the second Chern character\footnote{We use K\"unneth's theorem here.} ${\rm ch}_2(P)\in
H^4(C_g\times\Sigma_h,\IQ)=H^2(C_g,\IQ)\otimes H^2(\Sigma_h,\IQ)$ of
the principal $G$-bundle $P \to C_g \times \Sigma_h$ on which $A_H$ is
a connection. 
This reduction along the Seifert fibre of $M_3$ may be thought of as a lift of the technique of \cite{Closset2017,Closset2018a}.

Five-dimensional gauge theories also have a topological global $U(1)_{\rm inst}$ symmetry \cite{Seiberg1996}, with conserved current
\begin{equation}
	J_{\mathrm{inst}} = \ast \big( F \stackrel{\wedge}{,} F\big) \ .
\end{equation}
The $U(1)_{\rm inst}$ charge of the current $J_{\mathrm{inst}}$ is the instanton number computed from $F$. 
The derivation given above makes it clear that this topological
symmetry is related to the $U(1)$ invariance under rotation of the Seifert fibre.

\subsubsection*{Pushdown}

Let us now describe the pushdown of these solutions. The localization
of the supersymmetric gauge theory onto connections constant along the
fibre, whose curvature descends to four dimensions, is reminiscent of
the framework of~\cite{Bawane2014}, where the topologically twisted
theory on $S^2 \times S^2$ was studied. In particular, the vertical
component $A_{\theta}$ of the gauge field $A$ is covariantly constant on
the four-dimensional base manifold $C_g \times \Sigma_h$, thus
$A_{\theta}$ is a scalar field on $C_g \times \Sigma_h$ constrained in
exactly the same way as the scalar $\sigma$.
Following \cite{deWit2011}, we can redefine our vector multiplet and
hypermultiplets in terms of four-dimensional supersymmetry
multiplets. The real scalar $\sigma$ is combined with the
vertical component $A_\theta$ to give a complex scalar $\phi =
\mathrm{i}\, \sigma  - v \,\llcorner\, A$, together with a purely
four-dimensional gauge connection $A_H$. This reduction brings the
$\mathcal{N}=1$ five-dimensional vector multiplet down to the
$\mathcal{N}=2$ four-dimensional vector multiplet. Similar
manipulations can be done for the hypermultiplets. The
five-dimensional Majorana spinor $\varepsilon$ breaks down into one
left and one right chiral four-dimensional Killing spinor.

In \cite[Sections 3 and 4]{Bawane2014}, it is explained how to
topologically twist the $\mathcal{N}=2$ gauge theory on any
four-dimensional manifold admitting a $U(1)$ isometry. We can then simply borrow their results:
The fixed point equations in four dimensions are
\begin{equation}
	[F, \phi ] = [F, \phi^{\dagger} ] = [ \phi, \phi^{\dagger}] =
        0 \ , \quad w \,\llcorner\, \mathsf{D}_{H} \phi^{\dagger} = 0
        \qquad \mbox{and} \qquad w \,\llcorner\, F - \mathrm{i}\,
        \mathrm{d} \phi = 0 \ ,
\end{equation}
where $w$ is the vector field used for the twist.
The vanishing Lie brackets imply that we can conjugate all fields $A_H$,
$\sigma$ and $v \,\llcorner\, A$ into the same Cartan subalgebra of
$\frg$ simultaneously. This means that covariantly constant scalars
$\sigma$ can be taken to be constant. As pointed out in \cite[Section~4.3]{Bawane2014}, to obtain the full partition function one should
include not only the sum over gauge fluxes $\vec{m}$ and $\vec{n}$ through the
two surfaces $C_g$ and $\Sigma_h$ (which we have equivalently obtained
from direct computations in five dimensions), but also the Nekrasov
partition functions which sum over point-like instantons corresponding
to the fixed points of the action of the maximal torus of the symmetry
group
given by the direct product of the gauge group with the isometry group.

To explicitly compute the instanton contributions, we could take $C_0=\Sigma_0=S^2$,
where our result coincides with that of~\cite[Section~2]{Willett2018};
they proceed in the other direction, starting from the theory on $S^2 \times S^2$ with $\Omega$-background and then lifting it to $S^3 _b \times S^2$.
The other geometry where instanton contributions are tractable is a reduction onto $S^2 \times T^2$. 
In this case, the $\mathcal{N}=2$ four-dimensional theory was analyzed in
\cite{Honda2015b,Gadde2015} (see also \cite{Benini2011}), which in turn 
computes the elliptic genus of an $\mathcal{N}=(2,2)$ gauge theory on
the torus $T^2$~\cite{Benini2013}.\footnote{The case of interest to us
  is described in~\cite[Section 5.1]{Honda2015b}, but without gauging the flavour symmetry. 
In the two-dimensional theory, this corresponds to turning off the moduli associated to background fields, and in particular the
two-dimensional R-symmetry is not gauged.} 
While contributions from four-dimensional point-like instantons on
$C_g \times \Sigma_h$ are hard to compute in more general geometries, we may
hope to recover the full answer from a resurgent analysis, as
explained in \cite{Russo2012} for $S^4$. Borel summability of the
perturbation series for $\mathcal{N}=2$ theories on $S^4$ and
$\mathcal{N}=1$ theories on $S^5$ (both possibly squashed) has been
studied in \cite{Honda2016}. We show in Section \ref{sec:Borel}
below how to adapt this argument to some examples in the present setting. 

\subsubsection*{Irregular fibrations}
\label{sec:solFirr}
	
The irregular case is more subtle. In this case, we may attempt to
proceed as in the regular case, but now with the Killing vector $v$
no longer pointing along the fibre direction. In other words, the equation
\begin{equation}
\label{eq:localizF}
v \,\llcorner\, \ast F = F 
\end{equation}
cannot be interpreted in terms of horizontal and anti-self-dual
components. Instead, we can decompose the gauge field as
\begin{equation}
A = A_{\eta}\, \eta + A_{T} \ ,
\end{equation}
where $\eta$ is the one-form dual to $v$, $A_{\eta}$ is the component of
the gauge connection along the direction of the isometry generated by
$v$, and $A_T$ is the transverse gauge connection. That is, we replace
the notion of horizontal with that of transverse, which is natural in
the present context~\cite{Alday2015}. Then the condition
$v \,\llcorner\, F = 0$, necessary to fulfill \eqref{eq:localizF}, can
be solved in an analogous way as for the regular case, leading to
\begin{equation}
{\sf D}_TA_\eta=\mathrm{d} A_{\eta} +\ii [A_T,A_\eta]= 0 \qquad
\mbox{and} \qquad \mathcal{L}_v A_{\eta} =0 = \mathcal{L}_v A_T \ .
\end{equation}

From the analysis of the regular case above, it is clear that we can
again pushdown the theory to four dimensions. However, this time we do
not reduce to the base $C_g \times \Sigma_h$ of the Seifert fibration,
but instead to the submanifold transverse to $\eta$ (equivalently, the
submanifold orthogonal to $v$). This is what one expects by the
construction of \cite{Alday2015}.

\subsection{Borel summability}
\label{sec:Borel}

Consider the perturbative contribution to the partition function of
$\cN=1$ gauge theory on $S^3\times \Sigma_h$ with gauge group
$G=U(N)$ and massless hypermultiplets, with $N_{f}$ in the fundamental
representation and $N_{\bar{f}}$ in the anti-fundamental
representation. The perturbative partition function is formally the same as the
$\cN=2$ perturbative partition function computed in
Section~\ref{sec:pertN=2}, but now allowing a more general assignment
of charges $\Delta_f$ and $\Delta_{\bar{f}}$ for the
hypermultiplets. Changing integration variables to hyperspherical coordinates
\begin{equation}
	\tilde{\sigma}_j = \sqrt{\tau} \, x_j \qquad \mbox{with} \quad
        \sum_{j=1}^{N}\, x_j ^2 =1 \qquad \text{ and } \qquad 0\leq\tau<\infty
\end{equation}
as in \cite{Honda2016a}, we can rewrite the partition function \eqref{eq:N=2pertS3} in the form
\begin{equation}
Z_{\mathcal{N}=2} ^{\mathrm{pert}} (S^3 \times \Sigma_h) = \int_{0}
^{\infty}\, \dd \tau \, \e^{- \frac{\tau}{\gamma} } \ Y_N (\tau ) \ , 
\end{equation}
where $\gamma^{-1} = \frac{ 4 \pi^2 }{r\, g_{\textrm{\tiny YM}} ^2}\, \mathrm{vol} (\Sigma_h )$ and
\begin{align}
Y_N (\tau ) = \frac{ \tau^{(N^2 -2)/2}}{2}\, \int_{\mathbb{R}^N}\, \dd
  x \ & \delta \Big( \sum_{j=1}^{N}\, x_j ^2 -1 \Big)\ \prod_{1 \le j<k \le N}\, (x_j - x_k)^2 \notag \\
	& \times \ \Big( \prod_{1 \le j<k \le N}\, \frac{\sinh \big( \pi\, \sqrt{\tau}\, (x_j - x_k) \big)^2 }{\tau\, (x_j-x_k)^2 } \Big)^{1-h}  \\
	& \times \ \Big( \prod_{j=1} ^N \, \mathtt{s}_1 \big( \ii ( 1 -
          \Delta_f ) - \sqrt{\tau}\, x_j \big)^{N_{f}} \, \mathtt{s}_1
          \big( \ii ( 1 - \Delta_{\bar{f}} ) + \sqrt{\tau}\, x_j
          \big)^{N_{\bar{f}}} \Big)^{1-h} \ . \notag
\end{align}
This form is suitable to study the Borel summability of the partition
function $\mathcal{Z}_{\mathcal{N}=1} ^{\mathrm{pert}} (S^3 \times
\Sigma_h)$, and of the $\mathcal{N}=2$ theory as a particular case. In
fact, this expression is exactly as in \cite{Honda2016a}, except for
the powers $1-h$. For genus $h=0$ Borel summability follows from \cite{Honda2016a} and for $h=1$ it is straightforward. 
The proof of Borel summability at $h>1$ is almost exactly as in \cite{Honda2016a} except for minor changes in the proof of the uniform convergence. 
As in \cite{Honda2016a}, the argument can be extended to include $N_{\rm adj}$ adjoint hypermultiplets.\par

For the particularly simple case $\Sigma_0=S^2$, we can extend this
analysis of Borel summability beyond the perturbative sector. It
suffices to notice that non-perturbative contributions to the $\cN=1$
partition function on $S^3\times S^2$ come from four-dimensional
(anti-)instantons on $S^2 \times S^2$. The $\mathcal{N}=2$ instanton
partition function was obtained in \cite{Bawane2014}, and it coincides
with the instanton partition function on
$S^4$~\cite{Bawane2014,Honda2016}. Therefore for $U(N)$ gauge theory
with $N_f$ hypermultiplets in the fundamental representation and
$N_{\bar{f}}$ hypermultiplets in the anti-fundamental representation,
the Borel summability of each instanton sector is proven by combining the methods of \cite{Honda2016a} and \cite{Honda2016}.

\subsection{Extension to $S^1 \times C_g \times \Sigma_h$}
\label{sec:5dDirectProd}

As our final consideration of the general features of cohomological
localization in five dimensions, we discuss how the prescriptions of
this section should be adapted when the gauge theory is put on the
five-manifold $M_5=S^1 \times C_g \times \Sigma_h$; this is the five-dimensional lift of the discussion in Section~\ref{sec:ext3d}. 
The direct product $S^1 \times C_g \times \Sigma_h$ is a special case,
which is not included in our previous discussion for two reasons. 
Firstly, although it clearly corresponds to a regular fibration, this
geometry does not follow directly from a rigid supergravity
background; instead, we have to first put the theory on $\mathbb{R} \times C_g \times \Sigma_h$ using a supergravity background and then compactify. 
The second difference from the other $U(1)$ fibrations discussed in
this section is the localization locus. Supersymmetric localization on $S^1 \times S^4$ was studied in \cite{Kim2012b,Terashima2012,Kim2013a}. 
The $\mathsf{Q}_\varepsilon$-exact action of
\cite[eq. (4.12)]{Terashima2012} in our notation is proportional to
\begin{equation}
	\int_{M_5}\, \bigg( \frac{1}{2}\, \big(F \stackrel{\wedge}{,} \ast F\big)  -
        \big(\mathsf{D} \sigma \stackrel{\wedge}{,} \ast \mathsf{D}
        \sigma\big) + \kappa \wedge \Big( - (\sigma,\sigma)\, \omega_{C_g}
        \wedge \omega_{\Sigma_h} + \frac{1}{2}\, (\sigma,F)\wedge
        \omega_{C_g} - \frac{1}{4}\, \big(F \stackrel{\wedge}{,} F)
        \Big) \bigg)
\end{equation}
where we dropped the contributions from fields which must be set
to zero at the localization locus. Bearing in mind that we will have to rotate $\sigma \mapsto \mathrm{i}\, \sigma_0$ and integrate over $\sigma_0 \in \mathfrak{g}$, 
the saddle points of this action
require $\sigma$ to be covariantly constant and $F$ to vanish in the
direction of $S^1$. This implies that the gauge connection has a
constant component $a_0$ along $S^1$, and we can gauge rotate $a_0$ into a Cartan subalgebra of $\mathfrak{g}$.
Moreover, the curvature
\begin{equation}
	F_{C_g \times \Sigma_h} = f_{C_g}\, \omega_{C_g} + f_{\Sigma_h}\, \omega_{\Sigma_h} 
\end{equation}
is a saddle point if $f_{C_g}  = 2\, \sigma = f_{\Sigma_h}$, subject
to the additional condition $[a_0, f_{C_g}]=0 $. This vanishing Lie bracket guarantees that $\sigma$ is constant and can be conjugated into the same Cartan subalgebra as $a_0$. 
Therefore the full localization locus inside the Coulomb branch is parametrized by $a_0$ and
$\vec{m} \in \Lambda$, an element of the co-root lattice in
the same Cartan subalgebra as $a_0$, through
\begin{equation}
	 A = a_0 \,\kappa + \vec{m}\, ( A_{C_g} + A_{\Sigma_h}) \qquad
         \mbox{and} \qquad \sigma = \frac{\vec{m}}{r} \ ,
\end{equation}
where $A_{C_g} $ and $ A_{\Sigma_h}$ are monopole connections on $C_g$ and $\Sigma_h$, respectively, satisfying
\begin{equation}
	\int_{C_g}\, \mathrm{d} A_{C_g} = \frac{1}{2}\, \mathrm{vol}
        (C_g) \qquad \text{and} \qquad \int_{\Sigma_h} \, \mathrm{d}
        A_{\Sigma_h} = \frac{1}{2}\, \mathrm{vol} (\Sigma_h) \ .
\end{equation}
The pushdown to four dimensions in the present
geometry is the usual Kaluza-Klein dimensional reduction, in contrast to the reduction of Section~\ref{sec:continst-pushdown}.

The triviality of the $U(1)$ fibration in the present case implies
that the index of the twisted Dolbeault complex is given by
\begin{equation}
\label{eq:trivialDolb}
	\mathrm{index}\ \bar{\partial}^{(m)} = (1-g)\,(1-h) \ .
\end{equation}
With the same notation $q= \e^{- \frac{\epsilon}{r}}$ from
Section~\ref{sec:ext3d}, we immediately find
\begin{align}
	Z_{\rm vec} ( S^1 \times C_g \times \Sigma_h) & = Z_{\rm vec} ( S^1 \times C_g )^{1-h} \notag \\[4pt]
	& = \prod_{\alpha \in \triangle}\,  q^{ -\frac12\, (1-g)\,(1-h)\,
                  \lvert  ( \alpha , \vec{m} ) \rvert }\,  \Big(
                  1- \e^{- \mathrm{i}\, \epsilon\, (\alpha, a_0) }\,
                  q^{ \lvert  ( \alpha , \vec{m} ) \rvert }
                  \Big)^{(1-g)\,(1-h)} \ .
\end{align}
This lifts the result of Section~\ref{sec:ext3d} to the
five-dimensional manifold $M_5=S^1\times
C_g\times\Sigma_h$.\footnote{As in Section~\ref{sec:ext3d}, we omit a
  factor $\prod_{\alpha \in \triangle}\, (\alpha, a_0)^{-1}$, which
  cancels the Jacobian arising from the integration over $a_0$ after
  gauge rotation into the chosen Cartan subalgebra.}

\subsubsection*{$\boldsymbol{\cN=2}$ perturbative partition
  functions on $\boldsymbol{T^3\times\Sigma_h}$}

View the three-torus $T^3$ as the trivial circle
bundle over $C_1=T^2$. It follows from
\eqref{eq:trivialDolb} that in this case the index of the Dolbeault complex
is zero, and so from the localization formula \eqref{eq:ZvecM3xSigma} with
$A^{(0)}=0$ together with
\eqref{eq:5dhyperpert} we obtain
\begin{align}
	Z_{\rm vec}^{\rm pert} (T^3 \times \Sigma_h ) =  \prod_{\alpha \in
                                             \triangle_+}\, (\alpha,
                                             \sigma_0)^{-2} \qquad
  \mbox{and} \qquad
	Z_{\rm hyp}^{\rm pert} (T^3 \times \Sigma_h ) = 1 \ .
\end{align}
For the supersymmetric Yang-Mills theory on $T^3\times\Sigma_h$, the
only non-trivial contributions to the one-loop determinants come from the ghosts
in the vector multiplet, which produce a factor that exactly cancels
the Jacobian arising from conjugation of $\sigma_0$ into the Cartan
subalgebra $\frt$, as mentioned before. Hence the perturbative partition function is
given by a Gaussian integral
\begin{align}
Z_{\mathcal{N}=2} ^{\rm pert} (T^3 \times \Sigma_h) =
                                                      \int_{\mathfrak{t}_{\rm reg}}\,
                                                      \dd
                                                      \tilde{\sigma}\,
                                                      \e^{-\frac{
                                                      4\pi^3\,{\rm vol}(\Sigma_h)} {r\,g^2
                                                      _{\textrm{\tiny
                                                      YM}}}\,
  (\tilde\sigma,\tilde\sigma)} \ .
\end{align}
For a quiver gauge theory with unitary gauge group $G= U(N_1) \times U (N_2) \times \cdots  \times U(N_n)$, this reads
\begin{equation}
	Z_{\mathcal{N}=2} ^{\rm pert} (T^3 \times \Sigma_h) = \Big(
          \frac{r\,g^2_{\textrm{\tiny YM}}}{ 4\pi^3\,{\rm
              vol}(\Sigma_h)} \Big)^{\frac{1}{2}\, (N_1+\cdots+N_n)
        } \ .
\end{equation}
Note that the perturbative partition function on $T^3 \times \Sigma_h$
does not distinguish between a non-abelian theory with gauge group $G=
U(N)$ and an abelian quiver theory with gauge group $G=U(1)^{N}$.

\section{$q$-deformed Yang-Mills theories from cohomological
  localization}
\label{sec:qYMfrom5d}

This final section is devoted to the study of five-dimensional
supersymmetric Yang-Mills theory on $S^3 _b \times \Sigma_h$, where
$S^3 _b$ is either the squashed sphere of \cite{Imamura2011a} or the
ellipsoid of \cite{Hama2011} (see Appendix~\ref{app:squashedspheres}
for details); recall from Section~\ref{sec:SCFT6d} that these are the
five-dimensional theories that naturally descend from six-dimensional
superconformal field theories on squashed geometries. We show how $q$-deformations of Yang-Mills theory on a
Riemann surface
$\Sigma_h$ arise from our localization procedure. Our
formula for the partition function of the standard $q$-deformed
Yang-Mills theory in Section~\ref{sec:qYMS3xSigma} improves the result
of \cite{Fukuda2012,Kawano2015}, wherein the Gaussian term was not retained; as
we discuss, this
Boltzmann factor is important for applications to holography. Both
treatments of~\cite{Fukuda2012} and~\cite{Kawano2015} focus on the
zero area limit where ${\rm vol}(\Sigma_h)\to0$, which hides the fact that the resulting $q$-deformed Yang-Mills theory has $p=1$. We elucidate the geometric significance of this new $q$-deformation through an analysis of the resulting matrix model on the sphere $\Sigma_0=S^2$, by adapting the procedure of \cite{Caporaso2005} to the present case.

\subsection{Localization on $S^3 \times \Sigma_h$}
\label{sec:qYMS3xSigma}

Consider the $\cN=1$ partition function without matter on $S^3
\times \Sigma_h$, beyond the perturbative calculation of
Section~\ref{sec:pertN=2}. More precisely, we include the full set of
solutions obtained in Section~\ref{sec:continst-pushdown}, but discard
the four-dimensional point-like instantons.\footnote{This sector of
  the partition function was named ``perturbative'' in
  \cite{Willett2018}. However, in the present paper the term
  perturbative refers to the expansion around the trivial connection,
  while the subsector of the full partition function we are
  considering now includes a much wider class of solutions. A proper
  definition hinted at in \cite[Section 2.4]{Willett2018} might be
  ``partition function neglecting codimension four field
  configurations''. These are non-perturbative with respect to an
  expansion in the geometric area parameter $\mathrm{vol} (\Sigma_h)$.}  
The resulting partition function is given by
\begin{equation}
Z_{\mathcal{N}=1} (S^3 \times \Sigma_h) = \sum_{A^{(0)} } \
\int_{\mathfrak{t}_{\rm reg}}\, \mathrm{d} \sigma_0\,  \e^{- S_{\rm cl} (F,
  \sigma_0 ) } \ \prod_{\alpha \in \triangle_+}\, \sinh \big( \pi\,r\,
(\alpha, \phi)  \big)^{2-2h} \ ,
\end{equation}
with $S_{\rm cl}$ the action evaluated at the localization locus:
\begin{equation}
S_{\rm cl}(F,\sigma_0) = S_{\rm YM}(F) + \frac{1}{2\, g^2
  _{\textrm{\tiny YM}}}  \, \int_{S^3 \times \Sigma_h}\, \Big(\frac{1}{r^2}\, (\sigma_0,\sigma_0)\, \kappa\wedge\dd\kappa\wedge \omega_{\Sigma_h} + \frac{1}{r}\, (\sigma_0,F)\,\wedge \kappa \wedge \mathrm{d} \kappa\Big) \ .
\end{equation}
The first summand in the action, $S_{\rm YM}(F)$, is the five-dimensional Yang-Mills
action evaluated in Section \ref{sec:solFreg}:
\begin{equation}
\label{eq:SinstS2Sigma}
\begin{aligned}
S_{\rm YM}(F)  & = \frac{1}{2\, g^2 _{\textrm{\tiny YM}}} \, \int_{S^3 \times \Sigma_h}\, \kappa \wedge \big(F_{H,+} \stackrel{\wedge}{,} F_{H,+}\big)  \\[4pt]
	& = - \frac{\pi\, r}{g^2 _{\textrm{\tiny YM} }} \, \int_{S^2
            \times \Sigma_h}\, \big(F \vert_{S^2 \times \Sigma_h}
          \stackrel{\wedge}{,} F \vert_{S^2 \times \Sigma_h}\big)
          \\[4pt]
& = \frac{8 \pi^3\, r}{g^2 _{\textrm{\tiny YM} }}\, (\vec{m} ,
  \vec{n}) \ ,
\end{aligned}
\end{equation}
where in the second equality we used the defining equations, and in
the last equality $(\vec{m} , \vec{n}) \in \mathbb{Z}$ is proportional
to the second
Chern character ${\rm ch}_2$ associated to the pertinent principal $G$-bundle
over $S^2\times\Sigma_h$. 

The sum in the partition function runs over the $G$-connections $A^{(0)}$ whose curvature $F$ satisfies the fixed point equation
\begin{equation}
	v \,\llcorner\, \ast F = F \ .
\end{equation}
These solutions were studied in Section \ref{sec:solFreg}. It is
straightforward to check that connections whose curvature lives on
$\Sigma_h$, that is $ F = F \vert_{\Sigma_h}$, do not satisfy this
equation. In other words, connections that are flat when projected
onto $S^3$, and which thus belong to the localization locus of a
purely three-dimensional theory, do not belong to the localization
locus of the fully twisted five-dimensional theory. The flat connection on
$S^3$ does belong to the localization locus of the five-dimensional theory
which is partially twisted along $\Sigma_h$ (with arbitrary $A|_{\Sigma_h}$), but not to the
localization locus of the fully twisted theory; the partition function of the field theory topologically twisted along $\Sigma_h$ is important from the perspective of the six-dimensional theory and its reduction to four-dimensional theories of class $\cS$~\cite{Fukuda2012,Kawano2015,Willett2018}. The same argument can
be made for connections which are flat along $\Sigma_h$, which then only
belong to the localization locus of the five-dimensional theory which
is partially twisted along $S^3$ (with arbitrary $A|_{S^3}$)~\cite{Yagi2013,Lee2013,Cordova2013}.

We can write
\begin{equation}
\label{eq:ZS3xSigmatot}
Z_{\mathcal{N}=1}  (S^3 \times \Sigma_h) = \sum_{ \vec{m},\vec n \in
  \mathbb{Z}^{\mathrm{rank} (G)} } \ \int_{\mathfrak{t}_{\rm reg}}\, \mathrm{d}
\phi\, \e^{- S_{\rm cl}(\vec m,\vec n;\phi) } \ \prod_{\alpha \in \triangle_+}\, \sinh
\big( \pi\, r\, (\alpha, \phi)  \big)^{2-2h} \ .
\end{equation}
We shifted the integration variable to $\phi = - \sigma_0 - v
\,\llcorner\, A^{(0)}$. This is the natural variable to use when descending
to a four-dimensional description, as the sum in
\eqref{eq:ZS3xSigmatot} is taken over the gauge fluxes $\vec m$ and
$\vec n$ through $S^2$
and $\Sigma_h$, respectively; the first and second Chern characters associated to the principal
$G$-bundle over $S^2\times\Sigma_h$ are obtained by using the pairing $(\,\cdot\,,\,\cdot\,)$. The action evaluated at the fixed points consists of three terms:
\begin{equation}
	S_{\rm cl}(\vec m,\vec n;\phi) = S_{\rm YM}(\vec m,\vec n)
        + \frac{1}{2\, g_{\textrm{\tiny YM}} ^2} \,
        \int_{S^3\times\Sigma_h}  \, \Big(\frac{1}{r^2}\,
        (\phi,\phi)\, \kappa\wedge\dd\kappa \wedge \omega_{\Sigma_h} -
        \frac{1}{r}\, (\phi,F|_{S^2 \times \Sigma_h})\wedge \kappa
        \wedge \mathrm{d} \kappa\Big) \ ,
\end{equation}
where the first summand is written in \eqref{eq:SinstS2Sigma}.
Passing to the scaled variable $\tilde{\phi} = r\, \phi $, we get
\begin{align} \label{eq:Sclphitilde}
S_{\rm cl}(\vec m,\vec n;\tilde\phi) &=  S_{\rm YM}(\vec m,\vec n) \\
                                     & \qquad +
\frac{1}{g_{\textrm{\tiny YM}} ^2} \, \Big(\frac{8 \pi^2\,
  \mathrm{vol} (\Sigma_h)}{r} \, (\tilde{\phi},\tilde\phi)  +
\frac{2\pi\ii}{r^2} \, \int_{S^3 \times \Sigma_h}\, \frac{\mathrm{i}}{
  4 \pi}\, (\tilde\phi,F|_{S^2 \times \Sigma_h}) \wedge \kappa \wedge
\mathrm{d} \kappa \Big) \ . \nonumber
\end{align}

\subsubsection*{Zero flux sector}

We will now restrict ourselves to those connections for which $S_{\rm
  YM}(\vec m,\vec n)=0$. Contributions with non-vanishing second Chern character are
exponentially suppressed. 
We turn off the gauge fluxes through $S^2$, $\vec m=\vec0$, and follow standard
techniques from two-dimensional Yang-Mills theory, see in particular \cite{Blau93}. Using 
\begin{equation}
\frac{1}{2} \int_{S^3 \times \Sigma_h}\, \frac{\mathrm{i}}{ 2 \pi}\,
(\tilde\phi,F|_{S^2\times\Sigma_h}) \wedge \kappa \wedge \mathrm{d}
\kappa  = 4 \pi^2\, r^3\, (\tilde\phi,\vec n) \ ,
\end{equation}
and summing over $\vec n\in  \mathbb{Z}^{{\rm rank}(G)}$, the third
summand in the action \eqref{eq:Sclphitilde} produces the
delta-function constraint
\begin{equation}
	\frac{ 4 \pi^2\, r }{g_{\textrm{\tiny YM}} ^2} \,
        \tilde{\phi} = \vec k 
\end{equation}
for some integer vector $\vec k\in\IZ^{{\rm rank}(G)}$ which can be identified with a regular weight of $\frg$. Plugging this
into the remaining Gaussian part of the action gives
\begin{equation}
	\frac{ g_{\textrm{\tiny YM}} ^2 }{ 4 \pi\, r } \, \frac{
          \mathrm{vol} (\Sigma_h) }{ \pi\, r^2 } \, (\vec{k},\vec k\,) \ .
\end{equation}
In the one-loop determinant, we obtain
\begin{equation}
	\sinh \big( \pi\, r\, (\alpha, \phi) \big) = \sinh \big( \pi\, (\alpha, \tilde{\phi}) \big) = \sinh \left( \frac{ g_{\textrm{\tiny YM}} ^2 }{ 4 \pi\, r }\, (\alpha, \vec{k}\, ) \right) = - \big[ (\alpha, \vec{k}\,) \big]_{q} 
\end{equation}
where
\begin{equation}
\label{eq:defgstr}
	q= \e^{-g_{\rm str}} \qquad \mbox{with} \quad g_{\rm str}:=\frac{
          g_{\textrm{\tiny YM}} ^2 }{ 2 \pi\, r } \ .
\end{equation}

The final form of the partition function is then
\begin{equation}\label{eq:0instqZ}
	Z_{\mathcal{N}=1} ^{\vec{m}= \vec{0}} (S^3 \times \Sigma_h) =
        \sum_{\vec{k} \in (\mathbb{Z}^{\mathrm{rank}(G) }) _{\rm reg}} \
        \prod_{\alpha \in \triangle_{+}}\,  \big[ (\alpha, \vec{k}\,)
        \big]_{q}  ^{2-2h} \ q^{\frac{1}{2}\, (\vec{k},\vec k\,) \, \frac{ \mathrm{vol} (\Sigma_h) }{ \pi\, r^2 } } \ .
\end{equation}
From \eqref{eq:defgstr} we identify the six-dimensional radius $\beta = \frac{g_{\textrm{\tiny YM}} ^2 }{2\pi }$, and it is instructive to rewrite
\begin{equation}\label{eq:0instqZvol}
	q^{\frac{1}{2}\, (\vec{k},\vec k\,) \, \frac{ \mathrm{vol} (\Sigma_h) }{ \pi\, r^2 } } = \exp \Big( -(\vec{k},\vec k\,) \, \frac{ 2 \pi\, \beta ~ \mathrm{vol} (\Sigma_h)}{ \mathrm{vol}_{\kappa} ( S^3 ) }  \Big) \ ,
\end{equation}
where $\mathrm{vol}_{\kappa} ( S^3 )$ is the volume of $S^3$ taken with respect to the contact structure $\kappa$, in our normalization. 

The expression \eqref{eq:0instqZ} is the partition function of the standard $q$-deformed
two-dimensional Yang-Mills theory on $\Sigma_h$ (cf. Section~\ref{genasp}), at $p=1$ and with
Gaussian weight corrected by the ratio of volumes between the fibre
Riemann surface $\Sigma_h$ and the total space sphere $S^3$. This ratio of
volumes matches exactly with \cite{Tachikawa2012}. For unitary gauge
group $G=U(N)$, this is reminiscent
of the large $N$ free energy of gauge theories with a holographic dual. In
particular, for
$\Sigma_h=\Sigma_0=S^2$, the $N^3$ behaviour of the free energy at
large $N$ in the present theory follows immediately from the large $N$
limit of $q$-deformed Yang-Mills theory, which in turn is given by the
free energy of Chern-Simons theory on $S^3$ as we are in the case
$p=1$. This indeed gives the right answer for a gauge theory on a
Sasaki-Einstein five-manifold which has a holographic dual~\cite{Kallen2012b}.
See \cite{Giasemidis2013} for a discussion and a similar example where the $N^3$
behaviour of the free energy in five dimensions is extracted from
Chern-Simons gauge theory in three dimensions. See also
\cite{Mezei2018} for the localization
of five-dimensional maximally supersymmetric Yang-Mills theory to a
three-dimensional subsector, and the relation with Chern-Simons
theory.

We notice that the string coupling $g_{\rm str}$ depends only on the ratio
$\frac{ g^2 _{\textrm{\tiny YM}}}{r}$. This is consistent with
dimensional reduction from the $(2,0)$ theory on $S^1 \times S^3 \times \Sigma_h$ discussed in Section~\ref{sec:SCFT6d}. In the six-dimensional setting, 
$g^2 _{\textrm{\tiny YM}}$ plays the role of the circumference of the circle $S^1$ on which we
have reduced. 
On the other hand, the small area limit $\mathrm{vol} (\Sigma_h) \to
0$ gives the superconformal index of the four-dimensional gauge theory
on $S^1 \times S^3$, consistently with the conjecture of
\cite{Gadde2011}. From \eqref{eq:0instqZvol} it is evident that the small
${\rm vol}(\Sigma_h)$ limit is the same as the large ${\rm vol}_\kappa(S^3)$
limit which decompactifies the three-sphere $S^3$ to $\IR^3$.

\subsubsection*{Reinstating gauge fluxes}

We have so far restricted ourselves to the sector where $S_{\rm
  YM}(\vec m,\vec n) =0$. In general we have to consider additional
solutions which are given by connections whose curvature lives on $S^2
\times \Sigma_h$ with non-trivial second Chern character ${\rm ch}_2\neq0$, and
\begin{equation}
	S_{\rm YM}(\vec m,\vec n) = \frac{8 \pi^3\, r}{g^2 _{\textrm{\tiny YM}} } \,
        (\vec{m} , \vec{n}) = 2 \pi\, \mathrm{i}\, \Big( - \mathrm{i} \,
        \frac{4 \pi^2\, r}{g^2 _{\textrm{\tiny YM}} } \, (\vec{m} ,
        \vec{n})  \Big) \ ,
\end{equation}
where as above $\vec{n}$ is the gauge flux through $\Sigma_h$ and
$\vec{m}$ is the gauge flux through $S^2$. The full partition function
includes this term in the action, along with a sum over $\vec{m} \in
\mathbb{Z}^{\mathrm{rank}( G)}$. The bracketed term has exactly the
same coefficient as a BF-type term in the action. This means that the
procedure we used for the $\vec{m}=\vec0$ sector should be modified by
a shift $\tilde{\phi} \mapsto \tilde{\phi} - \mathrm{i}\, \vec m$
after the sum over all $\vec n\in \mathbb{Z}^{{\rm rank}(G)}$. We
finally arrive at
\begin{align}
	Z_{\mathcal{N}=1} (S^3 \times \Sigma_h) = \sum_{ \vec{m} \in
  \mathbb{Z}^{\mathrm{rank} (G)} } \ \sum_{ \vec{k} \in
  (\mathbb{Z}^{\mathrm{rank}(G)}) _{\rm reg}} \ & \prod_{\alpha \in \triangle_{+} }
                                      \, \Big( \big[ ( \alpha , \vec{
                                      k } - \mathrm{i}\, \vec{ m } )
                                      \big]_{q} \, \big[ (\alpha,
                                      \vec{k} + \mathrm{i}\, \vec{m} )
                                      \big]_{q} \Big)^{1-h} \nonumber
  \\ & \times \ q^{\frac{1}{2}\, ( \vec{k} - \mathrm{i}\, \vec{m},\vec
       k-\ii\vec m) \, \frac{ \mathrm{vol} (\Sigma_h) }{ \pi\, r^2 } } \ .
\end{align}
The $q$-deformed measure has in fact precisely the right form to
support non-trivial fluxes $\vec{m}$.

\subsection{Localization on squashed $S^3 _b \times \Sigma_h$}
\label{sec:qYMsquashed}

There exist two types of squashings of $S^3$ that can be lifted to
five dimensions with $\mathcal{N}=2$ supersymmetry. The first type is
the familiar case of \cite{Hama2011}. From the point of
view of cohomological localization, the squashing simply corresponds to a rescaling of the fibre
radius, and the results for the one-loop determinants on $S^3 _b \times
\Sigma_h$ are exactly the same as for the round sphere $S^3$, but with
fibre radius $\epsilon \ne r$, where $r$ is the radius of the base
$S^2$. See Appendix
\ref{app:squashedspheres} for further details.

The other squashed sphere, which is a regular fibration, is that of
\cite{Imamura2011a}. We did not give a formal derivation of the
one-loop determinants in the cohomological gauge theory for this background. However, we
know that the Killing vector field $v$ has closed orbits and is
parallel to the Reeb vector field $\xi$. We can therefore lift the
results from $S^3 _b $ to $S^3 _b \times \Sigma_h$, and the one-loop
determinants are given by
\begin{align}
	Z_{\rm vec} (S^3 _b \times \Sigma_h) & = \bigg(\prod_{\alpha
                                               \in \triangle_{+}}\,
                                               \frac{\sinh \big( \pi\,
                                               \epsilon_1\, (\alpha,
                                               \sigma_0) \big) \, \sinh
                                               \big( \pi\,
                                               \epsilon_2\, (\alpha,
                                               \sigma_0) \big)}{\pi^2\,\epsilon_1\,\epsilon_2\,(\alpha,\sigma_0)^2} \bigg)^{1-h}
                                               \ , \nonumber \\[4pt]
	Z_{\rm hyp} (S^3 _b \times \Sigma_h) & = \bigg( \prod_{\rho
                                               \in \Lambda_R} \, {\tt s}_{b}
                                               \big(
                                               \tfrac{\mathrm{i}}{2} \,
                                               (b + b^{-1})
                                               \, (1 - \Delta) -
                                               r\, ( \rho, \sigma_0 )
                                               \big)\bigg)^{1-h} \ ,
\end{align}
which agrees with \cite[Section 2]{Willett2018}. We adopted the notation
\begin{equation}
	\epsilon_1 = r\, b \qquad \mbox{and} \qquad \epsilon_2 = r\, b^{-1}
        \qquad \mbox{with} \quad b = \frac{ 1 - \mathrm{i}\, u}{
          \sqrt{ 1 + u^2}}
\end{equation}
from Appendix~\ref{app:squashedspheres}.

The computation of the full $\cN=1$ partition function without matter
on $S_b^3\times\Sigma_h$ proceeds exactly as in
Section~\ref{sec:qYMS3xSigma}, with only a modification in the
one-loop determinant. This results in a bi-orthogonalization of the
$q$-deformed measure, and the partition function in the sector of
vanishing second Chern character ${\rm ch}_2=0$ is given by
\begin{equation}
	Z_{\mathcal{N}=1} ^{\vec m=\vec 0} (S^3 _b \times
        \Sigma_h)
        = \sum_{\vec{k} \in (\mathbb{Z}^{\mathrm{rank} (G)})_{\rm reg} } \
        \prod_{\alpha \in \triangle_{+}} \, \Big( \big[ b\, (\alpha,
        \vec{k}\,) \big]_{q} \, \big[ b^{-1}\, (\alpha, \vec{k}\,)
        \big]_{q} \Big) ^{1-h} \ q^{\frac{1}{2}\, (\vec{k},\vec k\,)
          \, \frac{ \mathrm{vol} (\Sigma_h) }{ \pi\, r^2 } }
        \ .
\end{equation}

The extension to the full partition function including non-trivial
gauge fluxes $\vec m$ through the base $S^2$ is exactly as described
in Section~\ref{sec:qYMS3xSigma}, as we are presently working in the
regular case.
After inclusion of gauge fluxes $\vec{m} \ne \vec{0}$, our result appears to be only in partial agreement with \cite{Willett2018,Razamat2019}, where the field theory is first defined on $S_b^3\times\IR^2$, then dimensionally reduced, and finally put on the Riemann surface $\Sigma_h$ with a twist. 
That procedure allows for additional background fluxes for the flavour symmetry, which do not appear in our framework nor in \cite{Fukuda2012}.

\subsection{Localization on ellipsoid $S^3 _b \times \Sigma_h$}

We now consider the geometry $S^3 _b \times \Sigma_h$ for the
ellipsoid $S_b^3$ of~\cite{Hama2011}. Most of the steps are the same
as in the round $S^3$ case of Section~\ref{sec:qYMS3xSigma}.
The action at the localization locus is again given by
\begin{equation}
S_{\rm cl}(F,\sigma_0) = S_{\rm YM}(F) + \frac{1}{2\, g^2
  _{\textrm{\tiny YM}}}  \, \int_{S^3 \times \Sigma_h}\,
\Big(\frac{1}{r^2}\, (\sigma_0,\sigma_0)\, \kappa\wedge\dd\kappa\wedge
\omega_{\Sigma_h} + \frac{1}{r}\, (\sigma_0,F)\wedge \kappa \wedge
\mathrm{d} \kappa\Big) \ , 
\end{equation}
with
\begin{equation}
S_{\rm YM}(F)=\frac1{2\,g_{\textrm{\tiny YM}}^2}\,
\int_{S^3\times\Sigma_h}\, \big(F \stackrel{\wedge}{,} \ast F\big) \ .	
\end{equation}
The zero flux sector of the partition function is
\begin{align}\label{eq:0instellispsoid}
	Z_{\mathcal{N}=1} ^{\vec m=\vec 0} (S^3 _b \times \Sigma_h)
        = \sum_{\vec{n} \in \mathbb{Z}^{\mathrm{rank} (G)} } \
        \int_{\mathfrak{t}_{\rm reg}} \, \mathrm{d} \sigma_0\, & \e^{- S_{\rm
            cl}(\vec 0,\vec n;\sigma_0) } \\ & \times \ \prod_{\alpha \in
          \triangle_+}\, \Big( \sinh \big( \pi\, r\, b\, (\alpha,
        \sigma_0)  \big) \, \sinh \big( \pi\, r\, b^{-1}\, (\alpha,
        \sigma_0)  \big) \Big)^{1-h} \ . \nonumber
\end{align}
Changing variable $\tilde{\sigma}= r\, \sigma_0$ and
repeating the same steps used in Section~\ref{sec:qYMsquashed}, we arrive at
\begin{align}
	Z_{\mathcal{N}=1} ^{\vec m=\vec 0} (S^3 _b \times \Sigma_h)
        = \sum_{\vec{k} \in (\mathbb{Z}^{\mathrm{rank}( G)})_{\rm reg} }\, & \e^{-
          \frac{g_{\rm str}}{2}\, (\vec{k},\vec k\,) \, \frac{ \mathrm{vol}
            (\Sigma_h)}{\pi\,r^2} } \\ & \times \ \prod_{\alpha \in
          \triangle_+}\, \bigg( \sinh \Big( b\, \frac{ g_{\rm str}\, (\alpha,
          \vec{k}\,) }{2} \Big) \, \sinh \Big( b^{-1}\, \frac{g_{\rm str}\,
          (\alpha, \vec{k}\, ) }{2}  \Big) \bigg)^{1-h} \ , \notag
\end{align}
where as before we defined the string coupling $g_{\rm str} = \frac{
  g_{\textrm{\tiny YM}} ^2 }{ 2 \pi\, r } = \frac{ \beta}{r}$. The final expression can be rewritten in terms of $q= \e^{- g_{\rm str}}$ as
\begin{equation}
	Z_{\mathcal{N}=1} ^{\vec m=\vec 0} (S^3 _b \times \Sigma_h)
        = \sum_{\vec{k} \in (\mathbb{Z}^{\mathrm{rank}( G)})_{\rm reg} } \,
        q^{\frac{1}{2}\, (\vec{k},\vec k\,) \, \frac{ \mathrm{vol}
            (\Sigma_h)}{\pi\,r^2} } \ \prod_{\alpha \in
          \triangle_+}\, \Big( \big[ b \, (\alpha, \vec{k}\,)\big]_q
        \, \big[ b^{-1} \, (\alpha, \vec{k}\,)\big]_q  \Big)^{1-h} \ .
\end{equation}

From \eqref{eq:0instellispsoid} we see that the perturbative partition
function on $S^3 _b \times S^2$, retaining only the contribution from
the trivial flat connection $A^{(0)}=0$, coincides with the
perturbative partition function of Chern-Simons gauge theory on the
lens space $L(p,1)$, continued to arbitrary values $p = b^{2} \in \mathbb{R}$:
\begin{equation}
	Z_{\mathcal{N}=1} ^{\mathrm{pert}} (S^3 _b \times S^2) =
        \int_{\mathfrak{t}_{\rm reg}} \, \mathrm{d} {\sigma} \, \e^{- \frac{1}{2\,
            g_{\rm str}}\, ({\sigma},\sigma) \, \frac{ \mathrm{vol}
            (\Sigma_h)}{\pi\,r^2} } \ \prod_{\alpha \in \triangle_+} \, \sinh \Big( b\, \frac{  (\alpha, {\sigma}) }{2} \Big) \, \sinh \Big( b^{-1}\, \frac{(\alpha, {\sigma} ) }{2} \Big) \ ,
\end{equation}
where we rescaled $\sigma=g_{\rm str}\,\sigma_0$.
This becomes more evident if we use instead an asymmetric length
scaling to define the variable $\tilde{\sigma}=
\epsilon_1\, \sigma_0 = r\,b\, \sigma_0$. For general genus $h$ we
then arrive at the discrete matrix model
\begin{equation}
	Z_{\mathcal{N}=1} ^{\vec m=\vec 0} (S^3 _b \times \Sigma_h)
        = \frac{1}{b^{\mathrm{rank}( G)} } \, \sum_{\vec{k} \in
          (\mathbb{Z}^{\mathrm{rank} (G)})_{\rm reg} } \, q^{\frac{1}{2}\,
          (\vec{k},\vec k\,) \, \frac{ \mathrm{vol} (\Sigma_h)}{\pi\,r^2} } \ \prod_{\alpha \in \triangle_+}\, \Big(
        \big[ (\alpha, \vec{k}\,)\big]_q \, \big[ b^{-2} \, (\alpha,
        \vec{k}\,)\big]_q  \Big)^{1-h} \ .
\end{equation}
Further details and analysis of this matrix model for genus $h=0$ and
gauge group $G=U(N)$, along the
lines of \cite{Caporaso2005}, are provided in Section~\ref{app:mmqym}
below.

Reinstating the additional contributions of contact instantons, with
non-vanishing fluxes through the base $S^2$ of the $U(1)$ fibration of
$S^3 _b$, is a much more subtle issue. This is because the $U(1)$-action now
involves the Killing vector field $v$ which differs from the Reeb
vector field $\xi$ on $S^3 _b$, so that contractions with $v$ do not
separate the horizontal and vertical parts of the differential forms
involved.

\subsection{The matrix model}
\label{app:mmqym}

We focus now on the partition function $Z_{\mathcal{N}=1} ^{\vec m=\vec 0} (S^3 _b \times S^2)$ for gauge group $G=U(N)$. The partition function is formally the same for either the squashed sphere or the ellipsoid $S^3 _b$. Only the geometric meaning of the squashing parameter $b$ is different in the two cases, in particular $b$ is a complex number of unit modulus $\lvert b \rvert =1$ for the squashed sphere and $b>0$ is real for the ellipsoid. However, as the partition function can be analytically continued in both cases, there is no difference in practice.

Our goal is then to study the discrete random matrix ensemble with partition function
\begin{equation}
\label{eq:Z-b2YM}
	\mathcal{Z}_N (b) = \frac{1}{b^{N}\, N!}\, \sum_{ \vec{\ell}\, \in \zed ^N }\, \e^{- \frac{ g_{\rm str}}{2}\, \sum_{j=1} ^N\, \ell_j ^2 } \ \prod_{1 \le j < k \le N}\, 4 \sinh \Big( \frac{g_{\rm str}}{2}\, (\ell_j - \ell_k) \Big)\, \sinh \Big( \frac{g_{\rm str}}{2\, b^2}\, (\ell_j - \ell_k) \Big) \ ,
\end{equation}
which can be identified with the discrete version of the bi-orthogonal Stieltjes-Wigert ensemble studied in \cite{Dolivet2006}. If $p:=b^2 \in \zed$, the continuous version of the bi-orthogonal Stieltjes-Wigert ensemble provides the partition function of Chern-Simons theory on the lens space $L(p,1)$. In the limit $b \to 1$ we recover the partition function of $q$-deformed Yang-Mills theory constructed from the monopole bundle over $S^2$ with $p=1$, whose continuous counterpart is Chern-Simon theory on $S^3$ (with analytically continued level). In the present setting, $b^2$ may be any positive real number, not necessarily integer, and indeed the analysis of the bi-orthogonal Stieltjes-Wigert ensemble in \cite{Dolivet2006} does not rely on $p$ being integer. We shall now clarify the geometric significance of the dependence on the squashing parameter $b$ of this $q$-deformed Yang-Mills theory.

\subsubsection*{Semi-classical expansion}

Following~\cite{Caporaso2005}, we will begin by performing a modular inversion of the series \eqref{eq:Z-b2YM} to obtain the dual description of the $q$-deformed Yang-Mills matrix model in terms of instanton degrees of freedom. For this, we consider the function
\begin{equation}
	F_{b} (x_1, \dots, x_N) = \e^{- \frac{g_{\rm str}}{4}\, \sum_{j=1} ^N\, x_j ^2 } \  \prod_{1 \le j < k \le N}\, 2 \sinh \Big( \frac{g_{\rm str}}{2 \,b^2}\, (x_j - x_k) \Big) 
\end{equation}
of continuous variables $(x_1, \dots, x_N) \in \mathbb{R}^N$. Its Fourier transform is given by
\begin{align}
\widehat{F} _{b} (y_1, \dots, y_N) &:= \int_{\mathbb{R}^N}\, \dd x \, \e^{ \sum_{j=1} ^N\, (2 \pi \ii x_j\, y_j - \frac{ g_{\rm str}}{4}\, x_j^2) } \ \prod_{1 \le j < k \le N}\,  \Big( \e^{ \frac{(x_j - x_k)\, g_{\rm str}}{2\,b^2} } -  \e^{ - \frac{(x_j - x_k)\, g_{\rm str}}{2\,b^2} } \Big) \notag \\[4pt]
	& = \e^{- \frac{4 \pi^2}{g_{\rm str}}\, \sum_{j=1} ^N\, \big( y_j + \frac{ \ii (N-1)\, g_{\rm str} }{4 \pi \,b^2 } \big)^2 + \frac{N \,(N-1)^2 \, g_{\rm str} }{2 \,b^4} } \notag \\
& \qquad \times \ \int_{\mathbb{R}^N}\, \dd u \, \e^{ - \frac{g_{\rm str}}{4}\, \sum_{j=1} ^N\, u_j ^2 } \ \prod_{1 \le j < k \le N}\, \Big( \e^{\frac{g_{\rm str}\, u_j}{b^2}  + \frac{ 4 \pi\, \mathrm{i}\, y_j }{ b^2} } -  \e^{\frac{g_{\rm str}\, u_k}{b^2}  + \frac{ 4 \pi\, \mathrm{i}\, y_k }{b^2} } \Big) \ ,
\end{align}
where we completed squares and changed integration variables to $u_j = x_j - \frac{ 4 \pi \ii }{g_{\rm str}}\, y_j + \frac{N-1}{b^2}$. We now change integration variables again, in the usual way for matrix models with hyperbolic interactions, by defining
\begin{equation}
	z_j = \e^{ \frac{g_{\rm str}\, u_j}{b^2} + \frac{2\, g_{\rm str} }{b^4}} 
\end{equation}
to get
\begin{align}
	\widehat{F}_{b} (y_1, \dots, y_N) &= \Big(\frac{b^2}{g_{\rm str}} \Big)^N \, \e^{ - \frac{g_{\rm str}}{2\,b^4 }\, N\,(N^2+1) }\,\e^{- \frac{4 \pi^2}{g_{\rm str}}\, \sum_{j=1} ^N \big( y_j + \frac{ \ii (N-1)\, g_{\rm str} }{4 \pi\, b^2 } \big)^2  } \notag \\
& \qquad \times \ \int_{(0, \infty)^N} \dd z \, \e^{- \frac{ b^4}{4\, g_{\rm str}} \sum_{j=1} ^N\, ( \log z_j )^2 } \ \prod_{1 \le j <k \le N}\, \Big( z_j\, \e^{\frac{ 4 \pi\, \mathrm{i}\, y_j }{ b^2} } - z_k\, \e^{\frac{ 4 \pi\, \mathrm{i}\, y_k }{b^2} } \Big) \ .
\end{align}

The integral expression we have arrived at is exactly the same as in \cite[eq.~(3.14)]{Caporaso2005} under the identification of the string coupling constant $\tilde g_{\rm str}$ there as $\tilde{g}_{\rm str} = \frac{g_{\rm str}}{b^2} $, which as we have seen is the coupling that reproduces the standard $q$-deformed Yang-Mills theory. We can therefore evaluate the integral using Stieltjes-Wigert polynomials to get
\begin{align}
	\widehat{F}_{b} (y_1, \dots, y_N) = \Big( \frac{4 \pi}{ g_{\rm str}} \Big)^{\frac{N}{2}}\, & \e^{\frac{g_{\rm str}}{6\,b^4}\, N\, (N-1)\,(N-2)}\, \e^{- \frac{4 \pi^2}{g_{\rm str}}\, \sum_{j=1} ^N \,\big( y_j + \frac{ \ii (N-1)\, g_{\rm str} }{4 \pi\,b^2 } \big)^2  } \notag \\ & \times \  \prod_{1 \le j < k \le N }\, \Big( \e^{\frac{ 4 \pi\, \mathrm{i}\, y_j }{b^2 } }  - \e^{\frac{ 4 \pi\, \mathrm{i}\, y_k }{b^2} }  \Big) \ .
\end{align}
At this point, we apply the convolution theorem for Fourier transformations to get
\begin{align}
	Z _{b} (y_1, \dots, y_N) & := \int_{\mathbb{R}^N}\, \dd x \, \e^{2 \pi\, \mathrm{i}\, \sum_{j=1} ^N\, x_j\, y_j }\, F_{b} (x_1, \dots, x_N)\, F_{1} (x_1, \dots, x_N) \notag \\[4pt]
		&=  \int_{\mathbb{R}^N}\, \dd t \  \widehat{F}_{b} \Big( \frac{y_1 - t_1}{2}, \dots,  \frac{y_N - t_N}{2} \Big)\, \widehat{F}_{1} \Big( \frac{y_1 + t_1}{2}, \dots,  \frac{y_N + t_N}{2} \Big) \ .
\end{align}
After some calculation, we arrive finally at
\begin{align}
	Z _{b} (y_1, \dots, y_N) =  \e^{- \frac{ 2 \pi^2 }{g_{\rm str}}\, \sum_{j=1} ^N\, y_j ^2 } \ \cW _{b} (y_1, \dots, y_N) \ ,
\end{align}
with weight given by
\begin{align}
\cW_{b} (y_1, \dots, y_N)  = & \ \Big( \frac{ 4 \pi }{g_{\rm str} } \Big)^{N}\,  \e^{\frac{g_{\rm str}}{12 }\, N\, (N^2-1)\, ( 1+ b^{-4}) } \label{eq:cWb2} \\ & \times \ \int_{\mathbb{R}^N}\, \dd t \, \e^{ - \frac{2 \pi^2}{g_{\rm str}}\, \sum_{j=1} ^N\, t_j ^2 } 
 	 \ \prod_{1 \le j<k \le N}\, 2\, \Big( \cos \pi\, \big(  y_{jk}\, ( 1+b^{-2}) + t_{jk}\, ( 1-b^{-2}) \big) \notag \\
 	& \hspace{6.5cm} - \, \cos \pi\, \big(  y_{jk}\, ( 1-b^{-2}) + t_{jk}\, ( 1+ b^{-2}) \big) \Big) \ , \notag
\end{align}
where we adopted the shorthand notation $y_{jk} := y_j - y_k$ and $t_{jk} := t_j - t_k$. This correctly reproduces the weight of~\cite[eq.~(3.20)]{Caporaso2005} in the limit $b=1$.

The final step in developing the semi-classical expansion of the partition function \eqref{eq:Z-b2YM} is Poisson resummation, and we finally arrive at
\begin{equation}
\label{eq:Z_instanton}
	\mathcal{Z}_N (b) = \frac{1}{b^{N}\, N!}\, \sum_{ \vec{\ell}\, \in \zed ^N }\, Z_{b} ( \vec{\ell}\ )  = \frac{1}{b^{N}\, N!}\, \sum_{ \vec{\ell}\, \in \zed ^N }\, \e^{- \frac{ 2 \pi^2 }{g_{\rm str}}\, (\vec{\ell},\vec\ell\ ) } \ \cW_{b} (\vec{\ell}\ ) \ .
\end{equation}
This expression admits the standard interpretation as a sum over instanton solutions of the two-dimensional gauge theory: Since the $q$-deformation arises here through one-loop determinants in the initial five-dimensional field theory, at the classical level this theory is just ordinary Yang-Mills theory on the sphere $S^2$. The exponential prefactors in the series \eqref{eq:Z_instanton} are then the classical contributions to the gauge theory path integral from the Yang-Mills action evaluated on instantons of topological charge $\ell_j\in\IZ$ corresponding to a Dirac monopole of the $j$-th factor of the maximal torus $U(1)^N\subset U(N)$, while the integrals \eqref{eq:cWb2} are the fluctuation determinants around each instanton.

\subsubsection*{Rational limit and Chern-Simons theory on $L(p,s)$}

Up to now the derivation of \eqref{eq:Z_instanton} worked for every positive real value of the squashing parameter $b$. Let us now specialise the squashing parameter to the rational values
\begin{align}
b^2=\frac ps \ \in \ \IQ \ ,
\end{align}
where $p$ and $s$ are coprime positive integers with $1\leq s\leq p$. From the five-dimensional perspective that we started with, the ellipsoid Seifert manifold $S_b^3$ then has the topology of a lens space $L(p,s)$, viewed as a circle bundle over $S^2$ with two marked points~\cite{Closset2019}; the exceptional fibres over the marked points respectively makes them $\IZ_p$ and $\IZ_s$ orbifold points. The first Chern class of the line V-bundle $\scrL(p,s)$ over the $\PP^1$ orbifold associated to $L(p,s)$ is
\begin{align}
c_1\big(\scrL(p,s)\big) = \frac sp \, \omega_{\mathbb{P}^1} \ ,
\end{align}
which cancels the local curvatures at the marked points of $\PP^1$ to ensure
that the total degree of the Seifert fibration is zero.
This is also homeomorphic to the `fake' lens space which is the quotient $S^3/\IZ_p$ by the free $\IZ_p$-action
\begin{align}\label{eq:orbifold}
\big(z_1,z_2\big)\longmapsto \big(\e^{2\pi\ii s/p}\, z_1,\e^{2\pi\,\mathrm{i}/p}\,z_2\big) \ ,
\end{align}
where $S^3$ is regarded as the
unit sphere in $\IC^2$. From the two-dimensional perspective, we will now show that the instanton expansion \eqref{eq:Z_instanton} retains topological information reflecting its five-dimensional origin, by rewriting it in terms of flat connection contributions to $U(N)$ Chern-Simons gauge theory on the lens spaces $L(p,s)$. 

Since
\begin{align}
\pi_1 \big(L(p,s)\big) = \zed_p \ ,
\end{align} 
gauge inequivalent flat $U(N)$ connections are labelled by $N$-tuples $\vec{m} \in (\zed_p) ^N$, which are torsion magnetic charges coming from the pullback of a Yang-Mills instanton on the sphere $S^2$ to a flat connection on $L(p,s)$~\cite{Griguolo:2006kp}. Let us then rewrite the series variables $\vec\ell\in\IZ^N$ in \eqref{eq:Z_instanton} as
\begin{equation}
	\vec \ell = \vec m + p\,\vec l \ ,
\end{equation}
with $\vec l\in\IZ^N$.
We can rewrite the interactions among eigenvalues from \eqref{eq:cWb2} as
\begin{equation}
	4\, \sin \frac{\pi\,s}{p}\, \big(m_{jk} - t_{jk}+ p\, (l_j - l_k) \big) \, \sin \pi\, \big(  m_{jk} + t_{jk} + p\, (l_j - l_k) \big) \ ,
\end{equation}
and thus it depends only on the values of $\vec\ell\in\IZ^N$ modulo $p$, that is, on $\vec m\in(\IZ_p)^N$.\footnote{Strictly speaking, this is only true if the integers $p$ and $s$ have the same even/odd parity, which we tacitly assume.} They are also invariant under the Weyl symmetry group $S_N$ of the gauge group $U(N)$, so that we can reduce the sum over all $N$-tuples $\vec{m} \in (\zed_p)^N$ to the ordered ones with
\begin{equation}
	m_N \ge m_{N-1} \ge \cdots \ge m_1 \ .
\end{equation}
Thus the partition function \eqref{eq:Z_instanton} depends only on how many times the integers $k\in\{0,1,\dots,p-1\}$ appear in the string $(m_1, \dots, m_N)$. We denote these multiplicities as $\boldsymbol N=(N_0,N_1,\dots,N_{p-1})$, which by construction are $p$-component partitions of the rank $N$:
\begin{align}
N_k\geq0 \qquad \mbox{and} \qquad \sum_{k=0} ^{p-1}\, N_k = N \ .
\end{align}
Under this reordering the Weyl symmetry breaks according to
\begin{equation}
	S_N \longrightarrow S_{N_0} \times S_{N_1} \times \cdots \times S_{N_{p-1}} \ .
\end{equation}

The partition function \eqref{eq:Z_instanton} is then rewritten as
\begin{align}
\mathcal{Z}_N (p,s) = \Big(\frac{s}{p}\Big)^{N/2} \ \sum_{\boldsymbol N\,\vdash\,N} \ \frac1{\prod\limits_{k=0} ^{p-1}\, N_k !} \ & \cW_{p,s} \big( 0^{N_0} , 1^{N_1}, \dots , (p-1)^{N_{p-1}} \big) \\
	& \times \ \sum_{\vec{l}\, \in \zed^{N} } \ \prod_{k=0} ^{p-1}\, \exp \Big( - \frac{2 \pi^2}{g_{\rm str}}\, \sum_{j= N_0 + N_1\dots +N_{k-1} +1} ^{N_k}\, (k + p\,l_j)^2 \Big) \ , \notag
\end{align}
where here $k^{N_k}:=(k,\dots,k)$ is the $N_k$-vector whose entries are all equal to $k$.
As in \cite[Section~3.3]{Caporaso2005}, we identify in the second line a product of elliptic theta-functions
\begin{align}
\vartheta_3(\tau|z) = \sum_{l\in\IZ}\, \e^{\pi\ii\tau\,l^2+2\pi\ii l\,z}
\end{align}
which enables us to write
\begin{align}
	\mathcal{Z}_N (p,s) = \sum_{\boldsymbol N\,\vdash\,N}\, \e^{-
  \frac{ 2 \pi^2}{g_{\rm str}}\, \sum_{k=0} ^{p-1}\, N_k\, k^2} \ &
                                                                    \cW_{p,s}\big(
                                                                    0^{N_0}
                                                                    ,
                                                                    1^{N_1},
                                                                    \dots
                                                                    ,
                                                                    (p-1)^{N_{p-1}}
                                                                    \big)
                                                                    \nonumber
  \\ & \times \ \prod_{k=0}^{p-1}\, \frac{ \vartheta _3 \big( \frac{ 2 \pi \ii p^2 }{g_{\rm str}} \big\vert  \frac{ 2 \pi \ii p\, k }{g_{\rm str}} \big)^{N_k} }{ N_k ! } \ .
\label{eq:ZNpstheta}\end{align}
We can write the fluctuation weight $\cW_{p,s}\big( 0^{N_0} , 1^{N_1}, \dots , (p-1)^{N_{p-1}} \big)$ explicitly in its integral form \eqref{eq:cWb2} and reorganize the integration variables $t_j$ into subsets $t^{J} _j$ with $J \in \left\{ 0,1, \dots, p-1 \right\}$ and $j \in \left\{ 1, \dots , N_J \right\}$. We then shift integration variables as $u^{J} _j := t^{J} _j - j$ to get
\begin{align}
& \cW_{p,s}\big( 0^{N_0} , 1^{N_1}, \dots , (p-1)^{N_{p-1}} \big)	\notag \\[4pt] & \quad = \Big( \frac{ 4 \pi }{g_{\rm str} } \Big)^{N}\, \e^{\frac{g_{\rm str}}{12 }\, N\, (N^2-1)\, \big( 1+ \frac{s^2}{p ^2} \big) } \label{eq:cWps} \\ & \hspace{1cm} \times \ \prod_{J=0} ^{p-1} \ \int_{\mathbb{R}^{N_J}} \, \dd u^{J} \, \e^{ - \frac{2 \pi^2}{g_{\rm str}}\, \sum_{j=1} ^{N_{J}}\, (u_j ^{J} + j) ^2 } \ \prod_{1 \le j < k \le N_J }\, 4\, \sin  \frac{ \pi\,s}{p}\,  \big( u_j ^{J} - u_k  ^{J}  \big)\, \sin \pi\, \big(   u_j ^{J} - u_k  ^{J} \big) \notag \\
	& \hspace{3cm} \times \ \prod_{0\leq J<K\leq p-1} \ \prod_{j=1}^{ N_J } \ \prod_{k=1}^{N_K}\, 4\, \sin \frac{ \pi\,s }{p}\, \big(  u_j ^{J} - u_k  ^{K}  + J-K  \big) \, \sin \pi\, \big( u_j ^{J} - u_k  ^{K} +J-K  \big) \ . \notag
\end{align}
The products here which are independent of $(p,s)$ combine to give a standard Weyl determinant, while the $(p,s)$-dependent products carry the information about the surgery data of the Seifert homology sphere $X(s/p)$. 

In fact, if we drop the product of theta-functions from the sum \eqref{eq:ZNpstheta} and rescale the string coupling as before to $\tilde g_{\rm str}=s\,g_{\rm str}/p$, we can recognise the analytically continued partition function of $U(N)$ Chern-Simons gauge theory at level $k\in\IZ$ on the lens space $L(p,s)$: the exponential prefactor is recognized as the classical contribution to the path integral from the Chern-Simons action evaluated on the flat $U(N)$ connection labelled by $\boldsymbol N$~\cite{Marino2002,Griguolo:2006kp}, with the analytic continuation
\begin{align}
\tilde g_{\rm str} = \frac{s\,g_{\rm str}}p = \frac{2\pi\ii}{k+N} \ .
\end{align}
Moreover, after a straightforward change of integration variables (and subsequent analytic continuation), the integral expression \eqref{eq:cWps} is easily seen to agree with the multi-eigenvalue integral formula from~\cite[Theorem~7]{Brini:2008ik} for the contribution to the one-loop fluctuation determinant from the flat connection $\boldsymbol N$. Thus the full partition function \eqref{eq:ZNpstheta} can be written as
\begin{equation}
	\mathcal{Z}_N (p,s) = \sum_{\boldsymbol N\,\vdash\,N} \, Z_{p,s} ^{\textrm{\tiny CS}} (\boldsymbol N) \ \prod_{k=0}^{p-1}\, \frac{ \vartheta _3 \big( \frac{ 2 \pi \ii p^2 }{g_{\rm str}} \big\vert  \frac{ 2 \pi \ii p\, k }{g_{\rm str}} \big)^{N_k} }{ N_k ! } \ ,
\end{equation}
where
\begin{equation}
Z_{p,s} ^{\textrm{\tiny CS}} (\boldsymbol N) :=  \exp\bigg( \frac{ 2 \pi^2\,s}{\tilde g_{\rm str}\,p}\, \sum_{k=0} ^{p-1}\, N_k\, k^2 \bigg) \ \cW_{p,s}\big( 0^{N_0} , 1^{N_1}, \dots , (p-1)^{N_{p-1}} \big)
\end{equation}
is the contribution to the Chern-Simons partition function from the point of the moduli space of flat connections on the lens space $L(p,s)$ labelled by
\begin{equation}
\vec m = \big( 0^{N_0} , 1^{N_1}, \dots , (p-1)^{N_{p-1}} \big) \ .
\end{equation}

The connection between Chern-Simons theory on lens spaces $L(p,s)$
with $s>1$ and $q$-deformed Yang-Mills theory was also obtained
in~\cite{Griguolo:2006kp}, but in a much different and more
complicated fashion. There the two-dimensional gauge theory is defined
on the collection of exceptional divisors of the four-dimensional
Hirzebruch-Jung space $X_4(p,s)$, which is the minimal resolution of
the $A_{p,s}$ singularity defined by the same orbifold action
\eqref{eq:orbifold} on $\IC^2$. The corresponding partition function
depends explicitly on the intersection moduli $e_i$ of the exceptional
divisors, which in the three-dimensional case translate into framing
integers that enter the surgery construction of the Seifert space
$L(p,s)$; after stripping away the Chern-Simons fluctuation
determinants, the resulting partition function computes the
contribution of fractional instantons to the partition function of
topologically twisted $\cN=4$ Yang-Mills theory on
$X_4(p,s)$~\cite{Griguolo:2006kp}. This is \emph{not} the case
here. Like the topological Chern-Simons theory, our two-dimensional
gauge theory partition function \eqref{eq:ZNpstheta} is independent of
the framing integers $e_i$ and depends only on the pair of integers
$(p,s)$ which uniquely determine $L(p,s)$ up to homeomorphism. In
particular, stripping away the Chern-Simons fluctuation determinants
$\cW_{p,s}\big( 0^{N_0} , 1^{N_1}, \dots , (p-1)^{N_{p-1}} \big)$
would leave a $(p,s)$-independent partition function proportional to
\smash{$\vartheta_3\big(\frac{2\pi\ii}{g_{\rm
    str}}\big|0\big)^N$}~\cite{Griguolo:2006kp}, which is the
contribution of fractional instantons to the partition function of
$\cN=4$ gauge theory on $X_4(1,1)\cong\cO_{\PP^1}(-1)$. This suggests
that our squashing of the two-dimensional $q$-deformed gauge theory on
$S^2$ is, like the standard theory at $p=1$, also related to the Calabi-Yau geometry of the resolved conifold $\cO_{\PP^1}(-1)\oplus\cO_{\PP^1}(-1)$; the topological string interpretation of this theory is certainly worthy of further investigation.

\subsubsection*{Large $\boldsymbol N$ limit}

For any finite value of the rank $N$, the partition function
$\cZ_N(b)$ is a continuous function of the squashing parameter
$b>0$. In this sense our squashed $q$-deformations of two-dimensional
Yang-Mills theory are continuations of the lens space theories
analysed above: Since the set of rational $b^2$ is dense in the space
of all squashing parameters $b>0$, any partition function can be
expressed as a limit of the two-dimensional gauge theories whose
geometric meanings were explained above. It would be interesting to
understand more precisely what the underlying geometry means for generic
real values $b>0$.

However, we do expect the partition function \eqref{eq:Z-b2YM} to
experience a phase transition in the large~$N$ regime, triggered by
the discreteness of the matrix model, at least for large enough values
of the squashing parameter $b$. The standard $q$-deformed Yang-Mills
theory on $S^2$ undergoes a phase transition for $p >2$~\cite{Caporaso2005}, and we can
extrapolate this to our more general setting. The
eigenvalue distribution $\rho(\lambda)$ of a discrete random matrix ensemble is subject to the constraint
\begin{equation}
	\rho (\lambda) \le 1 \ ,
\end{equation}
which in the present case is always fulfilled at large $N$ when $b \le \sqrt2$. It would be
interesting to see how the phase transition appears at $b>\sqrt2$ in terms of the
bi-orthogonal Stieltjes-Wigert polynomials of~\cite[Section~4.1]{Dolivet2006}. It was argued in \cite{Dolivet2006},
and later proved in \cite{Szabo2013}, that around the trivial flat
connection the discrete and continuous versions of the
Stieltjes-Wigert ensemble are essentially the same, thus the
zero-instanton sector of our squashing of $q$-deformed Yang-Mills theory can be
obtained exactly via bi-orthogonal polynomials. For the case $b=1$ this
gives the full partition function of $q$-deformed Yang-Mills theory,
since the only flat connection on $L(1,1) \cong S^3$ is trivial.

Introduce the 't~Hooft coupling
\begin{equation}
	t:= g_{\rm str}\, N 
\end{equation}
and take the double scaling limit $N \to \infty, g_{\rm str} \to 0$
with $t$ fixed. In this limit, the partition function
\eqref{eq:Z-b2YM} is proportional to the Chern-Simons matrix model on
$L(p,1)$ around the trivial connection, continued to
$p=b^2\in\IR$. Equivalently, from the instanton expansion
\eqref{eq:Z_instanton} we infer that, as long as the fluctuations
$\cW_b(\vec\ell\ )$ give sub-leading contributions, all instanton
contributions are suppressed except for the trivial one. Taking the
large $N$ limit of~\cite[eq.~(2.26)]{Dolivet2006}, in the large $N$
regime we obtain
\begin{align}
	\mathcal{Z}_N^{(\infty)} (b) = 2^{N\,(N+1)}\, \Big( \frac{ 2
  \pi}{g_{\rm str}} \Big)^{\frac{N}{2} }\, \exp \bigg(  - \frac{ N^2\,
  t }{12\,b^4 } \, \big( 3\, b^8 + 6\, b^4 -13\big) - \frac{ N^2\,
  b^8}{t^2}\, F^{(0)} _{\textrm{\tiny CS}} \Big( \frac{ t}{b^4} \Big)
  \bigg) \ ,
\end{align}
where 
\begin{equation}
	F^{(0)} _{\textrm{\tiny CS}} (t) = \frac{t^3}{12}\, - \frac{
          \pi^2\, t }{6} - \mathrm{Li}_3 \big( \e^{-t}\,\big) + \zeta
        (3) 
\end{equation}
is the planar free energy of Chern-Simons theory on $S^3$ with
(analytically continued) 't Hooft coupling $t$. For $b\le \sqrt2$ this
solution is exact, and indeed the free energy of the supersymmetric
gauge theory on $S^3_b \times S^2$ exhibits the $N^3$ behaviour in the
strong coupling region for $t \to \infty$, as in the case of the five-sphere
$S^5$~\cite{Kallen2012b}. However, for higher values of $b$, this solution ceases to be
valid for large $t$ and we expect the strong coupling region to have a
different solution. 

\subsection{Localization from seven dimensions}

With the premise of obtaining more general two-dimensional theories, we conclude by briefly commenting on how the derivation of $q$-deformed
Yang-Mills theories would change if instead one started from a
seven-dimensional cohomological gauge theory; as we argue, the story
works in essentially the same way as in the five-dimensional case,
though the physical significance of such a two-dimensional theory is
not clear in this instance. For this, we
consider our cohomological localization procedure for a seven-dimensional manifold $M_7= M_5 \times \Sigma_h$.
In order to have the right amount of supercharges in seven dimensions, we ought to
start with a five-dimensional Seifert manifold admitting $\mathcal{N}=2$
supersymmetry, and henceforth we take the five-sphere $M_5= S^5 $ for definiteness.

The seven-manifold $M_5 \times \Sigma_h$ cannot come from a supergravity background, and 
we must pull back the Killing spinors on $M_5$ to seven dimensions. We define the theory on $S^5 \times \mathbb{R}^2$ 
and build the components of a seven-dimensional Killing spinor from the
tensor product of a five-dimensional Killing spinor with a constant
spinor along $\mathbb{R}^2$. Schematically, 
\begin{equation}
\varepsilon^{(7)} = \varepsilon^{(5)} \otimes \zeta^{(2)} _{\pm}  \ .
\end{equation}
Then we topologically twist the theory, and finally we put the
cohomological field theory on $S^5 \times \Sigma_h$.\footnote{Had we partially twisted the theory along $\Sigma_h$,
  we would have obtained a localization locus which is the same as for
  the purely five-dimensional theory. The main difference between the
  seven-dimensional and the five-dimensional setting with partial
  twist would be the following. For the case $S^3 \times \Sigma_h$ the
  localization locus consisted only of the trivial connection on
  $S^3$, thus effectively reducing to a two-dimensional theory. For
  $S^5 \times \Sigma_h$, on the contrary, we would obtain a sum of
  copies of two-dimensional theories, each one in a different
  background. Such backgrounds for the two-dimensional theory are
  contact instantons on $S^5$, which descend to anti-self-dual instantons on $\mathbb{P}^2$. We do not pursue the partial twist description here.} 
The cohomological gauge theory in seven dimensions was constructed in
\cite{MinahanZabzine2015} for the seven-sphere $S^7$, and extended in
\cite{Polydorou2017,Rocen2018} to other geometries. We do not review
it here and refer to \cite{MinahanZabzine2015,Polydorou2017} for the
details. Our construction in seven dimensions differs from that of
\cite{MinahanZabzine2015} in exactly the same manner as our
construction in five dimensions differs from that of
\cite{Kallen2012,Kallen2012a}. In particular, we pick a preferred
vector field $v$ which generates the rotations
along the $U(1)$ fibre of $M_5$.

Again, the details of the supersymmetry transformations and
localization locus do not depend on the specific details of the
geometry given (although the solutions do depend on it) and we simply reproduce step by step the
procedure of \cite[Section 5]{MinahanZabzine2015} and \cite[Section 4]{Polydorou2017}. 
Fortunately, most of the steps in the construction of a cohomological field theory in \cite{MinahanZabzine2015} rely only on the Seifert structure and not on the 
K-contact structure, and can be adapted to our framework with minor changes. 
The localization locus has a complicated expression, and we only focus here on the perturbative part, hence expanding around the trivial connection. 
For the vector multiplet, we have to study the same ratio of fluctuation
determinants as in~\cite[eqs.~(4.18)--(4.19)]{Polydorou2017}, which
has the form
\begin{equation}
h(\phi) = \frac{\det \mathrm{i}\, {\sf L}_\phi
          \vert_{\Omega^{(0,0)}_H(M_7,\frg) } \ \det \mathrm{i}\, {\sf
            L}_\phi \vert_{\Omega^{(0,2)}_H(M_7,\frg) } }{\det \mathrm{i}\,
          {\sf L}_\phi \vert_{\Omega^{(0,1)}_H(M_7,\frg) } \ \det
          \mathrm{i}\, {\sf L}_\phi \vert_{\Omega^{(0,3)}_H(M_7,\frg) } }  \ ,
\end{equation}
where the differential operator ${\sf L}_\phi$ is, as usual, the sum of a
Lie derivative $\cL_v$ along $v$ and a gauge transformation
$\cG_\phi$. This ratio of determinants can be computed using the same
strategy as in five dimensions. In fact, the product geometry $S^5
\times \Sigma_h$ is simpler than, for example $S^7$, since we know how to
decompose the vector space of differential forms into eigenmodes of the Lie
derivative $\cL_v$. For $M_7= S^5 \times \Sigma_h$, this is exactly
the decomposition of \cite{Kallen2012} on $S^5$, which is classified
by powers of the line bundle $\mathscr{L}$ on the projective plane
$\mathbb{P}^2$ associated to the Seifert fibration $S^5\to\PP^2$. 
At the end of the day, the number of remaining modes is counted by the index of
the twisted Dolbeault complex 
\begin{align}
	\Omega^{(0,0)} (\mathbb{P}^2 \times \Sigma_h)_{m,\alpha}
        \xrightarrow{ \bar{\partial} ^{(m)} }  \Omega^{(0,1)}
        (\mathbb{P}^2 \times \Sigma_h)_{m,\alpha} & \xrightarrow{
          \bar{\partial} ^{(m)} } \Omega^{(0,2)} (\mathbb{P}^2 \times
        \Sigma_h)_{m,\alpha} \nonumber \\ & \hspace{3cm} \xrightarrow{ \bar{\partial} ^{(m)} }
        \Omega^{(0,3)} (\mathbb{P}^2 \times \Sigma_h)_{m,\alpha} \ ,
\end{align}
where we adopted the shorthand notation 
\begin{equation}
\Omega ^{( \bullet, \bullet ) } (\mathbb{P}^2 \times \Sigma_h )_{m,\alpha}:= \Omega ^{( \bullet,  \bullet ) } ( \mathbb{P}^2 \times \Sigma_h, \mathscr{L} ^{\otimes m}\otimes \mathfrak{g}_{\alpha}) \ . 
\end{equation}
We then arrive at 
\begin{equation}
\label{eq:1loopdetS5xSigma}
Z_{\rm vec}^{\rm pert}(S^5\times\Sigma_h) =\prod_{\alpha \in \triangle} \ \prod_{m \ne 0}\, \Big(
        \frac{m}{r} - \mathrm{i}\, (\alpha, \phi)
        \Big)^{\mathrm{index} \, \bar{\partial} ^{(m)} } \ .
\end{equation}

The index is computed as follows. The total Chern class of the
holomorphic tangent bundle of the base $K_6=\PP^2\times\Sigma_h$ of
the seven-dimensional Seifert fibration is
\begin{align}
	c (T^{1,0} K_6) & = c ( T^{1,0} \mathbb{P}^2  ) \wedge c (
                          T^{1,0} \Sigma_h) \nonumber \\[4pt]
	&= \big( 1 + \omega_{\mathbb{P}^2} \big)^{\wedge 3} \wedge \big( 1 +
          \chi (\Sigma_h)\, \omega_{\Sigma_h} \big) \nonumber \\[4pt]
	&= 1 + \big(3\, \omega_{\mathbb{P}^2} + \chi (\Sigma_h)\,
          \omega_{\Sigma_h}  \big) + 3\, \big( \omega_{\mathbb{P}^2}
          \wedge \omega_{\mathbb{P}^2} + \chi (\Sigma_h)\,
          \omega_{\mathbb{P}^2} \wedge \omega_{\Sigma_h} \big) +
          3\,\omega_{\PP^2}\wedge\omega_{\PP^2}\wedge\omega_{\Sigma_h}
          \nonumber \\[4pt]
&= 1+c_1(T^{1,0}K_6)+c_2(T^{1,0}K_6)+c_3(T^{1,0}K_6) \ ,
\end{align}
which gives the corresponding Todd class
\begin{align}
	\mathrm{Td} ( T^{1,0}K_6) &= 1+\mbox{$\frac12$}\,
        c_1(T^{1,0}K_6)+ \mbox{$\frac1{12}$}\,
        \big(c_1(T^{1,0}K_6)\wedge
        c_1(T^{1,0}K_6)+c_2(T^{1,0}K_6)\big) \nonumber \\
& \qquad + \tfrac1{24}\,
        c_1(T^{1,0}K_6)\wedge c_2(T^{1,0}K_6) \nonumber \\[4pt]
&= 1 + \tfrac{3}{2}\, \omega_{\mathbb{P}^2} + (1-h)\,\omega_{\Sigma_h}
+  \omega_{\mathbb{P}^2}  \wedge \omega_{\mathbb{P}^2}  +
\tfrac32\,(1-h)\,\omega_{\mathbb{P}^2}  \wedge \omega_{\Sigma_h}
\nonumber \\ & \qquad + (1-h)\, \omega_{\mathbb{P}^2}  \wedge
\omega_{\mathbb{P}^2}  \wedge \omega_{\Sigma_h} \ .
\end{align}
Since $c_1(\scrL)=\omega_{\PP^2}$, the Chern character of the line bundle $\mathscr{L}^{\otimes m}$ is
\begin{align}
\mathrm{ch} ( \mathscr{L}^{\otimes m}) &= 1 + c_1  (
                                         \mathscr{L}^{\otimes m } ) +
                                         \tfrac12\,c_1  (
                                         \mathscr{L}^{\otimes m }
                                         )\wedge c_1  (
                                         \mathscr{L}^{\otimes m } )
                                         +\tfrac16\, c_1  (
                                         \mathscr{L}^{\otimes m }
                                         )\wedge c_1  (
                                         \mathscr{L}^{\otimes m }
                                         ) \wedge c_1  (
                                         \mathscr{L}^{\otimes m }
                                         ) \nonumber \\[4pt]
&= 1 + m\, \omega_{\mathbb{P}^2} + \tfrac12\, m^2\,
  \omega_{\mathbb{P}^2} \wedge \omega_{\mathbb{P}^2} \ .
\end{align}
A simple computation as in the five-dimensional setting then gives
\begin{equation}
	\mathrm{index} \, \bar{\partial} ^{(m)} \big|_{\mathbb{P}^2\times\Sigma_h } = (1-h)\,\big( 1+
        \tfrac32\, m + \tfrac12\, m^2\big) = (1-h) \ \mathrm{index} \,
        \bar{\partial} ^{(m)} \big|_{\mathbb{P}^2 } \ .
\end{equation}
We conclude that the computation of the one-loop determinant of the vector
multiplet on $S^5$ is lifted to $S^5 \times \Sigma_h$ in exactly the
same way as the computation on $S^3$ is lifted to $S^3 \times
\Sigma_h$. The resulting contribution to the partition function yields polylogarithmic corrections to the standard $q$-deformed measure~\cite{Kallen2012}, and in this case \eqref{eq:1loopdetS5xSigma} evaluates to
\begin{align}
Z_{\rm vec}^{\rm pert}(S^5\times\Sigma_h) = \prod_{\alpha\in\triangle_+} \, \bigg(\frac{\sinh\big(\pi\,r\,(\alpha,\phi)\big)^2}{\big(\pi\,r\,(\alpha,\phi)\big)^2} \, \e^{f(\pi\,r\,(\alpha,\phi))/\pi^2} \bigg)^{1-h} \ ,
\end{align}
where the function $f$ is defined by
\begin{align}
f(x)=-\frac{x^3}3-x^2\,\log\big(1-\e^{2\,x}\big)-x\,{\rm Li}_2\big(\e^{2\,x}\,\big)+\frac12\,{\rm Li}_3\big(\e^{2\,x}\,\big)-\frac{\zeta(3)}2 \ .
\end{align}

\paragraph*{Acknowledgements.}

We thank Luca Griguolo, Sara Pasquetti and Domenico Seminara for helpful discussions. 
The work of L.S. was supported by the Doctoral Scholarship
SFRH/BD/129405/2017 from the Funda\c{c}\~{a}o para a Ci\^{e}ncia e a
Tecnologia (FCT). The work of L.S. and M.T. was 
supported by the FCT Project PTDC/MAT-PUR/30234/2017. The work of R.J.S. was supported by the Consolidated Grant ST/P000363/1 from the UK Science and Technology Facilities Council (STFC). 

\begin{appendix}

\section{Spinor conventions}
\label{app:notation}

For field theories in three dimensions we follow the normalization and conventions of \cite{Kallen2011}. 
We work in the $\mathcal{N}=2$ formalism, with $SU(2)_R$ R-symmetry, and for theories with $\mathcal{N}=4$ supersymmetry only $SU(2)_R \subset SU(2)_C \times SU(2)_H$ is manifest. 
In five dimensions, we follow \cite{Kallen2012a}. We work in the
$\mathcal{N}=1$ formalism, with $SU(2)_R$ R-symmetry, using the
letters $I,J, \dots$ for the indices. In theories admitting
$\mathcal{N}=2$ supersymmetry, only $SU(2)_R \subset SU(2)_R \times
U(1)_{r} \subset SO(5)_R$ is manifest, where $SU(2)_R \times
U(1)_{r}$ is the maximal R-symmetry group preserved by the
product manifolds considered in the main text and $SO(5)_R$ the R-symmetry in five-dimensional flat space. 
In both three and five dimensions, $SU(2)_R$ indices are raised and lowered with the Levi-Civita symbol $\epsilon_{IJ}$ or $\epsilon^{IJ}$, with the convention $\epsilon_{12} = -1 = - \epsilon^{12}$.

We do not write Lorentz spin indices explicitly in the spinors, and understand that they are contracted using the charge conjugation matrix $C$, a real antisymmetric matrix satisfying
\begin{equation}
	C\, \Gamma^{\mu} = ( \Gamma^{\mu} )^{\top}\, C \ .
\end{equation}
With this choice, spinor components are taken to be Grassmann-even, and anticommutation is a consequence of $C^{\top}=-C$. 
This is also in agreement with the conventions in \cite{Pan2013,Alday2015}, where Killing spinors from rigid supergravity are taken to be Grassmann-even symplectic Majorana spinors. 
Our notation for Killing spinors is then as follows: $\varepsilon^{I}$ satisfies
\begin{equation}
\left( \epsilon_{IJ}\, C \varepsilon^{J} \right)^{\ast} =
\varepsilon^{I} \ ,
\end{equation}
with $C$ the charge conjugation matrix. This implies
\begin{equation}
\varepsilon^{\dagger} _I = \left( \epsilon_{IJ}\, \varepsilon^{J}
\right)^{\top}  \ .
\end{equation}

We impose the following reality conditions on the fields in the five-dimensional theories. 
The scalars $(\mathtt{q}^{\dagger})^{I}$ in a hypermultiplet are
related to $\mathtt{q}_I$ by complex conjugation and transposition. As
we are working in Euclidean space, there is no reality condition on
the spinor fields, and we have to choose a half-dimensional
integration cycle in the configuration space of fields. 
The gauginos $\lambda^{I}$ are symplectic Majorana spinors,
\begin{equation}
	\left( \epsilon_{IJ}\, C \lambda^{J} \right)^{\ast} =
        \lambda^{I} \ ,
\end{equation}
and we take as a \emph{definition} of the fields $\lambda^{\dagger}_I$ the equation
\begin{equation}
	\lambda^{\dagger}_I = \left( \epsilon_{IJ} \, \lambda^{J}
        \right)^{\top} \ .
\end{equation} 
The reasoning for the spinor $\psi$ in a hypermultiplet is analogous. See also \cite[Section 2.1]{Qiu2016} for discussion about the treatment of the reality condition for spinors. The sole difference between our conventions and those of \cite{Kallen2012a,Qiu2016} is a factor $\sqrt{-1}$ in the definition of the scalar $\sigma$ in the vector multiplet, as in those references the rotated field (which we denote $\sigma_0$) is taken from the very beginning.

In three dimensions, the gauginos $\lambda^{I}$ are not subject to additional constraints, and we impose $\lambda^{\dagger} _I$ to be related to $\lambda^{I}$ as in Minkowski signature, following \cite{Kapustin2009,Kallen2011}, and similarly for $\psi^{\dagger}$ and $\psi$.

\section{Squashed three-spheres}
	\label{app:squashedspheres}

Three types of squashed sphere $S_b^3$ that preserve at least
$\mathcal{N}=2$ supersymmetry (four supercharges) exist in the literature: the squashed
sphere called the ``familiar case'' in \cite{Hama2011}, the squashed
sphere of \cite{Imamura2011a}, and the ellipsoid of \cite{Hama2011}
which was originally called the ``less familiar case'' of squashed
sphere. We have ordered them according to their increasing deviation
from the standard
round sphere $S^3$. In the following we briefly describe and discuss them within the cohomological field theory formalism, see also \cite[Section 7]{Kawano2015} for related discussion.

The simplest case of squashed sphere is the ``familiar case'' of \cite{Hama2011}, 
for which the one-loop
determinants are the same as in the round case up to rescaling of
variables. This squashed sphere is obtained by simply changing the
radius of the Hopf fibre with respect to the radius of the base
$S^2$, so the background has isometry group $SU(2)\times U(1)$. The Killing spinor is covariantly constant, as on the round $S^3$. 
The Killing vector field $v$ has compact orbits and coincides
with the generator of rotations along the Hopf fibre, hence it 
is parallel to the Reeb vector field $\xi$. The computation of the one-loop determinants 
on this geometry is very simple in our setting: it is clear from construction (see \eqref{eq:zvec3d}--\eqref{eq:zvecsquash3d}) that 
only the radius of the circle fibre enters the one-loop determinant, and the result of \cite{Hama2011} follows immediately 
from K\"{a}ll\'{e}n's formula. More generally, the one-loop determinants on the squashed lens spaces 
$S^3 _b / \mathbb{Z}_p$, with $p\in \mathbb{Z}_{>0}$ and $S^3 _b$ the ``familiar'' squashed sphere of \cite{Hama2011}, are given by the same formula as for the round lens space, with the proper scaling of the size of the fibre.

Another squashing that preserves $\mathcal{N}=2$ supersymmetry is
the non-trivial squashed sphere of \cite{Imamura2011a}. This is obtained by
twisted dimensional reduction from $S^3 \times \mathbb{R}$ with round $S^3$. One employs a Scherk-Schwarz
compactification to put the theory on $S^3 \times S^1$ by identifying
\begin{align}
\exp\Big(2 \pi\, \beta\, \frac{\partial \ }{\partial t} + \pi\, \beta\,
  \mathcal{R}\Big) X \ \sim \ X
\end{align} 
for any field $X$, where $\beta$ is the radius of $S^1$,
$t$ is the coordinate along $\mathbb{R}$ and $\mathcal{R}$ is the
generator of an
R-symmetry transformation. This R-symmetry, which differs from the
R-symmetry at the superconformal fixed point, twists the
compactification by including a finite
rotation on $S^3$ in the periodic identification of fields along
$S^1$, and hence also twists the dimensional reduction when sending $\beta \to
0$. With this procedure, one is able to preserve an $SU(2) \times
U(1)$ isometry group and to obtain a metric on $S^3$ as for the ``familiar
case'' of \cite{Hama2011}, hence it does not modify the metric on the
base $S^2$ of the Hopf fibration. However, the Killing spinor now
becomes a non-constant field. In \cite[Section 5.2]{Closset2012} 
it was shown that the supergravity background of \cite{Imamura2011a} admits $\mathcal{N}=4$ supersymmetry. 
The Killing vector field $v$ has closed orbits, and therefore K\"{a}ll\'{e}n's localization directly applies. In this squashed sphere, 
the Killing spinor depends on a parameter $u$ related to the radius of the circle fibre through
\begin{equation}
	\frac{r^2}{\epsilon^2} = 1+u^2 \ ,
\end{equation}
where $r$ is the radius of the round base $S^2$ and $\epsilon$ is the radius of the $U(1)$ fibre. This relation 
arises from the twisted dimensional reduction. Since the Killing vector field is non-constant, although it points along the squashed Hopf fibre
one has to consider the complex dependence on $1 \pm \mathrm{i}\, u$:
at the two fixed points the fibre has radius, respectively, given by
\begin{equation}
	\epsilon_{1} = r\, b \qquad \mbox{and} \qquad \epsilon_2=r\,b^{-1} \qquad \text{with} \quad b =
        \sqrt{ \frac{1 - \mathrm{i}\, u}{1 + \mathrm{i}\, u } } \ .
\end{equation}
This justifies the applicability of the localization formula
\eqref{eq:zvecsquash3d}, and explains why the one-loop determinants with this non-trivial squashing 
are formally the same as for the ellipsoid of \cite{Hama2011}.

Finally we consider the ellipsoid of \cite{Hama2011}, which is the
squashed sphere considered mostly in the main text. 
It has only $U(1) \times U(1)$ isometry group as the squashing also deforms
the metric on the base $S^2$ of the fibration. It is defined as the locus in $\mathbb{C}^2$ satisfying
\begin{equation}
	b^2\, \vert z_1 \vert^2 + b^{-2}\, \vert z_2 \vert^2 = r^2 \ ,
\label{eq:ellipslocusC2}
\end{equation}
with $b= \sqrt{\epsilon_1/\epsilon_2}$ and $r= \sqrt{ \epsilon_1\, \epsilon_2}$.
The supergravity background only preserves four supercharges. 
As explained in \cite[Section 5.1]{Closset2012} and
highlighted in the main text, the orbits of the Killing vector field
$v$ are not closed, and $v$ does not point along the fibre
direction. It splits into two vector fields, generating two $U(1)$
isometries with closed orbits. The localized partition function on this geometry is discussed in Section~\ref{sec:3dappl}.

Notice the different geometric meaning of the squashing
parameter $b$ in the squashed sphere~\cite{Imamura2011a} and in the ellipsoid~\cite{Hama2011}. For the
ellipsoid $b>0$ is real, while for the squashed sphere $b \in \mathbb{C}$ with $\lvert b \rvert =1$. In
both cases the partition functions can be analytically continued to
arbitrary complex values $b \in \mathbb{C} \setminus \left\{ 0
\right\}$, and the expressions are given by
\eqref{eq:3dvecellipsoid} and \eqref{eq:3dhypellipsoid}. 
Thus, in practice, we can compute the one-loop determinants on the regular background $S^3 _b$ of \cite{Imamura2011a} 
and then continue the result to $b >0$, or \emph{vice versa}. 
The limit $b \to 0^{+}$ in the squashed sphere and ellipsoid also has different geometric meaning. 
For the squashed sphere, $b \to 0^{+}$ means $u \to -\mathrm{i}$, and
corresponds to blowing up the twisted Hopf fibre. In the ellipsoid, $b
\to 0^+$ reduces the geometric locus \eqref{eq:ellipslocusC2} to $\vert
z_2 \vert =0$, hence the $S^3 _b $ geometry degenerates to $\mathbb{C}$. 
In both cases the geometry becomes non-compact, and we do not expect
the cohomological localization to work in this limit.

\section{Sasaki-Einstein five-manifolds}
	\label{app:SE}

There exists a family of Sasaki-Einstein metrics on five-manifolds
$M_5$ which topologically are $U(1)$-fibrations over the product $S^2 \times
S^2$~\cite{Sasaki5d}. The simplest case is the familiar
conifold $T^{1,1}$, which is homeomorphic to $S^3\times S^2$ and so its associated line bundle $\scrL_{1,1}$ has 
Chern class $c_1(\scrL_{1,1})=\omega_1$ given by the generator $[\omega_1]\in H^2(S^2,\IZ)$ of
the first base factor. More generally, there is an infinite family of 
irregular backgrounds labeled by a pair of
integers $(p,s)$, and denoted $Y^{p,s}$
\cite{Gauntlett2004}. The first Chern class
associated to the circle bundle $Y^{p,s}\to
S^2\times S^2$ is
\begin{align}
c_1(\scrL_{p,s})=p\,\omega_1+s\,\omega_2
\end{align}
where $[\omega_1]$ and $[\omega_2]$ generate the second cohomology
$H^2(S^2,\IZ)$ of the respective factors of the base. When $p$ and
$s$ are coprime, $Y^{p,s}$ is again topologically $S^3\times S^2$.

Field theories with $\mathcal{N}=1$ supersymmetry on these manifolds have
been studied in \cite{Qiu2013,Qiufact2013} via application of
the index theorem. Except for the simplest
case $S^3 \times S^2$, the Killing vector field $v$ does not have
closed orbits. Moreover, the Reeb vector field $\xi$ does not act
freely on the base space of the fibration, which is a warped product
$S^2 \rtimes S^2$. These manifolds have a toric action, and in fact
admit a free $U(1)$ action, which however is generated by a different
vector field from the Reeb vector field. Therefore the formalism of this paper does not apply.
In \cite{Qiu2013}, the way around this problem was to use the same
idea that we did, but in the opposite direction. The manifolds $Y^{p,s}$
can be obtained as a quotient $S^3 \times S^3 /U(1)$, with $U(1)$
acting freely on the six-dimensional manifold $S^3\times S^3$. One can
compute the one-loop determinants using the index of the twisted
Dolbeault complex in six dimensions, and then use the fact that there
is a free $U(1)$ action to push it down to $Y^{p,s}$. In the case of
the conifold $T^{1,1}$, the vacuum
moduli spaces of instantons have been described in this way by~\cite{Geipel:2016uij}.

In fact, the most natural way to look at these geometries is as
follows. Consider $\mathbb{C}^4$ as the direct product
$\mathbb{C}^2\times\IC^2$, with a sphere $S^3 \subset \mathbb{C}^2$
embedded in each factor in the standard way. There is a $U(1)^{\times 4}$ action on $\mathbb{C}^4$,
where each $U(1)_i$ acts on the corresponding factor of $\mathbb{C}$ in
$\mathbb{C}^4$, with associated equivariant parameter $\epsilon_{i} ^{-1}$, $i=1, \dots, 4$. 
There is a further $U(1)_T$ action on $\mathbb{C}^4$
with charges which are functions of two integer parameters $p$ and
$s$; explicitly, $U(1)_T$ acts with charges $(p+s,p-s,-p,-s)$. Then the
action of $U(1)_T$ is free on $S^3 \times S^3 \subset
\mathbb{C}^4$, and taking the quotient $S^3 \times S^3/U(1)_T$ gives
$Y^{p,s}$ with a residual toric action by $U(1)^{\times 3}$, but none of the
remaining $U(1)$ actions is free.
The one-loop determinants then appear as products over four integers $m_i$,
$i=1,\dots,4$, each one corresponding to an eigenvalue
$\frac{m_i}{\epsilon_i}$ where $\epsilon_i^{-1}$ is the equivariant
parameter for rotation in the $i$-th plane $\mathbb{C}$ in $\mathbb{C}^4$. These
four integers are constrained by one linear relation, corresponding
to the quotient by the freely acting $U(1)_{T}$, which reads
\begin{equation}
 (p+s)\,m_1 + (p-s)\, m_2 -p\, m_3 -s\, m_4 = 0 \ .
\end{equation}

\end{appendix}

\end{document}